\documentclass[12pt]{article}
\usepackage{euscript,amsmath, amssymb, amsfonts}

\input{tcilatex}
\begin{document}

\title{{\Large Self-adjoint Schr\"{o}dinger and Dirac operators with
Aharonov-Bohm and magnetic-solenoid fields}}
\author{ D.M. Gitman\thanks{%
Institute of Physics, University of Sao Paulo, Brazil; e-mail:
gitman@dfn.if.usp.br}, A. Smirnov, I.V. Tyutin\thanks{%
Lebedev Physical Institute, Moscow, Russia; e-mail: tyutin@lpi.ru}, and B.L.
Voronov\thanks{%
Lebedev Physical Institute, Moscow, Russia; e-mail: voronov@lpi.ru}}
\date{}
\maketitle

\begin{abstract}
We study all the s.a. Schr\"{o}dinger and Dirac operators (Hamiltonians)
both with pure AB field and with magnetic-solenoid field. Then, we perform a
complete spectral analysis for these operators, which includes finding
spectra and spectral decompositions, or inversion formulas. In constructing
the Hamiltonians and performing their spectral analysis, we respectively
follow the von Neumann theory of s.a. extensions of symmetric differential
operators and the Krein method of guiding functionals. The examples of
similar consideration are given by us in arXiv:0903.5277, where a
nonrelativistic particle in the Calogero potential field is considered and
in Theor. Math. Phys. \textbf{150}(1) (2007) 34, where a Dirac particle in
the Coulomb field of arbitrary charge is considered. However, due to
peculiarities of the three-dimensional problems under consideration, we
elaborated a generalization of the approach used in the study of the Dirac
particle.
\end{abstract}

\section{Introduction}

Aharonov-Bohm (AB) effect \cite{AhaBo59} plays an important role in quantum
theory refining the status of electromagnetic potentials in this theory.
First this effect was discussed in relation to a study of interaction
between a non-relativistic charged particle and an infinitely long and
infinitesimally thin magnetic solenoid field (further \emph{AB field}) which
yields a magnetic flux $\Phi $ (a similar effect was discussed earlier by
Ehrenberg and Siday \cite{EhrSi49}). It was discovered that particle wave
functions vanish at the solenoid line. In spite of the fact that the
magnetic field vanishes out of the solenoid, the phase shift in the wave
functions is proportional to the corresponding magnetic flux \cite{WuYa75}.
A non-trivial particle scattering by the solenoid is interpreted as a
possibility for quantum particles to \textquotedblright
feel\textquotedblright\ potentials of the corresponding electromagnetic
field. Indeed, potentials of AB field do not vanish out of the solenoid. For
the first time, a construction of self-adjoint (s.a. in what follows) Schr%
\"{o}dinger operators with the AB field was given in \cite{Tyuti74}. First,
the need for s.a. extensions of the Dirac Hamiltonian with the AB field in $%
2+1$ dimensions was recognized in \cite{GJ89,G89,Hagen91}. S.a. extensions
of the Dirac Hamiltonian with the AB field in $3+1$ dimensions were found in 
\cite{VGS91}, see also \cite{CP94,ACP01}. The physically motivated boundary
conditions for the particle scattering \ by the AB field and a Coulomb
center were studied in \cite{CP93,H93}. A splitting of Landau levels in a
superposition of parallel uniform magnetic field and AB field (further \emph{%
magnetic-solenoid field}) gives an example of AB effect for bound states.
First, exact solutions of Schr\"{o}dinger equation with the
magnetic-solenoid field (non-relativistic case) were studied in \cite%
{Lewis83}. Exact solutions of the relativistic wave equations (Klein-Gorgon
and Dirac) with the magnetic-solenoid field were obtained in \cite%
{183,192,181} and used then to study AB effect in cyclotron and synchrotron
radiations, see \cite{192,181,198}. Later on the problem of self-adjointness
of the Dirac Hamiltonian with magnetic-solenoid field was studied in \cite%
{GavGi04,212}.

In the present work, we construct systematically all the s.a. Schr\"{o}%
dinger and Dirac operators both with pure AB field and with
magnetic-solenoid field. Then, we perform a complete spectral analysis for
these Hamiltonians, which includes finding spectra and spectral
decompositions, or inversion formulas. In constructing the Hamiltonians and
performing their spectral analysis, we respectively follow the theory of
s.a. extensions of symmetric differential operators \cite%
{AkhGl81,Naima69,VorGiT06} and the Krein method of guiding functionals \cite%
{AkhGl81,Naima69}. The examples of similar consideration are given in \cite%
{GitTyV09} where a nonrelativistic particle in the Calogero potential field
is considered and in \cite{240} where a Dirac particle in the Coulomb field
of arbitrary charge is considered. However, due to peculiarities of the
three-dimensional problem under consideration, we use a necessary
generalization of the approach \cite{240}.

We recall that the AB field of infinitely thin solenoid (with the constant
flux $\Phi $) along the axis $z=x^{3}$ can be described by electromagnetic
potentials $A_{\mathrm{AB}}^{\mu }$,$\ \mu =0,1,2,3$,\ 
\begin{align*}
A_{\mathrm{AB}}^{\mu }& =\left( 0,\mathbf{A}_{\mathrm{AB}}\right) ,\ \mathbf{%
A}_{\mathrm{AB}}=\left( A_{\mathrm{AB}}^{k},\ k=1,2,3\right) ,\;A_{\mathrm{AB%
}}^{3}=0, \\
A_{\mathrm{AB}}^{1}& ={-\frac{{\Phi }\sin \varphi }{{2\pi \rho }}},\;A_{%
\mathrm{AB}}^{2}={\frac{{\Phi }\cos \varphi }{{2\pi \rho }}},
\end{align*}%
where $\rho ,\varphi $ are cylindrical coordinates, $x^{1}=\rho \cos \varphi
,\ x^{2}=\rho \sin \varphi $. The magnetic field of AB solenoid has the form 
$\mathbf{B}_{\mathrm{AB}}=(0,0,B_{\mathrm{AB}})$. It is easy to see that
outside the $z$ axis the magnetic field $\mathbf{B}_{\mathrm{AB}}=\mathrm{rot%
}\mathbf{A}_{\mathrm{AB}}$ is equal to zero. Nevertheless, for any surface $%
\Sigma $ with a boundary $L$ being any contour (even an infinitely small
one) around the $z$ axis, the circulation of the vector potential along $L$
does not vanish and reads $\oint_{L}\mathbf{A}_{\mathrm{AB}}d\mathbf{l=}\Phi
.$ If one interprets this circulation as the flux of the magnetic field $%
\mathbf{B}_{\mathrm{AB}}$ through the surface $\Sigma $,%
\begin{equation*}
\int_{\Sigma }\mathbf{B}_{\mathrm{AB}}d\mathbf{\sigma }=\oint_{L}\mathbf{A}_{%
\mathrm{AB}}d\mathbf{l}=\Phi ,
\end{equation*}%
then we obtain an expression for the magnetic field,%
\begin{equation*}
B_{\mathrm{AB}}=\Phi \delta (x^{1})\delta (x^{2})\,,
\end{equation*}%
where the term infinitely thin solenoid comes from.

In the cylindrical coordinates, we have%
\begin{equation*}
\frac{e}{c\hbar }A_{\mathrm{AB}}^{1}=-\phi \rho ^{-1}\sin \varphi \,,\;\frac{%
e}{c\hbar }A_{\mathrm{AB}}^{2}=\phi \rho ^{-1}\cos \varphi \,,\ \ \phi =\Phi
/\Phi _{0},
\end{equation*}%
where $\Phi _{0}$ is a fundamental unit of magnetic flux, 
\begin{equation*}
\Phi _{0}=2\pi c\hbar /e=4,135\times 10^{-7}\,\mathrm{Gauss}\cdot \mathrm{cm}%
^{2}
\end{equation*}%
(we recall that $e>0$ is the absolute value of the electron charge).

The magnetic-solenoid field is a superposition of a constant uniform
magnetic field of strength $B$ directed along the axis $z$ and the AB field
with the flux $\Phi$\ in the same direction. The magnetic-solenoid field is
given by electromagnetic potentials by potentials $A^{\mu}=(0,\mathbf{A})$,\ 
$\mathbf{A}=\left( A^{k},\ k=1,2,3\right) $ of the form%
\begin{equation}
A^{1}=A_{\mathrm{AB}}^{1}-\frac{Bx^{2}}{2},\;A^{2}=A_{\mathrm{AB}}^{2}+\frac{%
Bx^{1}}{2},\ A^{3}=0.  \label{9.1.9}
\end{equation}
The potentials (\ref{9.1.9}) define the magnetic field $\mathbf{B\;}$of the
form$\mathbf{\ }$ 
\begin{equation*}
\mathbf{B}=(0,0,B+B_{\mathrm{AB}}).
\end{equation*}
In the cylindrical coordinates, the potentials of the magnetic-solenoid
field\ have the form 
\begin{align}
& \frac{e}{c\hbar}A^{1}=-\tilde{\phi}\rho^{-1}\sin\varphi\,,\;\frac{e}{c\hbar%
}A^{2}=\tilde{\phi}\rho^{-1}\cos\varphi\,,\ A^{3}=0,  \notag \\
& \tilde{\phi}=\phi+\frac{\epsilon_{B}\gamma\rho^{2}}{2},\;\gamma={\frac {{e}%
|B{|}}{{c\hbar}}}>0\,,\ \epsilon_{B}=\mathrm{sign}\ B.  \label{9.1.11}
\end{align}

For further consideration, it is convenient to introduce the following
representation: 
\begin{equation}
\phi =\epsilon _{B}\left( \phi _{0}+\mu \right) ,\;\phi _{0}=\left[ \epsilon
_{B}\phi \right] \in \mathbb{Z},\ \mu =\epsilon _{B}\phi -\phi _{0},\ 0\leq
\mu <1.\   \label{9.1.7}
\end{equation}%
The quantity $\mu $ is called the mantissa of the magnetic flux and, in
fact, determines all the physical effects in the AB field, see e.g. \cite%
{181}.

\section{S.a. Schr\"{o}dinger Hamiltonians}

In this section, we consider two-dimensional and three-dimensional
nonrelativistic motions of a particle of mass $m_{e}\;$and charge $%
q=\epsilon _{q}e$,$\ \epsilon _{q}=\mathrm{sign}q=\pm 1$ ( positron or
electron) in the magnetic-solenoid field. The canonical formulation of the
problem is the following. The starting point is the \textquotedblleft formal
Schr\"{o}dinger Hamiltonian $\check{H}"\,$with the magnetic-solenoid field%
\emph{\ }that is respectively a two- or three-dimensional s.a. differential
operation well-known from physical textbooks. In three dimensions, it is
given by 
\begin{equation}
\check{H}=\frac{1}{2m_{e}}\left( \mathbf{\check{p}}-\frac{q}{c}\mathbf{A}%
\right) ^{2},\ \mathbf{\check{p}}=-i\hbar \mathbf{\nabla ,}\ \mathbf{\nabla }%
=\left( \partial _{x},\ \partial _{y},\ \partial _{z}\right) .
\label{9.1.20}
\end{equation}%
\ It is convenient to represent $\check{H}$ as a a sum of two terms, $\check{%
H}^{\bot }$ and $\check{H}^{||}$,%
\begin{equation*}
\check{H}=\check{H}^{\bot }+\check{H}^{||},\ 
\end{equation*}%
where the two-dimensional s.a. differential operation $\check{H}^{\bot }$,
the \textquotedblleft formal two-dimensional Schr\"{o}dinger
Hamiltonian\textquotedblright\ with the magnetic-solenoid field, 
\begin{align}
& \check{H}^{\bot }=M^{-1}\mathcal{\check{H}}^{\bot },\ \mathcal{\check{H}}%
^{\bot }=\left( -i\mathbf{\nabla }^{\bot }-\frac{q}{c\hbar }\mathbf{A}^{\bot
}\right) ^{2},\   \notag \\
& M=2m_{e}\hbar ^{-2},\ \ \mathbf{\nabla }^{\bot }=\left( \partial _{x},\
\partial _{y}\right) ,\ \ \mathbf{A}^{\bot }=\left( A^{1},A^{2}\right) ,
\label{9.1.22}
\end{align}%
$A^{1}$ and $A^{2}$ are given by (\ref{9.1.11}), corresponds to a
two-dimensional motion in the $xy$ plane perpendicular to the $z$ axis,
while the one-dimensional differential operation $\check{H}^{||}$, 
\begin{equation*}
\check{H}^{||}=\mathcal{\check{H}}=\frac{\check{p}_{z}^{2}}{2m_{e}},\ \ 
\check{p}_{z}=-i\hbar \partial _{z},
\end{equation*}%
corresponds to a one-dimensional free motion along the z-axis.

The problem to be solved\textrm{\ }is to construct s.a. nonrelativistic two-
and three-dimensional Hamiltonians $\hat{H}^{\perp}$ and $\hat{H}$
associated with the respective s.a. differential operations $\check{H}%
^{\bot} $ and $\check{H}$ and to perform a complete spectral analysis for
these operators.

We begin with the two-dimensional problem. We successively consider the case
of pure AB field, with $B=0$, and then the case of the magnetic-solenoid
field, with $B\neq0$ In the subsequent subsection, we generalize obtained
results to three-dimensions.

\subsection{Two-dimensional case}

\subsubsection{Reduction to radial problem}

In the case of two dimensions, the space of particle quantum states is the
Hilbert space $\mathfrak{H}=L^{2}\left( \mathbb{R}^{2}\right) $ of
square-integrable functions $\psi(\mathbf{\rho}),\ \mathbf{\rho}=(x,y)$,
with the scalar product%
\begin{equation*}
(\psi_{1},\psi_{2}) =\int\overline{\psi_{1}(\mathbf{\rho})}\psi_{2}(\mathbf{%
\rho})d\mathbf{\rho},\,d\mathbf{\rho}=dxdy=\rho d\rho d\varphi.
\end{equation*}

A quantum Hamiltonian $\hat{H}^{\perp}$ should be defined as a s.a. operator
in this Hilbert space. It is \ more convenient to deal with a s.a. operator $%
\mathcal{\,}\widehat{\mathcal{H}}^{\bot}=M\,\hat{H}^{\perp}$ associated with
s.a. differential operation $\mathcal{\check{H}}^{\bot}=M\,\check{H}^{\perp}$
defined in (\ref{9.1.22}).

The construction is essentially based on the requirement of rotation
symmetry which certainly holds in a classical description of the system.
This requirement is formulated as the requirement of the invariance of a
s.a. Hamiltonian under rotations around the solenoid line, the $z$ axis. As
in classical mechanics, the rotation symmetry allows separating the polar
coordinates $\rho$ and $\varphi$ and reducing the two-dimensional problem to
a one-dimensional radial problem.

The group of rotations $SO(2)$ in $\mathbb{R}^{2}$ naturally acts in the
Hilbert space $\mathfrak{H}$ by unitary operators: if $S\in SO(2)$, then the
corresponding operator $\hat{U}_{S}$ is defined by the relation $(\hat{U}%
_{S}\psi)(\mathbf{\rho})=\psi(S^{-1}\mathbf{\rho})$, $\psi\in\mathfrak{H}$.

The Hilbert space $\mathfrak{H}$ is a direct orthogonal sum of subspaces $%
\mathfrak{H}_{m}$, that are the eigenspaces of the representation $\hat
U_{S} $, 
\begin{equation*}
\mathfrak{H}=\sideset{}{^{\,\lower1mm\hbox{$\oplus$}}} \sum_{m\in\mathbb{Z}}%
\mathfrak{H}_{m},\;\hat U_{S}\mathfrak{H}_{m}=e^{-im\theta}\mathfrak{H}_{m},
\end{equation*}
where $\theta$ is the rotation angle corresponding to $S$.

It is convenient to change indexing, $\ m\rightarrow l$, $\mathfrak{H}%
_{m}\rightarrow\mathfrak{H}_{l}$, as follows%
\begin{equation*}
m=\epsilon\left( \phi_{0}-l\right) ,\ l=\phi_{0}-\epsilon m,
\end{equation*}
where%
\begin{equation}
\epsilon=\epsilon_{q}\epsilon_{B},\ \phi_{0}=[\epsilon_{B}\phi].
\label{9.1.24}
\end{equation}
\emph{\ }

We define a rotationally-invariant initial symmetric operator $\widehat {%
\mathcal{H}}^{\bot}$ associated with the differential operation $\mathcal{%
\check{H}}^{\bot}$\ in the Hilbert space $\mathfrak{H}=L^{2}(\mathbb{R}^{2})$
as follows:%
\begin{equation*}
\widehat{\mathcal{H}}^{\bot}:\left\{ 
\begin{array}{l}
D_{\mathcal{H}^{\bot}}=\{\psi(\mathbf{\rho}):\ \psi\in\mathcal{D}(\mathbb{R}%
^{2}\backslash\{0\})\} \\ 
\widehat{\mathcal{H}}^{\bot}\psi=\mathcal{\check{H}}^{\bot}\psi,\ \forall
\psi\in D_{\mathcal{H}^{\bot}}\ 
\end{array}
\right. ,
\end{equation*}
where $\mathcal{D}(\mathbb{R}^{2}\backslash\{0\})$ is the space of smooth
and compactly supported functions vanishing in a neighborhood of the point $%
\mathbf{\rho=}0$. The domain $D_{\mathcal{H}^{\bot}}$ is dense in $\mathfrak{%
H}$ and the symmetricity of $\widehat{\mathcal{H}}^{\bot}$ is obvious.

In the polar coordinates $\rho$ and $\varphi$, $\mathcal{\check{H}}^{\bot}$
becomes%
\begin{equation}
\mathcal{\check{H}}^{\bot}=-\partial_{\rho}^{2}-\rho^{-1}\partial_{\rho}+%
\rho^{-2}(i\partial_{\varphi}+\epsilon_{q}\tilde{\phi})^{2},  \label{9.1.23}
\end{equation}
where $\tilde{\phi}$ is given by (\ref{9.1.11}).

For every $l$, the relation 
\begin{equation}
(S_{l}f)(\rho ,\varphi )=\frac{1}{\sqrt{2\pi \rho }}\mathrm{e}^{i\epsilon
\left( \phi _{0}-l\right) \varphi }f_{l}\left( \rho \right)  \label{9.1.23b}
\end{equation}%
determines a unitary operator $S_{l}\colon L^{2}(\mathbb{R}_{+})\rightarrow 
\mathfrak{H}_{l}$, where $L^{2}(\mathbb{R}_{+})$ is the Hilbert space of
square-integrable functions on the semi-axis $\mathbb{R}_{+}$ with scalar
product 
\begin{equation*}
\left( f,g\right) =\int_{\mathbb{R}_{+}}\overline{f\left( \rho \right) }%
g\left( \rho \right) d\rho .\ 
\end{equation*}

For every $l$, we define the linear operator $V_{l}$ from $\mathfrak{H}$ to $%
L^{2}(\mathbb{R}_{+})$ by setting 
\begin{equation}
(V_{l}\psi )(\rho )=\frac{\sqrt{\rho }}{\sqrt{2\pi }}\int_{0}^{2\pi }\psi
(\rho ,\varphi )\mathrm{e}^{-i\epsilon (\phi _{0}-l)\varphi }\,\mathrm{d}%
\varphi .  \label{operator_V}
\end{equation}%
If $\psi =\sum_{l\in \mathbb{Z}}\psi _{l}\in \mathfrak{H}$, then we have $%
\psi _{l}=S_{l}V_{l}\psi $ for all $l$. In other words, $%
V_{l}=S_{l}^{-1}P_{l}$, where $P_{l}$ is the orthogonal projector onto the
subspace $\mathfrak{H}_{l}$. However, we prefer to work with $V_{l}$ rather
than $P_{l}$ because the latter cannot be reasonably defined in the
three-dimensional case, where the Hilbert state space should be decomposed
into a direct integral instead of a direct sum (see below). Clearly, $%
V_{l}\psi \in \mathcal{D}(\mathbb{R}_{+})$ for any $\psi \in \mathcal{D}(%
\mathbb{R}^{2}\setminus \{0\})$, and it follows from~(\ref{9.1.23}) and~(\ref%
{operator_V}) that 
\begin{equation}
V_{l}\widehat{\mathcal{H}}^{\bot }\psi =\hat{h}(l)V_{l}\psi ,\;\psi \in 
\mathcal{D}(\mathbb{R}^{2}\setminus \{0\}),  \label{vlhl}
\end{equation}%
where the symmetric operators $\hat{h}(l)$ in $L^{2}(\mathbb{R}_{+})$
defined on $D_{h(l)}=\mathcal{D}(\mathbb{R}_{+}),$ where it acts as%
\begin{equation}
\check{h}(l)=-\partial _{\rho }^{2}+\rho ^{-2}\left[ \left( l+\mu +\gamma
\rho ^{2}/2\right) ^{2}-1/4\right] ,  \label{9.2.1}
\end{equation}%
with $\mu =\epsilon _{B}\phi -\phi _{0}$ defined in (\ref{9.1.7}).

In view of~(\ref{vlhl}), for any $\psi\in\mathcal{D}(\mathbb{R}%
^{2}\setminus\{0\})$, the $\mathfrak{H}_{l}$-component $(\widehat{\mathcal{H}%
}^{\bot}\psi)_{l}$ of $\widehat{\mathcal{H}}^{\bot}\psi$ can be written as 
\begin{equation}
(\widehat{\mathcal{H}}^{\bot}\psi)_{l}=S_{l}V_{l}\widehat{\mathcal{H}}^{\bot
}\psi=S_{l}\hat{h}(l)S_{l}^{-1}S_{l}V_{l}\psi=S_{l}\hat{h}%
(l)S_{l}^{-1}\psi_{l}.  \label{component}
\end{equation}

Suppose we have a (not necessarily closed) operator $\hat{f}_{l}$ in $%
\mathfrak{H}_{l}$ for each $l$. We define the operator 
\begin{equation}
\hat{f}=\sideset{}{^{\,\lower1mm\hbox{$\oplus$}}} \sum_{l\in\mathbb{Z}}\hat {%
f}_{l}  \label{dirsum}
\end{equation}
in $\mathfrak{H}$ by setting 
\begin{equation*}
\hat{f}\psi=\sum_{l\in\mathbb{Z}}\hat{f}_{l}\psi_{l},\;\psi=\sum _{l\in%
\mathbb{Z}}\psi_{l}.
\end{equation*}
The domain $D_{f}$ of $\hat{f}$ consists of all $\psi=\sum_{l\in\mathbb{Z}%
}\psi_{l}\in\mathfrak{H}$ such that $\psi_{l}\in D_{f_{l}}$ for all $l$ and
the series $\sum_{l\in\mathbb{Z}}\hat{f}_{l}\psi_{l}$ converges in $%
\mathfrak{H}$. The operator $\hat{f}$ is closed (self-adjoint) if and only
if all $\hat{f}_{l}$ are closed (resp., self-adjoint). For every $l$, we
have $D_{f_{l}}=D_{f}\cap\mathfrak{H}_{l}$.

We say that a closed operator $\hat{f}$ in $\mathfrak{H}$ is rotationally
invariant if it can be represented in form~(\ref{dirsum}) for some family of
operators $\hat f_{l}$ in $\mathfrak{H}_{l}$.

By~(\ref{component}), the direct sum of the operators $S_{l}\hat{h}%
(l)S_{l}^{-1}$ is an extension of $\widehat{\mathcal{H}}^{\bot}$: 
\begin{equation}
\widehat{\mathcal{H}}^{\bot}\subset\sideset{}{^{\,\lower1mm\hbox{$\oplus$}}}
\sum_{l\in\mathbb{Z}}S_{l}\hat{h}(l)S_{l}^{-1}.  \label{extensionH}
\end{equation}
Let $\hat{h}_{\mathfrak{e}}(l)$ be s.a. extensions of the symmetric
operators $\hat{h}(l)$. Then the operators 
\begin{equation}
\widehat{\mathcal{H}}_{\mathfrak{e}}^{\bot}(l)=S_{l}\hat{h}_{\mathfrak{e}%
}(l)S_{l}^{-1}  \label{R.6}
\end{equation}
are s.a. extensions of $S_{l}\hat{h}(l)S_{l}^{-1}$, and it follows from~(\ref%
{extensionH}) that the orthogonal direct sum 
\begin{equation}
\widehat{\mathcal{H}}_{\mathfrak{e}}^{\perp}=\sideset{}{^{\,\lower1mm\hbox{$%
\oplus$}}} \sum_{l\in\mathbb{Z}}\widehat{\mathcal{H}}_{\mathfrak{e}%
}^{\bot}(l)\ ,  \label{9R.7}
\end{equation}
is a rotationally invariant s.a. extension of the initial operator $\widehat{%
\mathcal{H}}^{\bot}$.

Conversely, let $\widehat{\mathcal{H}}_{\mathfrak{e}}^{\perp}$ be a
rotationally invariant s.a. extension of $\widehat{\mathcal{H}}^{\bot}$.
Then it has the form~(\ref{9R.7}), where $\widehat{\mathcal{H}}_{\mathfrak{e}%
}^{\bot}(l)$ are s.a. operators in $\mathfrak{H}_{l}$. Set $\hat {h}_{%
\mathfrak{e}}(l)=S_{l}^{-1} \widehat{\mathcal{H}}_{\mathfrak{e}}^{\bot }(l)
S_{l}$. For all $l$, $\hat{h}_{\mathfrak{e}}(l)$ are s.a. operators in $%
L^{2}(\mathbb{R}_{+})$. If $f\in\mathcal{D}(_{+})$, then $S_{l} f\in\mathcal{%
D}(\mathbb{R}^{2}\setminus\{0\})\cap\mathfrak{H}_{l}$ and~(\ref{extensionH})
and~(\ref{9R.7}) imply that 
\begin{equation*}
S_{l} \hat{h}(l) f = S_{l} \hat{h}(l) S_{l}^{-1} S_{l} f =\widehat {\mathcal{%
H}}^{\bot} S_{l} f = \widehat{\mathcal{H}}_{\mathfrak{e}}^{\perp }S_{l} f= 
\widehat{\mathcal{H}}_{\mathfrak{e}}^{\bot}(l)S_{l} f = S_{l} \hat{h}_{%
\mathfrak{e}}(l) f.
\end{equation*}
Hence, $\hat{h}(l) f = \hat{h}_{\mathfrak{e}}(l) f$, i.e., $\hat {h}_{%
\mathfrak{e}}(l)$ is a s.a. extension of $\hat{h}(l)$. We thus conclude that 
$\widehat{\mathcal{H}}_{\mathfrak{e}}^{\perp}$ can be represented in form~(%
\ref{9R.7}), where $\widehat{\mathcal{H}}_{\mathfrak{e}}^{\bot}(l)$ are
given by~(\ref{R.6}) and $\hat{h}_{\mathfrak{e}}(l)$ are a s.a. extensions
of $\hat h(l)$.

The problem of constructing a rotationally invariant s.a. Hamiltonian $%
\widehat{\mathcal{H}}_{\mathfrak{e}}^{\perp}$ is thus reduced to
constructing s.a. radial Hamiltonians $\hat{h}_{\mathfrak{e}}\left( l\right) 
$.

We first consider the case of pure AB field where $B=0.$

\subsubsection{S.a. Hamiltonians with AB field}

In this case, we have $\gamma =0$, and s.a. radial differential operation $%
\check{h}(l)$ (\ref{9.2.1}) becomes%
\begin{equation}
\check{h}\left( l\right) =-\partial _{\rho }^{2}+\alpha \rho ^{-2},\ \
\alpha =\varkappa _{l}^{2}-1/4,\ \ \varkappa _{l}=|l+\mu |,\ l\in \mathbb{Z}.
\label{9.1.??}
\end{equation}%
It is easy to see that this differential operation and the corresponding
initial symmetric operator $\hat{h}\left( l\right) $ are \ actually
identical to the respective operation and operator encountered in studying
the Calogero problem, see \cite{GitTyV09}. We can therefore directly carry
over the previously obtained results to s.a. extensions of $\hat{h}\left(
l\right) $.

\paragraph{First region: $\protect\alpha\geq3/4,$}

In this region, we have $\left( l+\mu\right) ^{2}\geq1$, which is equivalent
to 
\begin{equation*}
l\geq1-\mu\ \mathrm{or}\ l\leq-1-\mu.
\end{equation*}
Because $l\in\mathbb{Z}$ and $0\leq\mu<1$, we has to distinguish the cases
of $\mu=0$ and $\mu>0$:%
\begin{align*}
\mu & =0:l\leq-1\ \mathrm{or}\ l\geq1,\ \mathrm{i.e.},\ l\neq0, \\
\mu & >0:l\leq-2\ \mathrm{or}\ l\geq1,\ \mathrm{i.e.},\,l\neq0,-1.
\end{align*}

For such $l$, the initial symmetric operator $\hat{h}\left( l\right) $ has
zero deficiency indices, is essentially s.a., and its unique s.a. extension
is $\hat{h}_{\mathfrak{e}}\left( l\right) =\hat{h}_{\left( 1\right) }\left(
l\right) =\hat{h}^{+}\left( l\right) $ with the domain%
\begin{equation*}
D_{\hat{h}\left( l\right) }=\{\psi _{\ast }:\psi _{\ast },\psi _{\ast
}^{\prime }\;\mathrm{are\;absolutely\;continuous\;(a.c.)\;in}\;\mathbb{R}%
_{+},\;\psi _{\ast },\check{h}\left( l\right) \psi _{\ast }\in L^{2}(\mathbb{%
R}_{+})\}.
\end{equation*}
The spectrum of $\hat{h}_{\left( 1\right) }\left( l\right) $ is simple and
continuous and coincides with the positive semiaxis, $\mathrm{spec}\hat{h}%
_{\left( 1\right) }\left( l\right) =\mathbb{R}_{+}$.

The generalized eigenfunctions $U_{\mathcal{E}}$, 
\begin{equation*}
U_{\mathcal{E}}(\rho )=\left( \rho /2\right) ^{1/2}J_{\varkappa _{l}}(\sqrt{%
\mathcal{E}}\rho ),\ \ \hat{h}_{(1)}\left( l\right) U_{\mathcal{E}}=\mathcal{%
E}U_{\mathcal{E}},\ \ \mathcal{E}\in \mathbb{R}_{+},
\end{equation*}%
of $\hat{h}_{(1)}\left( l\right) $ form a complete orthonormalized set in
each Hilbert space $\mathcal{L}_{l}.$

\paragraph{Second region: $-1/4<\protect\alpha<3/4$}

In this region, we have $0<\left( l+\mu\right) ^{2}<1$, which is equivalent
to 
\begin{equation}
-\mu<l<1-\mu\ \mathrm{or\ }-1-\mu<l<-\mu\ .  \label{9.1.42a}
\end{equation}

If $\mu=0,$ inequalities (\ref{9.1.42a}) have no solutions for $l\in \mathbb{%
Z}.$ If $\mu>0$, these inequalities have two solutions $l=l_{a}$, where, for
brevity, we introduce the notation%
\begin{equation*}
l_{a}=a,\ a=0,-1.
\end{equation*}
So, in the second region, we remain with the case of $\mu>0$..

For each $l=l_{a}$\ ($a=0,-1)$ there exists a one-parameter $U\left(
1\right) $-family of s.a. Hamiltonians $\hat{h}_{\mathfrak{e}}\left(
l_{a}\right) =\hat{h}_{\lambda _{a}}\left( l_{a}\right) $ parametrized by
the real parameter $\lambda _{a}\in \mathbb{S}\left( -\pi /2,\pi /2\right) $%
\textrm{, }where $\mathbb{S}\left( -\pi /2,\pi /2\right) $ denotes an
interval with identified ends (a circle). These Hamiltonians are specified
by\ the asymptotic s.a. boundary conditions at the origin,%
\begin{align}
& \psi _{\lambda _{a}}(\rho )=c\left[ (\kappa _{0}\rho )^{1/2+\varkappa
_{a}}\cos \lambda _{a}+(\kappa _{0}\rho )^{1/2-\varkappa _{a}}\sin \lambda
_{a}\right] +O(\rho ^{3/2}),\mathrm{\,}\rho \rightarrow 0,  \label{9.1.43} \\
& D_{h_{\lambda _{a}}\left( l_{a}\right) }=\{\psi \in D_{\check{h}\left(
l_{a}\right) }^{\ast }\left( \mathbb{R}_{+}\right) ,\ \psi \text{ }\mathrm{%
satisfies\ }\text{(\ref{9.1.43})}\}.  \label{9.1.43a}
\end{align}%
where $\varkappa _{a}\equiv \varkappa _{l_{a}}=|\mu +a|$, $0<\varkappa
_{a}<1 $,$\ $and $c$ is an arbitrary constant, whereas $k_{0}$ is a constant
of dimension of inverse length.

For $\lambda_{a}\not \in (-\pi/2,0)$, the spectrum of each of $\hat {h}%
_{\lambda_{a}}\left( l_{a}\right) $ is simple and continuous and \textrm{spec%
}$\hat{h}_{\lambda_{a}}\left( l_{a}\right) =\mathbb{R}_{+}.$

The generalized eigenfunctions $U_{\mathcal{E}}$,%
\begin{align}
& U_{\mathcal{E}}(\rho )=\sqrt{\frac{\rho }{2Q_{a}}}\left[ J_{\varkappa
_{a}}\left( \sqrt{\mathcal{E}}\rho \right) +\tilde{\lambda}_{a}\left( \sqrt{%
\mathcal{E}}/2\kappa _{0}\right) ^{2\varkappa _{a}}J_{-\varkappa _{a}}\left( 
\sqrt{\mathcal{E}}\rho \right) \right] ,  \notag \\
& Q_{a}=1+2\tilde{\lambda}_{a}\left( \mathcal{E}/4\right) ^{\varkappa
_{a}}\cos (\pi \varkappa _{a})+\left( \tilde{\lambda}_{a}\right) ^{2}\left( 
\mathcal{E}/4\right) ^{2\varkappa _{a}}>0,  \notag \\
& \tilde{\lambda}_{a}=\Gamma (1-\varkappa _{a})\Gamma ^{-1}(1+\varkappa
_{a})\tan \lambda _{a}\ ,  \notag \\
& \check{h}_{\lambda _{a}}\left( l_{a}\right) U_{\mathcal{E}}=\mathcal{E}U_{%
\mathcal{E}}\ ,\ \ \mathcal{E}\in \mathbb{R}_{+},  \label{9.1.44}
\end{align}%
of the Hamiltonian $\hat{h}_{\lambda _{a}}\left( l_{a}\right) $ form a
complete orthonormalized set in each Hilbert space $\mathcal{L}_{l_{a}}.$

For $\lambda_{a}\in(-\pi/2,0)$, the spectrum of each of $\hat{h}%
_{\lambda_{a}}\left( l_{a}\right) $ is simple, but in addition to the
continuous part of the spectrum, there exists one negative level $\mathcal{E}%
_{\lambda_{a}}^{\left( -\right) }=-4k_{0}^{2}|\tilde{\lambda}%
_{a}|^{-\varkappa_{a}^{-1}}$, such that \textrm{spec}$\hat{h}%
_{\lambda_{a}}\left( l_{a}\right) =\mathbb{R}_{+}\cup\{\mathcal{E}%
_{\lambda_{a}}^{\left( -\right) }\}$.

The generalized eigenfunctions $U_{\mathcal{E}}$ of the continuous spectrum, 
$\mathcal{E}\geq0$, are given by the same (\ref{9.1.44}), while the
eigenfunction $U^{\left( -\right) }$ corresponding to the discrete level is%
\begin{equation*}
U^{\left( -\right) }(\rho)=\sqrt{\frac{2\rho\left| \mathcal{E}_{\lambda
_{a}}^{\left( -\right) }\right| \sin(\pi\varkappa_{a})}{\pi\varkappa_{a}}}%
K_{\varkappa_{a}}\left( \sqrt{\left| \mathcal{E}_{\lambda_{a}}^{\left(
-\right) }\right| }\rho\right) \ ,\ \ \hat{h}_{\lambda_{a}}\left(
l_{a}\right) U^{\left( -\right) }=\mathcal{E}_{\lambda_{a}}^{\left( -\right)
}U^{\left( -\right) }\ ,
\end{equation*}
they together form a complete orthonormalized set in each Hilbert space $%
\mathcal{L}_{l_{a}}.$

\paragraph{Third region: $\protect\alpha=-1/4$}

In this region, we have $l+\mu=0$. If $\mu=0$, this equation has a unique
solution $l=l_{0}=0$, while if $\mu>0,$ there are no solutions, and we
remain with the only case of $\mu=0$.

For $l=l_{0}$, there exists a one-parameter $U\left( 1\right) $-family of
s.a. Hamiltonians $\hat{h}_{\mathfrak{e}}\left( l_{0}\right) =\hat {h}%
_{\lambda}\left( l_{0}\right) $.parametrized by the real parameter $%
\lambda\in\mathbb{S}\left( -\pi/2,\pi/2\right) $) These Hamiltonians are
specified by the asymptotic s.a. boundary conditions at the origin%
\begin{equation}
\psi_{\lambda}(\rho)=c\left[ \rho^{1/2}\ln\left( \kappa_{0}\rho\right)
\cos\lambda+\rho^{1/2}\sin\lambda\right] +O(\rho^{3/2}\ln\rho),\ \rho
\rightarrow0,  \label{9.1.46}
\end{equation}
($\ $the constants $c$ and $k_{0}$ are of the same meaning as in (\ref%
{9.1.43})), and their domains are given by%
\begin{align}
D_{h_{\lambda}\left( l_{0}\right) } & =\{\psi:\ \psi\in D_{\check {h}\left(
l_{0}\right) }^{\ast}\left( \mathbb{R}_{+}\right) ,\ \psi\text{ }\mathrm{%
satisfies\ }\text{(\ref{9.1.46})}\},  \label{9.1.46a} \\
D_{\check{h}\left( l_{0}\right) }^{\ast}\left( \mathbb{R}_{+}\right) &
=\{\psi_{\ast}:\psi_{\ast},\psi_{\ast}^{\prime}\;\mathrm{are\;a.c.\;in}\;%
\mathbb{R}_{+},\;\psi_{\ast},\check{h}\left( l_{0}\right) \psi_{\ast}\in
L^{2}(\mathbb{R}_{+})\}.  \notag
\end{align}

The spectrum of $\hat{h}_{\lambda}\left( l_{0}\right) $ is simple. For $%
|\lambda|=\pi/2$, the spectrum is continuous and nonnegative, $\mathrm{spec}%
\hat{h}_{\pm\pi/2}\left( l_{0}\right) =\mathbb{R}_{+}$. For $|\lambda
|<\pi/2,$ in addition to the continuous part of the spectrum, $\mathcal{E}%
\geq0$, there exists one negative level $\mathcal{E}_{\lambda}^{\left(
-\right) }=-4\kappa_{0}^{2}\exp\left[ 2(\tan\lambda-\mathbf{C})\right] $,
where $\mathbf{C}$ is the Euler constant, such that%
\begin{equation*}
\mathrm{spec}\hat{h}_{\lambda}\left( l_{0}\right) =\left\{ \mathcal{E}%
_{\lambda}^{\left( -\right) }\right\} \cup\mathbb{R}_{+},\ |\lambda |<\pi/2.
\end{equation*}

The generalized and normalized eigenfunctions $U_{\mathcal{E}}$ of the
continuous spectrum are%
\begin{align*}
& U_{\mathcal{E}}(\rho )=\sqrt{\frac{\rho }{2\left( \tilde{\lambda}^{2}+\pi
^{2}/4\right) }}\left[ \tilde{\lambda}J_{0}\left( \sqrt{\mathcal{E}}\rho
\right) +\frac{\pi }{2}N_{0}\left( \sqrt{\mathcal{E}}\rho \right) \right] ,
\\
& \tilde{\lambda}=\tan \lambda -\mathbf{C}-\ln \left( \sqrt{\mathcal{E}}%
/2\kappa _{0}\right) ,\ \ \hat{h}_{\lambda }\left( l_{0}\right) U_{\mathcal{E%
}}=\mathcal{E}U_{\mathcal{E}}\ ,\ \ \mathcal{E}\in \mathbb{R}_{+},\ |\lambda
|\leq \pi /2,
\end{align*}%
while the normalized eigenfunction $U^{\left( -\right) }$ corresponding to
the discrete level is%
\begin{equation*}
U^{\left( -\right) }(\rho )=\sqrt{2\rho |\mathcal{E}_{\lambda }^{\left(
-\right) }|}K_{0}\left( \sqrt{|\mathcal{E}_{\lambda }^{\left( -\right) }|}%
\rho \right) ,\ \ \hat{h}_{\lambda }\left( l_{0}\right) U^{\left( -\right) }=%
\mathcal{E}_{\lambda }^{\left( -\right) }U^{\left( -\right) }\ ,\ |\lambda
|<\pi /2,
\end{equation*}%
they together form a complete orthonormalized set in the Hilbert space $%
\mathcal{L}_{l_{0}}.$

\paragraph{Complete spectrum and inversion formulas}

In the previous subsubsecs., we constructed all s.a. radial Hamiltonians $%
\hat{h}_{\mathfrak{e}}\left( l\right) $ associated with the s.a.
differential operation $\check{h}\left( l\right) $ as s.a. extensions of the
symmetric operator $\hat{h}\left( l\right) $ for any $l$ $\in \mathbb{Z\,\ \ 
}$and for any any $\phi_{0}$ and $\mu$. We assemble our previous results
into two groups.

For $\mu=0$, we have%
\begin{align}
& \hat{h}_{\mathfrak{e}}\left( l\right) =\hat{h}_{\left( 1\right) }\left(
l\right) ,\ \ l\neq l_{0},\ \ D_{h_{\left( 1\right) }\left( l\right) }=D_{%
\check{h}\left( l\right) }^{\ast}\left( \mathbb{R}_{+}\right) ,  \notag \\
& \hat{h}_{\mathfrak{e}}\left( l_{0}\right) =\hat{h}_{\lambda}\left(
l_{0}\right) \ \ \lambda\in\mathbb{S}\left( -\pi/2,\pi/2\right) \ ,
\label{9.1.47}
\end{align}
the domain $D_{h_{\lambda}\left( l_{0}\right) }$\ is given by (\ref{9.1.46a}%
);

For $\mu>0$, we have%
\begin{align}
& \hat{h}_{\mathfrak{e}}\left( l\right) =\hat{h}_{\left( 1\right) }\left(
l\right) ,\ l\neq l_{a}=a=0,-1,\ D_{h_{\left( 1\right) }\left( l\right) }=D_{%
\check{h}\left( l\right) }^{\ast}\left( \mathbb{R}_{+}\right) ,  \notag \\
& \hat{h}_{\mathfrak{e}}\left( l_{a}\right) =\hat{h}_{\lambda_{a}}\left(
l_{a}\right) ,\ \lambda_{a}\in\mathbb{S}\left( -\pi/2,\pi/2\right) \ ,
\label{9.1.48}
\end{align}
the domain $D_{h_{\lambda_{a}}\left( l_{a}\right) }$ is given by (\ref%
{9.1.43a}).

As a final result, we find a family of all s.a. rotationally-invariant
two-dimensional nonrelativistic Hamiltonians $\hat{H}_{\mathfrak{e}}^{\bot}=$
$M^{-1}\widehat{\mathcal{H}}_{\mathfrak{e}}^{\perp}$ associated with the
s.a. differential operation $\check{H}^{\bot}$ (\ref{9.1.22}) with $B=0$.
Each set of possible s.a. radial Hamiltonians $\hat{h}_{\mathfrak{e}}\left(
l\right) $ generates a s.a. rotationally-invariant Hamiltonian $\hat{H}_{%
\mathfrak{e}}^{\bot}$ in accordance with the relations (\ref{R.6}) and (\ref%
{9R.7}).

When presenting the spectrum and inversion formulas for $\hat{H}_{\mathfrak{e%
}}^{\bot }$, we also consider the case of $\mu =0$ and the case of $\mu >0$
separately . We let $E$ denote the spectrum points of $\hat{H}_{\mathfrak{e}%
}^{\bot }$ and let $\Psi _{E\text{ }}$ denote the corresponding
(generalized) eigenfunctions. The spectrum points of the operators $\hat{h}_{%
\mathfrak{e}}\left( l\right) $ and $\hat{H}_{\mathfrak{e}}^{\bot }$ are
evidently related by $\mathcal{E}=M\,E$. Therefore, when writing formulas
for eigenfunctions $\Psi _{E}$ of the operator $\hat{H}_{\mathfrak{e}}^{\bot
}$ in terms of eigenfunctions $U_{\mathcal{E}}$ of the operators $\hat{h}_{%
\mathfrak{e}}(l)$, we has to introduce the factor $1/\sqrt{2\pi \rho }$ $%
\mathrm{e}^{i\epsilon _{q}\left( \phi _{0}-l\right) \varphi }$ in accordance
with eq.(\ref{9.1.23b}) with $\epsilon =\epsilon _{q}$ (because $\epsilon
_{B}=1$), to make the substitutions $\mathcal{E}=ME$ and $\mathcal{E}%
_{\lambda _{a}}^{\left( -\right) }=ME_{\lambda _{a}}^{\left( -\right) }$, $%
\mathcal{E}_{\lambda }^{\left( -\right) }=ME_{\lambda }^{\left( -\right) }$
for the respective points of the continuous spectrum and discrete spectrum,
and, in addition, to multiply eigenfunctions of the continuous spectrum of
the operators $\hat{h}_{\mathfrak{e}}\left( l\right) $ by the factor $\sqrt{M%
}$ because of the change of the spectral measure $d\mathcal{E}$ to the
corresponding spectral measure\footnote{%
From the physical standpoint, the latter is related to the change of the
\textquotedblleft normalization of the eigenfunctions of the continuous
spectrum to $\delta $ function\textquotedblright\ from $\delta (\mathcal{E}-%
\mathcal{E}^{\prime })$ to $\delta (E-E^{\prime })$.} $dE$.

For $\mu=0$, there is a family of s.a. two-dimensional nonrelativistic
Hamiltonians $\hat{H}_{\mathfrak{e}}^{\bot}$ parametrized by the real
parameter $\lambda\in\mathbb{S}\left( -\pi/2,\pi/2\right) ,\,\hat {H}_{%
\mathfrak{e}}^{\bot}=\hat{H}_{\lambda}^{\bot}$,%
\begin{align*}
& \hat{H}_{\lambda}^{\bot}=\sideset{}{^{\,\lower1mm\hbox{$\oplus$}}}
\sum_{l\in\mathbb{Z},l\neq l_{0}}\hat{H}^{\bot}\left( l\right) \oplus\hat {H}%
_{\lambda}^{\bot}\left( l_{0}\right) , \\
& \hat{H}^{\bot}\left( l\right) =M^{-1}S_{l}\hat{h}_{(1)}(l)S_{l}^{-1},\,l%
\neq l_{0}, \\
& \hat{H}_{\lambda}^{\bot}\left( l_{0}\right) =M^{-1}S_{l_{0}}\hat {h}%
_{\lambda}(l_{0})S_{l_{0}}^{-1}.
\end{align*}

The spectrum of $\hat{H}_{\lambda}^{\bot}$ is given by%
\begin{equation*}
\mathrm{spec}\hat{H}_{\lambda}^{\bot}=\left\{ 
\begin{array}{c}
E_{\lambda}^{\left( -\right) }=-4M^{-1}\kappa_{0}^{2}\exp\left[
2(\tan\lambda-\mathbf{C})\right] ,\ |\lambda|<\pi/2 \\ 
\varnothing,\ \lambda=\pm\pi/2%
\end{array}
\right\} \cup\mathbb{R}_{+}.
\end{equation*}

The complete set of orthonormalized eigenfunctions of $\hat{H}_{\lambda
}^{\bot }$ consists of the generalized eigenfunctions $\Psi _{l,E}(\mathbf{%
\rho })$ of the continuous spectrum,$\ E\geq 0,$%
\begin{align*}
& \Psi _{l,E}(\mathbf{\rho })=\left( M/4\pi \right) ^{1/2}\mathrm{e}%
^{i\epsilon _{q}\left( \phi _{0}-l\right) \varphi }J_{\varkappa _{l}}\left( 
\sqrt{ME}\rho \right) ,\ \ l\neq l_{0}\ , \\
& \Psi _{l_{0},E}^{\lambda }(\mathbf{\rho )=}\sqrt{\frac{M}{4\pi \left( 
\tilde{\lambda}^{2}+\pi ^{2}/4\right) }}\mathrm{e}^{i\epsilon _{q}\phi
_{0}\varphi }\left[ \tilde{\lambda}J_{0}\left( \sqrt{ME}\rho \right) +\frac{%
\pi }{2}N_{0}(\sqrt{ME}\rho )\right] , \\
& \tilde{\lambda}=\tan \lambda -\mathbf{C}-\ln \left( \sqrt{ME}/2\kappa
_{0}\right) ,
\end{align*}%
and the eigenfunction $\Psi _{l_{0}}^{\lambda }(\mathbf{\rho })$
corresponding to the discrete level $E_{\lambda }^{\left( -\right) }$ in the
case of$\ |\lambda |<\pi /2$,%
\begin{equation*}
\Psi _{l_{0}}^{\lambda }(\mathbf{\rho })=M\sqrt{\left\vert E_{\lambda
}^{\left( -\right) }\right\vert /\pi }\mathrm{e}^{i\epsilon _{q}\phi
_{0}\varphi }K_{0}\left( \sqrt{M\left\vert E_{\lambda }^{\left( -\right)
}\right\vert }\rho \right) \mathbf{\ },
\end{equation*}%
such that 
\begin{align*}
\check{H}^{\bot }\Psi _{l,E}(\mathbf{\rho })& =E\Psi _{l,E}(\mathbf{\rho }%
),\ \ \check{H}^{\bot }\Psi _{l_{0},E}^{\lambda }(\mathbf{\rho )}=E\Psi
_{l_{0},E}^{\lambda }(\mathbf{\rho )},\ \ E\geq 0, \\
\hat{H}_{\lambda }^{\bot }\Psi _{l_{0}}^{\lambda }(\mathbf{\rho })&
=E_{\lambda }^{\left( -\right) }\Psi _{l_{0}}^{\lambda }(\mathbf{\rho }).
\end{align*}

The corresponding inversion formulas are%
\begin{align*}
& \Psi(\mathbf{\rho})=\sum_{l\in\mathbb{Z},l\neq
l_{0}}\int_{0}^{\infty}\Phi_{l}(E)\Psi_{l,E}(\mathbf{\rho)}%
dE+\int_{0}^{\infty}\Phi_{l_{0}}(E)\Psi_{l_{0},E}^{\lambda}(\mathbf{\rho)}%
dE+\Phi_{l_{0}}\Psi_{l_{0}}^{\lambda}(\mathbf{\rho)}, \\
& \Phi_{l}(E)=\int d\mathbf{\rho}\overline{\Psi_{l,E}(\mathbf{\rho)}}\Psi(%
\mathbf{\rho),\ }\Phi_{l_{0}}(E)=\int d\mathbf{\rho}\overline{\Psi
_{l_{0},E}^{\lambda}(\mathbf{\rho)}}\Psi(\mathbf{\rho),\ }\Phi_{l_{0}}=\int d%
\mathbf{\rho}\overline{\Psi_{l_{0}}^{\lambda}(\mathbf{\rho)}}\Psi (\mathbf{%
\rho),} \\
& \int d\mathbf{\rho}\left| \Psi(\mathbf{\rho)}\right| ^{2}=\sum _{l\in%
\mathbb{Z}}\int_{0}^{\infty}\left| \Phi_{l}(E)\right| ^{2}dE+\left|
\Phi_{l_{0}}\right| ^{2},\ \forall\Psi\in L^{2}(\mathbb{R}^{2}),
\end{align*}
the terms including $\Phi_{l_{0}}$ and $\Psi_{l_{0}}^{\lambda}(\mathbf{\rho}%
) $ are absent in the case of $|\lambda|=\pi/2$. \ 

For $\mu>0$, there is a family of s.a. two-dimensional nonrelativistic
Hamiltonians $\hat{H}_{\mathfrak{e}}^{\bot}$ parametrized by two real
parameters $\lambda_{a}\in\mathbb{S}\left( -\pi/2,\pi/2\right) $)$\ ,\hat {H}%
_{\mathfrak{e}}^{\bot}=\hat{H}_{\left\{ \lambda_{a}\right\} }^{\bot}$, $%
a=0,-1,$ 
\begin{align*}
& \hat{H}_{\left\{ \lambda_{a}\right\} }^{\bot}=\sideset{} {^{\,\lower
1mm\hbox{$\oplus$} }} \sum_{l\in\mathbb{Z},l\neq l_{a}}\hat{H}^{\bot}\left(
l\right) \oplus \sideset{} {^{\,\lower 1mm\hbox{$\oplus$} }} \sum_{a}\hat{H}%
_{\lambda_{a}}^{\bot}\left( l_{a}\right) , \\
& \hat{H}^{\bot}\left( l\right) =M^{-1}S_{l}\hat{h}_{(1)}(l)S_{l}^{-1},\,l%
\neq l_{a}, \\
& \hat{H}_{\lambda_{a}}^{\bot}\left( l_{a}\right) =M^{-1}S_{l_{a}}\hat {h}%
_{\lambda_{a}}\left( l_{a}\right) S_{l_{a}}^{-1}.
\end{align*}

The spectrum of $\hat{H}_{\left\{ \lambda_{a}\right\} }^{\bot}$ is 
\begin{equation*}
\mathrm{spec}\hat{H}_{\left\{ \lambda_{a}\right\} }^{\bot}=\left\{ 
\begin{array}{l}
E_{\lambda_{a}}^{\left( -\right) }=-4M^{-1}k_{0}^{2}|\tilde{\lambda}%
_{a}|^{-\varkappa_{a}^{-1}},\ \ \lambda_{a}\in(-\pi/2,0) \\ 
\varnothing,\ \ \lambda_{a}\notin(-\pi/2,0)%
\end{array}
\right\} \cup\mathbb{R}_{+},
\end{equation*}
where $\varkappa_{a}=|\mu+a|\,,\tilde{\lambda}_{a}=\Gamma(1-\varkappa
_{a})\Gamma^{-1}(1+\varkappa_{a})\tan\lambda_{a}$ .

The complete set of orthonormalized eigenfunctions of $\hat{H}_{\left\{
\lambda _{a}\right\} }^{\bot }$ consists of the generalized eigenfunctions $%
\Psi _{l,E}(\mathbf{\rho }),\ l\neq l_{a}$, and$\ \Psi _{l_{a},E}^{\lambda
_{a}}(\mathbf{\rho )}$ of the continuous spectrum,$\ E\geq 0,$%
\begin{align*}
& \Psi _{l,E}(\mathbf{\rho })=\left( M/4\pi \right) ^{1/2}\mathrm{e}%
^{\epsilon _{q}(\phi _{0}-l)\varphi }J_{\varkappa _{l}}\left( \sqrt{ME}\rho
\right) ,\,\varkappa _{l}=|l+\mu |\,,\ l\neq l_{a}\,, \\
& \Psi _{l_{a},E}^{\lambda _{a}}(\mathbf{\rho )=}\sqrt{\frac{1}{4\pi Q_{a}}}%
\mathrm{e}^{i\epsilon _{q}(\phi _{0}-l_{a})\varphi }\left[ J_{\varkappa
_{a}}\left( \sqrt{ME}\rho \right) +\tilde{\lambda}_{a}\left( \sqrt{ME}%
/2\kappa _{0}\right) ^{2\varkappa _{a}}J_{-\varkappa _{a}}\left( \sqrt{ME}%
\rho \right) \right] , \\
& Q_{a}=1+2\tilde{\lambda}_{a}\left( ME/4\right) ^{\varkappa _{a}}\cos (\pi
\varkappa _{a})+\left( \tilde{\lambda}_{a}\right) ^{2}\left( ME/4\right)
^{2\varkappa _{a}}\ ,
\end{align*}%
and the eigenfunctions $\Psi _{l_{a}}^{\lambda _{a}}(\mathbf{\rho })$
corresponding to the discrete levels $E_{\lambda _{a}}^{\left( -\right) }$
in the case of $\lambda _{a}\in (-\pi /2,0)$ 
\begin{equation*}
\Psi _{l_{a}}^{\lambda _{a}}(\mathbf{\rho })=\sqrt{\frac{M^{2}\left\vert
E_{\lambda _{a}}^{\left( -\right) }\right\vert \sin (\pi \varkappa _{a})}{%
\pi ^{2}\varkappa _{a}}}\mathrm{e}^{i\epsilon _{q}(\phi _{0}-l_{a})\varphi
}K_{\varkappa _{a}}\left( \sqrt{\left\vert ME_{\lambda _{a}}^{\left(
-\right) }\right\vert }\rho \right) \ ,
\end{equation*}%
such that 
\begin{align*}
& \check{H}^{\bot }\Psi _{l,E}(\mathbf{\rho })=E\Psi _{l,E}(\mathbf{\rho }%
),\ l\neq l_{a},\ \ \check{H}^{\bot }\Psi _{l_{a},E}^{\lambda _{a}}(\mathbf{%
\rho )}=E\Psi _{l_{a},E}^{\lambda _{a}}(\mathbf{\rho )},\ \ E\geq 0, \\
& \hat{H}_{\left\{ \lambda _{a}\right\} }^{\bot }\Psi _{l_{b}}^{\lambda
_{b}}(\mathbf{\rho })=E_{\lambda _{b}}^{\left( -\right) }\Psi
_{l_{b}}^{\lambda _{b}}(\mathbf{\rho }),\ \ b=0,-1.
\end{align*}

The corresponding inversion formulas are%
\begin{align*}
& \Psi(\mathbf{\rho})=\sum_{l\in\mathbb{Z},\ l\neq l_{a}}\int_{0}^{\infty
}\Phi_{l}(E)\Psi_{l,E}(\mathbf{\rho)}dE+\sum_{a}\left[ \int_{0}^{\infty}%
\Phi_{l_{a}}(E)\Psi_{l_{a},E}^{\lambda_{a}}(\mathbf{\rho)}%
dE+\Phi_{l_{a}}\Psi_{l_{a}}^{\lambda_{a}}(\mathbf{\rho)}\right] , \\
& \Phi_{l}(E)=\int d\mathbf{\rho}\overline{\Psi_{l,E}(\mathbf{\rho)}}\Psi(%
\mathbf{\rho)},\ l\neq l_{a}\ , \\
& \Phi_{l_{a}}(E)=\int d\mathbf{\rho}\overline{\Psi_{l_{a},E}^{\lambda_{a}}(%
\mathbf{\rho)}}\Psi(\mathbf{\rho)},\ \Phi_{l_{a}}=\int d\mathbf{\rho }%
\overline{\Psi_{l_{a}}^{\lambda_{a}}(\mathbf{\rho)}}\Psi(\mathbf{\rho),} \\
& \int d\mathbf{\rho}\left| \Psi(\mathbf{\rho)}\right| ^{2}=\sum _{l\in%
\mathbb{Z}}\int_{0}^{\infty}\left| \Phi_{l}(E)\right| ^{2}dE+\sum _{a}\left|
\Phi_{l_{a}}\right| ^{2},\ \forall\Psi\in L^{2}(\mathbb{R}^{2}),
\end{align*}
the terms with $\Phi_{l_{a}}$ and $\Psi_{l_{a}}^{\lambda_{a}}(\mathbf{\rho)}$
are absent in the case of $\lambda_{a}\notin(-\pi/2,0)$.

We now consider the case of the magnetic-solenoid field\ where $B\neq0$.

\subsubsection{S.a. Hamiltonians with magnetic-solenoid field}

In this case, the radial differential operation $\check{h}\left( l\right) $
is given by (\ref{9.2.1}) with $\gamma =\frac{e|B|}{c\hbar }\neq 0$, or%
\begin{align*}
& \check{h}\left( l\right) =-\partial _{\rho }^{2}+g_{1}\rho ^{-2}+g_{2}\rho
^{2}+\mathcal{E}_{l}^{\left( 0\right) }, \\
& g_{1}=\varkappa _{l}^{2}-1/4,\ \varkappa _{l}=|l+\mu |\,,\,g_{2}=\gamma
^{2}/4,\ \mathcal{E}_{l}^{\left( 0\right) }=\gamma (l+\mu ).
\end{align*}

Up to the constant term $\mathcal{E}_{l}^{\left( 0\right) }$, this s.a.
differential operation is identical to the one-dimensional Schr\"{o}dinger
operation $-d_{x}^{2}+g_{1}x^{-2}+g_{2}x^{2}$ studied by us recently (to be
published). We therefore directly carry over the obtained results to s.a.
extensions of $\hat{h}\left( l\right) $. We note that as in the case of pure
AB field, a division to different regions of $g_{1}$is actually determined
by the same term $g_{1}\rho ^{-2}=($ $\varkappa _{l}^{2}-1/4)/\rho ^{-2}$
singular at the origin and independent of the value of $B.$

\paragraph{First region: $g_{1}\geq3/4$}

In this region, we have $\left( l+\mu\right) ^{2}\geq1$ and as before, we
distinguish the cases of$\,\mu=0\,$and $\mu>0:$ 
\begin{align*}
\mu & =0:l\leq-1\ \mathrm{or}\ l\geq1,\ \mathrm{i.e.,\,}l\neq l_{0}\ , \\
\mu & >0:l\leq-2\ \mathrm{or}\ l\geq1,\mathrm{i.e.},\,l\neq l_{a}.
\end{align*}

For such $l$, the initial symmetric operator $\hat{h}\left( l\right) $ has
zero deficiency indices, is essentially s.a., and its unique s.a. extension
is $\hat{h}_{\mathfrak{e}}\left( l\right) =\hat{h}_{\left( 1\right) }\left(
l\right) =\hat{h}^{+}\left( l\right) $ with the domain $D_{\check{h}\left(
l\right) }^{\ast}\left( \mathbb{R}_{+}\right) .$ The spectrum of $\hat {h}%
_{\left( 1\right) }\left( l\right) $ is simple and discrete,

\begin{equation}
\mathrm{spec}\hat{h}_{\left( 1\right) }\left( l\right) =\left\{ \mathcal{E}%
_{l,m}=\gamma\left( 1+|l+\mu|+(l+\mu)+2m\right) ,\ \ m\in \mathbb{Z}%
_{+}\right\} .  \label{9.2.2}
\end{equation}

Eigenfunctions $U_{l,m}^{\left( 1\right) }\,$of the Hamiltonian $\hat {h}%
_{(1)}\left( l\right) $ are 
\begin{align}
& U_{l,m}^{\left( 1\right) }(\rho)=Q_{l,m}\left( \gamma/2\right)
^{1/4+\varkappa_{l}/2}\rho^{1/2+\varkappa_{l}}\mathrm{e}^{-\gamma\rho^{2}/4}%
\Phi(-m,1+\varkappa_{l};\gamma\rho^{2}/2),  \notag \\
& Q_{l,m}=\left( \frac{\sqrt{2\gamma}\Gamma(1+\varkappa_{l}+m)}{m!\Gamma
^{2}(1+\varkappa_{l})}\right) ^{1/2},  \notag \\
& \hat{h}_{\left( 1\right) }\left( l\right) U_{l,m}^{\left( 1\right) }=%
\mathcal{E}_{l,m}U_{l,m}^{\left( 1\right) }\ ,\ m\in\mathbb{Z}_{+},
\label{9.2.2a}
\end{align}
they form a complete orthonormalized set in each Hilbert space $\mathcal{L}%
_{l}.$

\paragraph{Second region :$-1/4<\protect\alpha<3/4$}

In this region, we have $0<\left( l+\mu\right) ^{2}<1$,or equivalently (\ref%
{9.1.42a}). We know that if $\mu=0$, these inequalities have no solutions
for $l\in\mathbb{Z}$, while if $\mu>0$ there are the two solutions $%
l=l_{a}=a,\,a=0,-1$. Therefore, we again remain with the case of $\mu>0$.

For each $l=l_{a}$\ , there exists a one-parameter $U\left( 1\right) $%
-family of s.a. Hamiltonians $\hat{h}_{\mathfrak{e}}\left( l_{a}\right) =%
\hat{h}_{\lambda_{a}}\left( l_{a}\right) $ parametrized by the real
parameter $\lambda_{a}\in\mathbb{S}\left( -\pi/2,\pi/2\right) $\ These
Hamiltonians are specified by the asymptotic s.a. boundary conditions at the
origin%
\begin{equation}
\psi_{\lambda_{a}}(\rho)=c\left[ \left( \sqrt{\gamma/2}\rho\right)
^{1/2+\varkappa_{a}}\sin\lambda_{a}+\left( \sqrt{\gamma/2}\rho\right)
^{1/2-\varkappa_{a}}\cos\lambda_{a}\right] +O(\rho^{3/2}),  \label{9.1.43'}
\end{equation}
where $\varkappa_{a}=|\mu+a|$, $0<\varkappa_{a}<1$,$\ $and $c$ is an
arbitrary constant\footnote{%
In comparison with (\ref{9.1.43}), we fix the dimensional parameter $k_{0%
\text{ }}$ by $k_{0\text{ }}=\sqrt{\gamma/2}$.}, and their domains are given
by%
\begin{align}
& D_{h_{\lambda_{a}}\left( l_{a}\right) }=\{\psi\in D_{\check{h}\left(
l_{a}\right) }^{\ast}\left( \mathbb{R}_{+}\right) ,\ \psi\text{ }\mathrm{%
satisfies\ }\text{(\ref{9.1.43'})}\},  \label{9.1.44a} \\
& D_{\check{h}\left( l_{a}\right) }^{\ast}\left( \mathbb{R}_{+}\right)
=\{\psi_{\ast}:\psi_{\ast},\psi_{\ast}^{\prime}\;\mathrm{are\;a.c.\;in}\;%
\mathbb{R}_{+},\;\psi_{\ast},\check{h}\left( l_{a}\right) \psi_{\ast}\in
L^{2}(\mathbb{R}_{+})\}.  \notag
\end{align}

The spectrum of $\hat{h}_{\lambda_{a}}\left( l_{a}\right) $ is simple and
discrete and is bounded from below,%
\begin{equation*}
\mathrm{spec}\hat{h}_{\lambda_{a}}\left( l_{a}\right) =\left\{ \mathcal{E}%
_{a,m}=\tau_{a,m}+\mathcal{E}_{l_{a}}^{\left( 0\right) },\ m\in\mathbb{Z}%
_{+}\right\} ,
\end{equation*}
where $\tau_{a,m}$ are solutions of the equation $\omega_{\lambda_{a}}(%
\tau_{a,m})=0,$%
\begin{align}
&
\omega_{\lambda_{a}}(W)=\omega_{+}(W)\sin\lambda_{a}+\omega_{-}(W)\cos%
\lambda_{a},  \notag \\
& \omega_{\pm}(W)=\Gamma(1\pm\varkappa_{a})/\Gamma(1/2\pm\varkappa
_{a}/2-W/2\gamma).  \label{9.1.45}
\end{align}

The eigenfunctions $U_{\lambda_{a},m}^{\left( 2\right) }$of the Hamiltonian $%
\hat{h}_{\lambda_{a}}\left( l_{a}\right) $ are 
\begin{align}
& U_{\lambda_{a},m}^{\left( 2\right) }(\rho)=Q_{a,m}\left[ u_{+}(\rho
;\tau_{a,m})\sin\lambda_{a}+u_{-}(\rho;\tau_{a,m})\cos\lambda_{a}\right] , 
\notag \\
& Q_{a,m}=\left( \frac{\tilde{\omega}_{_{\lambda_{a}}}(\tau_{a,m})}{\sqrt{%
2\gamma}\varkappa_{a}\omega_{\lambda_{a}}^{\prime}(\tau_{a,m})}\right)
^{1/2},\;\tilde{\omega}_{\lambda_{a}}(W)=\omega_{+}(W)\cos\lambda_{a}-%
\omega_{-}(W)\sin\lambda_{a},  \notag \\
& u_{\pm}(\rho;W)=\left( \gamma/2\right)
^{1/4\pm\varkappa_{a}/2}\rho^{1/2\pm\varkappa_{a}}\mathrm{e}%
^{-\gamma\rho^{2}/4}\Phi(1/2\pm
\varkappa_{a}/2-W/2\gamma,1\pm\varkappa_{a};\gamma\rho^{2}/2),  \notag \\
& \hat{h}_{\lambda_{a}}\left( l_{a}\right) U_{\lambda_{a},m}^{\left(
2\right) }=\mathcal{E}_{a,m}U_{\lambda_{a},m}^{\left( 2\right) }\ ,
\label{9.2.4}
\end{align}
they form a complete orthonormalized set in each Hilbert space $\mathcal{L}%
_{l_{a}}.$

Explicit expressions for the spectrum and eigenfunctions can be easily
obtained in the cases of $\lambda _{a}=\pm \pi /2$ and $\lambda _{a}=0.$ In
the case of $\lambda _{a}=\pm \pi /2,$ they are given by the respective
formulas (\ref{9.2.2}) and (\ref{9.2.2a}) with the substitutions $%
l\rightarrow l_{a}$ and $\varkappa _{l}\rightarrow \varkappa _{a}$. In the
case of $\lambda _{a}=0,$ these formulas are modified by the additional
substitution $\varkappa _{a}\rightarrow -\varkappa _{a}$.

\paragraph{Third region: $\protect\alpha=-1/4$}

In this region, we have $l+\mu =0$. As we know, we remain with the only case
of $\mu =0$ and with $l=l_{0}=0$.

For $l=l_{0}$, there exists a one-parameter $U\left( 1\right) $-family of
s.a. Hamiltonians $\hat{h}_{\mathfrak{e}}\left( l_{0}\right) =\hat {h}%
_{\lambda}\left( l_{0}\right) $.parametrized by the real parameter $%
\lambda\in\mathbb{S}\left( -\pi/2,\pi/2\right) $) These Hamiltonians are
specified by the asymptotic s.a. boundary conditions at the origin%
\begin{equation}
\psi_{\lambda}(\rho)=c\left[ \rho^{1/2}\ln\left( \sqrt{\gamma/2}\rho\right)
\cos\lambda+\rho^{1/2}\sin\lambda\right] +O(\rho^{3/2}\ln\rho),\ \rho
\rightarrow0,  \label{9.1.46'}
\end{equation}
where$\ c$ is an arbitrary constant, and their domains are given by%
\begin{align}
D_{h_{\lambda}\left( l_{0}\right) } & =\{\psi:\ \psi\in D_{\check {h}\left(
l_{0}\right) }^{\ast}\left( \mathbb{R}_{+}\right) ,\ \psi\text{ }\mathrm{%
satisfies\ }(\text{\ref{9.1.46'})}\},  \label{9.2.5a} \\
D_{\check{h}\left( l_{0}\right) }^{\ast}\left( \mathbb{R}_{+}\right) &
=\{\psi_{\ast}:\psi_{\ast},\psi_{\ast}^{\prime}\;\mathrm{are\;a.c.\;in}\;%
\mathbb{R}_{+},\;\psi_{\ast},\check{h}\left( l_{0}\right) \psi_{\ast}\in
L^{2}(\mathbb{R}_{+})\}.  \notag
\end{align}

The spectrum of $\hat{h}_{\lambda }\left( l_{0}\right) $ is simple and
discrete and is bounded from below, \textrm{spec}$\hat{h}_{\lambda }\left(
l_{0}\right) =\left\{ \mathcal{E}_{m},\ m\in \mathbb{Z}_{+}\right\} $, where 
$\mathcal{E}_{m}$ are solutions of the equation $\omega _{\lambda }(\mathcal{%
E}_{m})=0,$%
\begin{equation}
\omega _{\lambda }(W)=\cos \lambda \lbrack \psi (\alpha _{0})-2\psi
(1)]-\sin \lambda ,\ \ \alpha _{0}=1/2-W/2\gamma \ .  \label{9.2.6}
\end{equation}%
In the case of $\lambda \neq \pm \pi /2$, the limit $\lambda \rightarrow \pm
\pi /2$ in these solutions yields the spectrum in the case of $\lambda =\pm
\pi /2.$

The eigenfunctions $U_{\lambda,m}^{\left( 3\right) }$ of the Hamiltonians $%
\hat{h}_{\lambda}\left( l_{0}\right) $ are 
\begin{align}
& U_{\lambda,m}^{\left( 3\right) }=Q_{\lambda,m}\left[ u_{1}(\rho;\mathcal{E}%
_{m})\sin\lambda+u_{3}(\rho;\mathcal{E}_{m})\cos \lambda\right] ,  \notag \\
& u_{1}(\rho;W)=\left( \gamma/2\right) ^{1/4}\rho^{1/2}\mathrm{e}%
^{-\gamma\rho^{2}/4}\Phi(\alpha_{0},1;\gamma\rho^{2}/2),  \notag \\
& u_{3}(\rho;W)=u_{1}(\rho;W)\ln\left( \sqrt{\gamma/2}\rho\right) +\left(
\gamma/2\right) ^{1/4}\rho^{1/2}\mathrm{e}^{-\gamma\rho^{2}/4}\left.
\partial_{\mu}\Phi(1/2+\mu-W/2\gamma,1+2\mu;\gamma\rho^{2}/2)\right| _{\mu
=0},  \notag \\
& Q_{\lambda,m}=\left[ -\frac{\tilde{\omega}_{\lambda}(\mathcal{E}_{m})}{%
\sqrt{2\gamma}\omega_{\lambda}^{\prime}(\mathcal{E}_{m})}\right] ^{1/2},\;%
\tilde{\omega}_{\lambda}(W)=\sin\lambda\lbrack\psi(\alpha_{0})-2\psi(1)]+%
\cos\lambda,  \notag \\
& \hat{h}_{\lambda}\left( l_{0}\right) U_{\lambda,m}^{\left( 3\right) }=%
\mathcal{E}_{m}U_{\lambda,m}^{\left( 3\right) }\ ,\ \ m\in\mathbb{Z}_{+},
\label{9.2.5}
\end{align}
they form a complete orthonormalized set in the Hilbert space $\mathcal{L}%
_{l_{0}}.$

We note that spectrum and eigenfunctions in the case of $\lambda=\pm\pi/2$
can be obtained from the respective formulas for the first region in the
formal limit $l\rightarrow0$.

\paragraph{Complete spectrum and inversion formulas}

In the previous subsubsecs., we constructed all s.a. radial Hamiltonians $%
\hat{h}_{\mathfrak{e}}\left( l\right) $ associated with the s.a.
differential operation $\check{h}\left( l\right) $ as s.a. extensions of the
symmetric operator $\hat{h}\left( l\right) $ for any $l$ $\in\mathbb{Z\,\ }$%
and for any any $\phi_{0}$, $\mu$, and $B$. We assemble our previous results
into two groups.

For $\mu =0$, we have%
\begin{align}
& \hat{h}_{\mathfrak{e}}\left( l\right) =\hat{h}_{\left( 1\right) }\left(
l\right) ,\ \ l\neq l_{0}=0\ ,\ \ D_{h_{\left( 1\right) }\left( l\right)
}=D_{\check{h}\left( l\right) }^{\ast }\left( \mathbb{R}_{+}\right) ,  \notag
\\
& \hat{h}_{\mathfrak{e}}\left( l_{0}\right) =\hat{h}_{\lambda }\left(
l_{0}\right) ,\ \ \lambda \in \mathbb{S}\left( -\pi /2,\pi /2\right) \ ,
\label{9.2.32}
\end{align}%
and the domain $D_{h_{\lambda }\left( l_{0}\right) }$ is given by eq. (\ref%
{9.2.5a});

For $\mu>0$, we have%
\begin{align}
& \hat{h}_{\mathfrak{e}}\left( l\right) =\hat{h}_{\left( 1\right) }\left(
l\right) ,\ \ l\neq l_{a}=a=0,-1\ ,\ \ D_{h_{\left( 1\right) }\left(
l\right) }=D_{\check{h}\left( l\right) }^{\ast}\left( \mathbb{R}_{+}\right) ,
\notag \\
& \hat{h}_{\mathfrak{e}}\left( l_{a}\right) =\hat{h}_{\lambda_{a}}\left(
l_{a}\right) ,\ \ \lambda_{a}\in\mathbb{S}\left( -\pi/2,\pi/2\right) ,
\label{9.2.33}
\end{align}
and the domain $D_{h_{\lambda_{a}}\left( l_{a}\right) }$ is given by eq. (%
\ref{9.1.44a}).

As a result, we find a family of all s.a. rotationally-invariant
two-dimensional nonrelativistic Hamiltonians $\hat{H}_{\mathfrak{e}}^{\bot}=$
$M^{-1}\widehat{\mathcal{H}}_{\mathfrak{e}}^{\perp}$ associated with the
s.a. differential operation $\check{H}^{\bot}$ (\ref{9.1.22}) with $B\neq0$.
Each set of possible s.a. radial Hamiltonians $\hat{h}_{\mathfrak{e}}\left(
l\right) $ generates a s.a. rotationally-invariant Hamiltonian $\hat {H}_{%
\mathfrak{e}}^{\bot}$ in accordance with the relations (\ref{R.6}) and (\ref%
{9R.7}). As in the case of pure AB field where $B=0$, we let $E$ denote the
spectrum points of $\hat{H}_{\mathfrak{e}}^{\bot}$.

It is convenient to change the indexing $l$, $m$ of the spectrum points and
eigenfunctions to $l$, $n$, as follows:%
\begin{align}
n & =n(l,m)=\left\{ 
\begin{array}{l}
m,\;l\leq-1 \\ 
m+l,\;l\geq0%
\end{array}
\right. ,\;m\in\mathbb{Z}_{+},\;l\in\mathbb{Z};  \notag \\
m & =m(n,l)=\left\{ 
\begin{array}{l}
n,\;l\leq-1 \\ 
n-l,\;0\leq l\leq n%
\end{array}
\right. ,\;n\in\mathbb{Z}_{+},\;l\in\mathbb{Z},  \label{9.2.33a}
\end{align}
and then to interchange their position, such that finally, the indices $l$, $%
m$ are replaced by indices $n$, $l$.

When writing formulas for eigenfunctions $\Psi _{n,l}$ of the operator $\hat{%
H}_{\mathfrak{e}}^{\bot }$ in terms of eigenfunctions $U_{l,m}$ of the
operators $\hat{h}_{\mathfrak{e}}(l)$, we has to introduce the factor $1/%
\sqrt{2\pi \rho }$ $\mathrm{e}^{i\epsilon \left( \phi _{0}-l\right) \varphi
} $ in accordance with eq. (\ref{9.1.23b}) and to make the substitution $%
\mathcal{E}_{l,m}=ME_{n,l}$ for the corresponding spectrum points.

The final result is the following.

There is a family of s.a. two-dimensional nonrelativistic Hamiltonians $\hat{%
H}_{\mathfrak{e}}^{\bot}$ parametrized by real parameters $\lambda _{\ast},%
\hat{H}_{\mathfrak{e}}^{\bot}=\hat{H}_{\lambda_{\ast}}^{\bot}$, 
\begin{align}
& \hat{H}_{\lambda_{\ast}}^{\bot}=\sideset{}{^{\,\lower1mm\hbox{$\oplus$}}}
\sum_{l\in\mathbb{Z},l\neq l_{\ast}}\hat{H}^{\bot}\left( l\right) \oplus%
\sideset{}{^{\,\lower1mm\hbox{$\oplus$}}} \sum_{l=l_{\ast}}\hat {H}%
_{\lambda_{\ast}}^{\bot}\left( l_{\ast}\right) ,  \notag \\
& \hat{H}^{\bot}\left( l\right) =M^{-1}S_{l}\hat{h}_{(1)}(l)S_{l}^{-1},\,l%
\neq l_{\ast},  \notag \\
& \hat{H}_{\lambda_{\ast}}^{\bot}\left( l_{\ast}\right) =M^{-1}S_{l_{\ast}}%
\hat{h}_{\lambda_{\ast}}\left( l_{a}\right) S_{l_{\ast}}^{-1}\,,  \notag \\
& l_{\ast}=\left\{ 
\begin{array}{l}
l_{0},\;\mu=0 \\ 
l_{a},\;\mu>0%
\end{array}
\right. ,\;\lambda_{\ast}=\left\{ 
\begin{array}{l}
\lambda\in\mathbb{S}\left( -\pi/2,\pi/2\right) ,\;\mu=0 \\ 
\lambda_{a}\in\mathbb{S}\left( -\pi/2,\pi/2\right) ,\;\mu>0%
\end{array}
\right. .  \label{9.2.33b}
\end{align}
The spectrum of $\hat{H}_{\lambda_{\ast}}^{\bot}$is given by (remember that $%
E_{n,l}=M^{-1}\mathcal{E}_{l,m(n,l)}$)%
\begin{align}
& \mathrm{spec}\hat{H}_{\lambda}^{\bot}=\left\{ \cup_{l\in\mathbb{Z},l\neq
l_{\ast}}\left\{ E_{n,l},\ n\in\mathbb{Z}_{+}\right\} \right\} \cup\left\{
\cup_{l=l_{\ast}}\left\{ E_{n}^{\left( \lambda_{\ast}\right) },\ n\in\mathbb{%
Z}_{+}\right\} \right\} ,  \notag \\
& E_{n,l}=\gamma M^{-1}[1+2n+2\theta(l)\mu],\;l\leq n,\;l\neq l_{\ast
},\;\theta(l)=\left\{ 
\begin{array}{c}
1,\;l\geq0 \\ 
0,\;l<0%
\end{array}
\right. ,  \label{9.2.34a} \\
& E_{n}^{\left( \lambda\right) }:\left\{ 
\begin{array}{l}
\omega_{\lambda}(ME_{n}^{\left( \lambda\right) })=0,\ |\lambda|<\pi/2 \\ 
E_{n}^{\left( \pm\pi/2\right) }=\gamma M^{-1}(1+2n)\ 
\end{array}
\right. ,\;\mu=0,  \notag \\
& \left\{ 
\begin{array}{l}
E_{n}^{(\lambda_{a})}=M^{-1}\left[ \tau_{a,n}+\gamma(a+\mu)\right] ,\
\omega_{\lambda_{a}}(\tau_{a,n})=0 \\ 
E_{n}^{(\pm\pi/2)}=\gamma M^{-1}[1+2n+2\theta(a)\mu]%
\end{array}
\right. ;\ n\in\mathbb{Z}_{+},\;\mu>0,  \label{9.2.34b}
\end{align}
where $\omega_{\lambda}\left( W\right) $ and $\omega_{\lambda_{a}}\left(
W\right) $ are given by the respective eqs. (\ref{9.2.6}) and (\ref{9.1.45}).

The complete set of orthonormalized eigenfunctions of $\hat{H}%
_{\lambda_{\ast }}^{\bot}$consists of the functions $\Psi_{n,l}(\mathbf{\rho}%
),\ l\neq l_{\ast},\ $and $\Psi_{n,l_{\ast}}^{\lambda_{\ast}}(\mathbf{\rho)}$
,%
\begin{equation}
\Psi_{n,l}(\mathbf{\rho})=\frac{1}{\sqrt{2\pi\rho}}\mathrm{e}^{i\epsilon
(\phi_{0}-l)\varphi}U_{l,m(n,l)}^{\left( 1\right) }(\rho),  \label{9.2.35}
\end{equation}
where $U_{l,m}^{\left( 1\right) }(\rho)$ are given by eqs. (\ref{9.2.2a}),
and (we note that $m(n,l_{\ast})=n$)%
\begin{align*}
\Psi_{n,l_{0}}^{\lambda}(\mathbf{\rho)} & =\frac{1}{\sqrt{2\pi\rho}}\mathrm{e%
}^{i\epsilon\phi_{0}\varphi}U_{\lambda,n}^{\left( 3\right) }(\rho),\;\mu=0,
\\
\Psi_{n,l_{a}}^{\lambda_{a}}(\mathbf{\rho)} & =\frac{1}{\sqrt{2\pi\rho}}%
\mathrm{e}^{i\epsilon(\phi_{0}-l_{a})\varphi}U_{\lambda_{a},n}^{\left(
2\right) }(\rho),\;\mu>0,
\end{align*}
where $U_{\lambda,n}^{\left( 3\right) }(\rho)$ and $U_{\lambda_{a},n}^{%
\left( 2\right) }(\rho)$ are given by the respective eqs. (\ref{9.2.5}) and (%
\ref{9.2.4}), such that%
\begin{equation*}
\hat{H}_{\lambda_{\ast}}^{\bot}\Psi_{n,l}(\mathbf{\rho})=E_{n,l}\Psi _{n,l}(%
\mathbf{\rho}),\ l\neq l_{\ast},\ \ \hat{H}_{\lambda_{\ast}}^{\bot }(\mathbf{%
\rho)}\Psi_{n,l_{\ast}}^{\lambda_{\ast}}=E_{n}^{\left( \lambda_{\ast}\right)
}\Psi_{n,l_{\ast}}^{\lambda_{\ast}}(\mathbf{\rho}).
\end{equation*}
We note, that for the case of $\lambda=\pm\pi/2$ where $l=l_{0}=0$ and for
the case of $\lambda_{a}=\pm\pi/2$ where $l=l_{a}=a=0,-1$,.the energy
eigenvalues $E_{n}^{(\lambda)}$ and $E_{n}^{(\lambda_{a})}$and corresponding
eigenfunctions $\Psi_{n}^{\lambda}$ and $\Psi_{n}^{\lambda_{a}}$ are given
by the respective eqs. (\ref{9.2.34a}) and (\ref{9.2.35}) extended to all
values of $l$.

The corresponding inversion formulas are%
\begin{align*}
& \Psi(\mathbf{\rho})=\sum_{l\in\mathbb{Z},\ l\neq l_{\ast}}\sum _{n\in%
\mathbb{Z}_{+}}\Phi_{n,l}\Psi_{n,l}(\mathbf{\rho)+}\sum_{l=l_{\ast},n\in%
\mathbb{Z}_{+}}\Phi_{n,l_{\ast}}\Psi_{n,l_{\ast}}^{\lambda_{\ast}}(\mathbf{%
\rho)}, \\
& \Phi_{n,l}=\int d\mathbf{\rho}\overline{\Psi_{n,l}(\mathbf{\rho)}}\Psi(%
\mathbf{\rho)},\ l\neq l_{\ast},\ \Phi_{n,l_{\ast}}=\int d\mathbf{\rho }%
\overline{\Psi_{n,l_{\ast}}^{\lambda_{\ast}}(\mathbf{\rho)}}\Psi (\mathbf{%
\rho),} \\
& \int d\mathbf{\rho}\left| \Psi(\mathbf{\rho)}\right| ^{2}=\sum _{l\in%
\mathbb{Z}}\sum_{n\in\mathbb{Z}_{+}}\left| \Phi_{n,l}\right| ^{2},\
\forall\Psi\in L^{2}(\mathbb{R}^{2}).
\end{align*}

\subsection{Three-dimensional case}

In three dimensions, we start with the differential operation $\check{H}$ (%
\ref{9.1.20}). The initial symmetric operator $\hat{H}$ associated with $%
\check{H}$ is defined in the Hilbert space $\mathfrak{H}=L^{2}(\mathbb{R}%
^{3})$ by%
\begin{equation}
D_{H}=\mathcal{D}(\mathbb{R}^{3}\backslash\mathbb{R}_{z}),\;\hat{H}\psi =%
\check{H}\psi,\;\forall\psi\in D_{H},  \notag
\end{equation}
where $\mathcal{D}(\mathbb{R}^{3}\backslash\mathbb{R}_{z})$ is the space of
smooth and compactly supported functions vanishing in a neighborhood of the $%
z$-axis. The domain $D_{H}$ is dense in $\mathfrak{H}$, and the symmetricity
of $\hat{H}$ is obvious. A s.a. quantum Hamiltonian must be defined as a
s.a. extension of $\hat{H}$.

There is an evident space symmetry in the classical description of the
system, the symmetry with respect to rotations around the $z$-axis and
translations along this axis, which is manifested as the invariance of the
classical Hamiltonian under these space transformations. The key point in
constructing a quantum description of the system is the requirement of the
invariance of the quantum Hamiltonian under the same transformations.

Namely, let $\mathbb{G}$ be the group of the above space transformations $S$%
: $\mathbf{r}\rightarrow S\mathbf{r}$. This group is unitarily represented
in $\mathfrak{H}$: if $S\in \mathbb{G}$, then the corresponding operator $%
U_{S}$ is defined by 
\begin{equation*}
(U_{S}\psi )(\mathbf{r})=\psi (S^{-1}\mathbf{r}),\forall \psi \in \mathfrak{H%
}.
\end{equation*}%
The operator $\hat{H}$ evidently commutes\footnote{%
We remind the reader of the notion of commutativity in this case (where one
of the operators, $U_{S}$, is bounded and defined everywhere): we say that
the operators $\hat{H}$ and $U_{S}$ commute if $U_{S}\hat{H}$ $\subseteq 
\hat{H}U_{S}$, i.e., if $\psi \in D_{H}$, then also $U_{S}\psi \in D_{H}$
and $U_{S}\hat{H}\psi =\hat{H}U_{S}\psi $} with $U_{S}$ for any $S$. We
search only for s.a. extensions $\hat{H}_{\mathfrak{e}}$ of $\hat{H}$ that
also commute with $U_{S}$ for any $S$. This condition is the explicit form
of the invariance, or symmetry, of a quantum Hamiltonian under the space
transformations. As in classical mechanics, this symmetry allows separating
the cylindrical coordinates $\rho ,$ $\varphi ,$ and $z$ and reducing the
three-dimensional problem to a one-dimensional radial problem. Let $L^{2}(%
\mathbb{R}\times \mathbb{R}_{+})$ denote the space of square-integrable
functions with respect to the Lebesgue measure $dp_{z}d\rho $ on $\mathbb{R}%
\times \mathbb{R}_{+}$, and let $V\colon \sum_{l\in \mathbb{Z}}^{\oplus
}L^{2}(\mathbb{R}\times \mathbb{R}_{+})\rightarrow \mathfrak{H}$ be the
unitary operator defined by the relation 
\begin{equation*}
(Vf)(\rho ,\varphi ,z)=\frac{1}{2\pi \sqrt{\rho }}\int_{\mathbb{R}%
_{z}}dp_{z}\sum_{l\in \mathbb{Z}}\mathrm{e}^{i(\epsilon (\phi _{0}-l)\varphi
+p_{z}z)}f(l,p_{z},\rho ).
\end{equation*}

In the two-dimensional case considered in the preceding subsection, all s.a.
Hamiltonians were represented as direct sums over $l$ of s.a. extensions of
suitable radial symmetric operators. In the three-dimensional case, a new
continuous parameter $p_{z}$ arises in addition to the discrete parameter $l$%
, and analogous representations should be written in terms of direct
integrals rather than direct sums. More precisely, any s.a. Hamiltonian $%
\hat{H}_{\mathfrak{e}}$ can be represented in the form 
\begin{equation*}
\hat{H}_{\mathfrak{e}}=V\int_{\mathbb{R}_{z}}^{\oplus }dp_{z}%
\sideset{}{^{\,\lower1mm\hbox{$\oplus$}}}\sum_{l\in \mathbb{Z}}\hat{h}_{%
\mathfrak{e}}(l,p_{z})V^{-1},
\end{equation*}%
where $\hat{h}_{\mathfrak{e}}(l,p_{z})$ for fixed $l$ and $p_{z}$ is a s.a.
extension of the symmetric operator $\hat{h}(l,p_{z})=\hat{h}%
(l)+p_{z}^{2}/2m_{e}$ in $L^{2}(\mathbb{R}_{+})$, and the operator $\hat{h}%
(l)$ in $L^{2}(\mathbb{R}_{+})$ is given by 
\begin{align}
& D_{h(l)}=\mathcal{D}(\mathbb{R}_{+}),  \notag \\
& \hat{h}(l)f(\rho )=\left( -\partial _{\rho }^{2}+\rho ^{-2}\left[ \left(
l+\mu +\gamma \rho ^{2}/2\right) ^{2}-1/4\right] \right) f(\rho ),\;f\in
D_{h(l)}.  \label{hhhhh}
\end{align}%
A detailed derivation of the above representation for $\hat{H}_{\mathfrak{e}%
} $ will be published in the short run.

The inversion formulas in three dimensions are obtained by the following
modifications in the two dimensional inversion formulas:

1) $\sum_{m\in\mathbb{Z}}\int dE\rightarrow\int dp_{z}\sum_{l\in\mathbb{Z}%
}\int dE^{\bot}$, where $E^{\bot}$ are spectrum points of two-dimensional
s.a. Hamiltonians $\hat{H}_{\mathfrak{e}}^{\bot}$, whereas the eigenvalues
(spectrum points) $E$ of three-dimensional s.a. Hamiltonians $\hat {H}_{%
\mathfrak{e}}$ are $E=E^{\bot}+p_{z}^{2}/2m$ $\left( \int
dp_{z}=\int_{-\infty}^{\infty}dp_{z}\right) $.

2) The contribution of discrete spectrum points of two-dimensional s.a.
Hamiltonian $\hat{H}_{\mathfrak{e}}^{\bot}$ have to be multiplied by $\int
dp_{z}$.

3) Eigenfunctions of two-dimensional s.a. Hamiltonians $\hat{H}_{\mathfrak{e}%
}^{\bot}$have to be multiplied by $\left( 2\pi\hbar\right) ^{-1/2}\mathrm{e}%
^{ip_{z}z/\hbar}$ in order to obtain eigenfunctions of three-dimensional
s.a. Hamiltonians $\hat{H}_{\mathfrak{e}}$.

4) The extension parameters $\lambda_{a}$ and $\lambda$ have to be replaced
by functions $\lambda_{a}(p_{z})$ and $\lambda(p_{z})$.

\subsubsection{S.a. nonrelativistic Hamiltonians with AB field}

For the case of $\mu=0$, there is a family of s.a. three-dimensional
Hamiltonians parametrized by a real-valued function$\ \lambda(p_{z})\,\left(
\lambda(p_{z})\in\mathbb{S}\left( -\pi/2,\pi/2\right) ,\ p_{z}\in \mathbb{R}%
\right) $.

The corresponding spectrum of $\hat{H}_{\lambda(p_{z})}$\ is%
\begin{equation*}
\mathrm{spec}\hat{H}_{\left\{ \lambda(p_{z})\right\} }=\left\{ 
\begin{array}{l}
p_{z}^{2}/2m_{e}-4M^{-1}\kappa_{0}^{2}\exp\left[ 2(\tan\lambda(p_{z})-%
\mathbf{C})\right] ,\ |\lambda(p_{z})|<\pi/2 \\ 
\varnothing,\ \ \lambda(p_{z})=\pm\pi/2%
\end{array}
\right\} \cup\mathbb{R}_{+}.
\end{equation*}

The complete set of orthonormalized generalized eigenfunctions of $\hat {H}%
_{\lambda(p_{z})}$ consists of functions $\Psi_{l,p_{z},E^{\bot}}(\mathbf{r}%
),\ l\neq l_{0},$ and$\ \Psi_{l_{0},p_{z},E^{\bot}}^{\lambda (p_{z})}(%
\mathbf{r)}$,%
\begin{align*}
& \Psi_{l,p_{z},E^{\bot}}(\mathbf{r})=\left( 8\pi^{2}\hbar/M\right) ^{-1/2}%
\mathrm{e}^{ip_{z}z/\hbar+i\epsilon_{q}(\phi_{0}-l)\varphi}J_{\varkappa_{l}}(%
\sqrt{ME^{\bot}}\rho), \\
& \Psi_{l_{0},p_{z},E^{\bot}}^{\lambda(p_{z})}(\mathbf{r)=}\left( 8\pi
^{2}\hbar\left( \tilde{\lambda}^{2}+\pi^{2}/4\right) /M\right) ^{-1/2}%
\mathrm{e}^{ip_{z}z/\hbar+i\epsilon_{q}\phi_{0}\varphi}\left[ \tilde{\lambda}%
J_{0}\left( \sqrt{ME^{\bot}}\rho\right) +\frac{\pi}{2}N_{0}(\sqrt{ME^{\bot}}%
\rho)\right] , \\
& \tilde{\lambda}=\tan\lambda(p_{z})-\mathbf{C}-\ln\left( \sqrt{ME^{\bot}}%
/2\kappa_{0}\right) ,
\end{align*}
and functions $\Psi_{l_{0},p_{z}}^{\lambda(p_{z})}(\mathbf{r})$,

\begin{align*}
& \Psi_{l_{0},p_{z}}^{\lambda(p_{z})}(\mathbf{r})=\frac{1}{2\pi\sqrt{\hbar}}%
\mathrm{e}^{ip_{z}z/\hbar+i\epsilon_{q}\phi_{0}\varphi}\left\{ 
\begin{array}{l}
\sqrt{2M^{2}\left\vert E_{\lambda(p_{z})}^{\bot\left( -\right) }\right\vert }%
K_{0}\left( \sqrt{M\left\vert E_{\lambda(p_{z})}^{\bot\left( -\right)
}\right\vert }\rho\right) \mathbf{,\ }\ |\lambda(p_{z})|<\pi/2 \\ 
0,\ \ \lambda(p_{z})=\pm\pi/2%
\end{array}
\right. , \\
& E_{\lambda(p_{z})}^{\bot\left( -\right)
}=-4M^{-1}\kappa_{0}^{2}\exp2(\tan\lambda(p_{z})-\mathbf{C}),
\end{align*}
such that%
\begin{align*}
& \check{H}\Psi_{l,p_{z},E^{\bot}}(\mathbf{r})=\left(
p_{z}^{2}/2m_{e}+E^{\bot}\right) \Psi_{l,p_{z},E^{\bot}}(\mathbf{r}),\
E^{\bot}\geq0, \\
& \check{H}\Psi_{l_{0},p_{z},E^{\bot}}^{\lambda(p_{z})}(\mathbf{r)=}\left(
p_{z}^{2}/2m_{e}+E^{\bot}\right)
\Psi_{l_{0},p_{z},E^{\bot}}^{\lambda(p_{z})}(\mathbf{r),\ }E^{\bot}\geq0, \\
& \check{H}\Psi_{l_{0},p_{z}}^{\lambda(p_{z})}(\mathbf{r})=\left(
p_{z}^{2}/2m_{e}+E_{\lambda(p_{z})}^{\bot\left( -\right) }\right) \Psi
_{l_{0},p_{z}}^{\lambda(p_{z})}(\mathbf{r})\ .
\end{align*}

The corresponding inversion formulas are%
\begin{align*}
& \Psi(\mathbf{r})=\int dp_{z}\left[ \sum_{l\in\mathbb{Z},\ l\neq0}\int
_{0}^{\infty}\Phi_{l,p_{z}}(E^{\bot})\Psi_{l,p_{z},E^{\bot}}(\mathbf{r)}%
dE^{\bot}+\int_{0}^{\infty}\Phi_{l_{0},p_{z}}(E^{\bot})\Psi_{l_{0},p_{z},E^{%
\bot}}^{\lambda(p_{z})}(\mathbf{r)}dE^{\bot}+\Phi_{l_{0},p_{z}}%
\Psi_{l_{0},p_{z}}^{\lambda(p_{z})}(\mathbf{r)}\right] , \\
& \Phi_{l,p_{z}}(E^{\bot})=\int\overline{\Psi_{l,p_{z},E^{\bot}}(\mathbf{r)}}%
\Psi(\mathbf{r)}d\mathbf{\mathbf{r},}\ l\neq l_{0}, \\
& \Phi_{l,p_{z}}(E^{\bot})=\int\overline{\Psi_{l_{0},p_{z},E^{\bot}}^{%
\lambda(p_{z})}(\mathbf{r)}}\Psi(\mathbf{r)}d\mathbf{\mathbf{r},}\
\Phi_{l_{0},p_{z}}=\int\overline{\Psi_{l_{0},p_{z}}^{\lambda(p_{z})}(\mathbf{%
r)}}\Psi(\mathbf{r)}d\mathbf{\mathbf{r},} \\
& \int\left| \Psi(\mathbf{r)}\right| ^{2}d\mathbf{r}=\int dp_{z}\left[
\sum_{l\in\mathbb{Z}}\int_{0}^{\infty}\left| \Phi_{l,p_{z}}(E^{\bot})\right|
^{2}dE^{\bot}+\left| \Phi_{l_{0},p_{z}}\right| ^{2}\right] ,\ \forall
\Psi\in L^{2}(\mathbb{R}^{3})\mathfrak{.}
\end{align*}

For the case of $\mu>0$, there is a family of s.a. three-dimensional
Hamiltonians $\hat{H}_{\{\lambda_{a}(p_{z})\}}$ parametrized by two
real-valued functions $\lambda_{a}(p_{z})\,(\lambda_{a}\in\mathbb{S}\left(
-\pi/2,\pi/2\right) ,\,a=0,-1,\ p_{z}\in\mathbb{R}$.

The spectrum of $\hat{H}_{\{\lambda_{a}(p_{z})\}}$ is 
\begin{align*}
\mathrm{spec}\hat{H}_{\{\lambda_{a}(p_{z})\}} & =\left\{ 
\begin{array}{c}
p_{z}^{2}/2m_{e}-4M^{-1}k_{0}^{2}|\tilde{\lambda}_{a}|^{-%
\varkappa_{a}^{-1}},\ \lambda_{a}(p_{z})\ \in(-\pi/2,0) \\ 
\varnothing,\ \lambda_{a}(p_{z})\notin(-\pi/2,0)%
\end{array}
\right\} \cup\mathbb{R}_{+}, \\
\varkappa_{a} & =|\mu+a|,\,\tilde{\lambda}_{a}=\Gamma(1-\varkappa_{a})%
\Gamma^{-1}(1+\varkappa_{a})\tan\lambda_{a}(p_{z}),\,\varkappa_{a}=|\mu+a|.
\end{align*}

The complete set of orthonormalized generalized eigenfunctions of $\hat {H}%
_{\{\lambda_{a}(p_{z})\}}$ consists of functions $\Psi_{l,p_{z},E^{\bot}}(%
\mathbf{r}),\ l\neq l_{a},$ and$\ \Psi_{l_{a},p_{z},E^{\bot}}^{\lambda
_{a}(p_{z})}(\mathbf{r)}$,%
\begin{align*}
& \Psi_{l,p_{z},E^{\bot}}(\mathbf{r})=\left( 8\pi^{2}\hbar/M\right) ^{-1/2}%
\mathrm{e}^{ip_{z}z/\hbar+i\epsilon_{q}(\phi_{0}-l)\varphi}J_{\varkappa_{l}}(%
\sqrt{ME^{\bot}}\rho), \\
& \Psi_{l_{a},p_{z},E^{\bot}}^{\lambda_{a}(p_{z})}(\mathbf{r)=}\left(
8\pi^{2}\hbar Q_{a}\right) ^{-1/2}\mathrm{e}^{ip_{z}z/\hbar+i\epsilon
_{q}(\phi_{0}-l_{a})\varphi}\left[ J_{\varkappa_{a}}\left( \sqrt{ME^{\bot}}%
\rho\right) +\left( \sqrt{ME^{\bot}}/2\kappa_{0}\right) ^{2\varkappa_{a}}%
\tilde{\lambda}_{a}\,J_{-\varkappa_{a}}\left( \sqrt{ME^{\bot}}\rho\right) %
\right] , \\
& Q_{a}=1+2\,\left( ME^{\perp}/4\right) ^{\varkappa_{a}}\tilde{\lambda}%
_{a}\cos(\pi\varkappa_{a})+\,\left( ME/4\right) ^{2\varkappa_{a}}\tilde{%
\lambda}_{a}^{2},
\end{align*}
and functions $\Psi_{l_{a},p_{z}}^{\lambda_{a}(p_{z})}(\mathbf{r})$,

\begin{align*}
& \Psi_{l_{a},p_{z}}^{\lambda_{a}(p_{z})}(\mathbf{r})=\left(
2\pi^{2}\hbar\right) ^{-1}\mathrm{e}^{ip_{z}z/\hbar+i(l_{a}+\epsilon_{q}\phi
_{0})\varphi}\left\{ 
\begin{array}{l}
\sqrt{\frac{M^{2}\left\vert E_{\lambda_{a}(p_{z})}^{\bot\left( -\right)
}\right\vert \sin(\pi\varkappa_{a})}{2\pi\varkappa_{a}}}K_{\varkappa_{a}}%
\left( \sqrt{M\left\vert E_{\lambda_{a}(p_{z})}^{\bot\left( -\right)
}\right\vert }\rho\right) \mathbf{,}\ \lambda_{a}(p_{z})\in(-\pi/2,0) \\ 
0,\ \lambda_{a}(p_{z})\notin(-\pi/2,0)%
\end{array}
\right. , \\
& E_{\lambda_{a}(p_{z})}^{\bot\left( -\right)
}=-4M^{-1}\kappa_{0}^{2}\exp2(\tan\lambda_{a}(p_{z})-\mathbf{C}),
\end{align*}
such that%
\begin{align*}
& \check{H}\Psi_{l,p_{z},E^{\bot}}(\mathbf{r})=\left(
p_{z}^{2}/2m_{e}+E^{\bot}\right) \Psi_{l,p_{z},E^{\bot}}(\mathbf{r}),\
E^{\bot}\geq0, \\
& \check{H}\Psi_{l_{a},p_{z},E^{\bot}}^{\lambda_{a}(p_{z})}(\mathbf{r)=}%
\left( p_{z}^{2}/2m_{e}+E^{\bot}\right)
\Psi_{l_{a},p_{z},E^{\bot}}^{\lambda_{a}(p_{z})}(\mathbf{r),\ }E^{\bot}\geq0,
\\
& \check{H}\Psi_{l_{a},p_{z}}^{\lambda_{a}(p_{z})}(\mathbf{r})=\left(
p_{z}^{2}/2m_{e}+E_{\lambda_{a}(p_{z})}^{\bot\left( -\right) }\right)
\Psi_{l_{a},p_{z}}^{\lambda_{a}(p_{z})}(\mathbf{r})\ .
\end{align*}

The corresponding inversion formulas are%
\begin{align*}
& \Psi(\mathbf{r})=\int dp_{z}\left[ \sum_{l\in\mathbb{Z},\ l\neq
l_{a}}\int_{0}^{\infty}\Phi_{l,p_{z}}(E^{\bot})\Psi_{l,p_{z},E^{\bot}}(%
\mathbf{r)}dE^{\bot}\right. \\
& +\left. \sum_{a}\int_{0}^{\infty}\Phi_{l_{a},p_{z}}(E^{\bot})\Psi
_{l_{a},p_{z},E^{\bot}}^{\lambda_{a}(p_{z})}(\mathbf{r)}dE^{\bot}+\Phi
_{l_{a},p_{z}}\Psi_{l_{a},p_{z}}^{\lambda_{a}(p_{z})}(\mathbf{r)}\right] , \\
& \Phi_{l,p_{z}}(E^{\bot})=\int d\mathbf{r}\overline{\Psi_{l,p_{z},E^{\bot}}(%
\mathbf{r)}}\Psi(\mathbf{r),\ }E^{\bot}\geq0,\ l\neq l_{a}, \\
& \Phi_{l_{a},p_{z}}(E^{\bot})=\int d\mathbf{r}\overline{%
\Psi_{l_{a},p_{z},E^{\bot}}^{\lambda_{a}(p_{z})}(\mathbf{r)}}\Psi(\mathbf{%
r),\ }E^{\bot }\geq0,\ \Phi_{l_{a},p_{z}}=\int d\mathbf{r}\overline{%
\Psi_{l_{a},p_{z}}^{\lambda_{a}(p_{z})}(\mathbf{r)}}\Psi(\mathbf{r),} \\
& \int d\mathbf{r}\left| \Psi(\mathbf{r)}\right| ^{2}=\int dp_{z}\left[
\sum_{l\in\mathbb{Z}}\int_{0}^{\infty}\left| \Phi_{l,p_{z}}(E^{\bot})\right|
^{2}dE^{\bot}+\sum_{a}\left| \Phi_{l_{a},p_{z}}\right| ^{2}\right] ,\ \
\forall\Psi\in L^{2}(\mathbb{R}^{3}).
\end{align*}

\subsubsection{S.a. nonrelativistic Hamiltonians with magnetic-solenoid field%
}

There is a family of s.a. three-dimensional Hamiltonians $\hat{H}%
_{\lambda_{\ast}(p_{z})}$ parametrized by real-valued functions $\lambda
_{\ast}(p_{z})\,(\lambda_{\ast}(p_{z})\in\mathbb{S}\left( -\pi/2,\pi
/2\right) $, $p_{z}\in\mathbb{R\,)}$, $\lambda_{\ast}$ are defined by eq. (%
\ref{9.2.33b}).

The spectrum of $\hat{H}_{\lambda_{\ast}(p_{z})}$ is 
\begin{equation*}
\mathrm{spec}\hat{H}_{\lambda_{\ast}(p_{z})}=\left\{
p_{z}^{2}/2m_{e}+E_{n}^{\bot\lambda_{\ast}(p_{z})},\ n\in\mathbb{Z}%
_{+}\right\} \cup \lbrack\gamma M^{-1},\infty),
\end{equation*}
where $E_{n}^{\bot\lambda_{\ast}(p_{z})}$ are defined by eqs. (\ref{9.2.34a}%
) and (\ref{9.2.34b}) with the substitution $\lambda_{\ast}\rightarrow
\lambda_{\ast}(p_{z})$.

The complete set of generalized orthonormalized eigenfunctions of $\hat {H}%
_{\lambda_{\ast}(p_{z})}$ consists of functions $\Psi_{p_{z},l,n}(\mathbf{r)}
$\textbf{,}$\mathbf{\ }l\neq l_{\ast}$, and $\Psi_{p_{z},l_{\ast
},,n}^{\lambda_{\ast}(p_{z})}(\mathbf{r)}$\textbf{,} $\mathbf{\ }n\in\mathbb{%
Z}_{+}$,\textbf{\ } 
\begin{equation}
\Psi_{p_{z},l,n}(\mathbf{r)}=\frac{1}{2\pi\sqrt{\hbar\rho}}\mathrm{e}%
^{ip_{z}z/\hbar+\epsilon(\phi_{0}-l)\varphi}U_{l,m(n,l)}^{\left( 1\right)
}(\rho),\ l\neq l_{\ast},  \label{9.2.38}
\end{equation}
where $l_{\ast}$ are defined by eq. (\ref{9.2.33b}), $m(n,l)$ is given by (%
\ref{9.2.33a}), and $U_{l,m}^{\left( 1\right) }(\rho)$ are given by eqs. (%
\ref{9.2.2a}),%
\begin{align*}
& \Psi_{p_{z},l_{0},n}^{\lambda(p_{z})}(\mathbf{r)}=\frac{1}{2\pi\sqrt {%
\hbar\rho}}\mathrm{e}^{ip_{z}z/\hbar+i\epsilon\phi_{0}\varphi}U_{\lambda
(p_{z}),n}^{\left( 3\right) }(\rho),\ \ \mu=0, \\
& \Psi_{p_{z},l_{a},n}^{\lambda_{a}(p_{z})}(\mathbf{\rho)}=\frac{1}{2\pi 
\sqrt{\hbar\rho}}\mathrm{e}^{ip_{z}z/\hbar+i\epsilon(\phi_{0}-l_{a})\varphi
}U_{\lambda_{a}(p_{z}),n}^{\left( 2\right) }(\rho),\;\mu>0,
\end{align*}
where $U_{\lambda(p_{z}),n}^{\left( 3\right) }(\rho)$ and $U_{\lambda
_{a}(p_{z}),n}^{\left( 2\right) }(\rho)$ are given by the respective eqs. (%
\ref{9.2.5}) and (\ref{9.2.4}) with the substitution $\lambda_{\ast
}\rightarrow\lambda_{\ast}(p_{z})$, such that%
\begin{align}
& \check{H}\Psi_{p_{z},l,n}(\mathbf{r)}=\left(
p_{z}^{2}/2m_{e}+E_{n,l}^{\bot}\right) \Psi_{p_{z},l,n}(\mathbf{r)},\ l\neq
l_{\ast },\   \label{9.2.36} \\
& \check{H}\Psi_{p_{z},l_{\ast},,n}^{\lambda_{\ast}(p_{z})}(\mathbf{r)}%
=\left( p_{z}^{2}/2m_{e}+E_{n}^{\bot\lambda_{\ast}(p_{z})}\right)
\Psi_{p_{z},l_{\ast},,n}^{\lambda_{\ast}(p_{z})}(\mathbf{r)},  \notag
\end{align}
where%
\begin{equation}
E_{n,l}^{\bot}=\gamma M^{-1}[1+2n+2\theta(l)\mu],\;l\leq n,\;l\neq l_{\ast
},\ n\in\mathbb{Z}_{+}\text{.}\   \label{9.2.37}
\end{equation}
We note that for $\lambda(p_{z})=\lambda_{a}(p_{z})=\pm\pi/2$, the energy
eigenvalues and corresponding eigenfunctions $\Psi_{p_{z},l,n}(\mathbf{r)}$
are given by eqs. (\ref{9.2.36}), (\ref{9.2.37}), and (\ref{9.2.38})
extended to all values of $l$.

The corresponding inversion formulas are%
\begin{align*}
& \Psi(\mathbf{r})=\int dp_{z}\sum_{n\in\mathbb{Z}_{+}}\left[ \sum _{l\in%
\mathbb{Z},\ l\neq l_{\ast}}\Phi_{p_{z},l,n}\Psi_{p_{z}l,m(n,l)}(\mathbf{r)+}%
\sum_{l=l_{\ast}}\Phi_{p_{z},l_{\ast},n}\Psi_{p_{z},l_{\ast},n}^{\lambda_{%
\ast}(p_{z})}(\mathbf{r)}\right] ,\  \\
& \Phi_{p_{z},l,n}=\int d\mathbf{r}\overline{\Psi_{p_{z},l,n}(\mathbf{r)}}%
\Psi(\mathbf{r),\ }l\neq l_{\ast},\ \Phi_{p_{z},l_{\ast},n}=\int d\mathbf{r}%
\overline{\Psi_{p_{z},l_{\ast},n}^{\lambda(p_{z})}(\mathbf{r)}}\Psi(\mathbf{%
r),} \\
& \int d\mathbf{r}\left| \Psi(\mathbf{r)}\right| ^{2}=\int dp_{z}\sum _{l\in%
\mathbb{Z}}\sum_{n\in\mathbb{Z}_{+}}\left| \Phi_{p_{z},l,n}\right| ^{2},\
\forall\Psi\in L^{2}(\mathbb{R}^{3}).
\end{align*}

\section{S.a. Dirac Hamiltonians with magnetic-solenoid field}

\subsection{Generalities}

In this section, we set $c=\hbar=1$. Written in the form of the Schr\"{o}%
dinger equation, the Dirac equation with the magnetic-solenoid field for a
relativistic particle of mass $m_{e}$, spin $1/2$,$\;$and charge $%
q=\epsilon_{q}e$ ( positron or electron) .is 
\begin{equation*}
i\frac{\partial\Psi\left( x\right) }{\partial t}=\check{H}\Psi\left(
x\right) ,\;x=\left( x^{0},\mathbf{r}\right) ,\ \mathbf{r}=\left( x^{k},\
k=1,2,3\right) ,\ x^{0}=t,
\end{equation*}
where $\Psi\left( x\right) =\{\psi_{\alpha}(x),\ \alpha=1,...,4\}$ is a
four-spinor and $\check{H}$ is the s.a. Dirac differential operation, the ''
formal Dirac Hamiltonian'', 
\begin{equation*}
\check{H}=\mathbf{\alpha}\left( \mathbf{\check{p}-\epsilon}_{q}e\mathbf{A}%
\right) +m_{e}\beta,
\end{equation*}
where\ the vector potential $\mathbf{A}$ is given by (\ref{9.1.11}), $\,%
\mathbf{\alpha}=\left( \gamma^{0}\gamma^{k},\ k=1,2,3\right) ,$ $%
\beta=\gamma^{0}$, and $\gamma^{\mu},\,\mu=0,1,2,3,$ are the Dirac $\gamma$
matrices.

The space of quantum states for a particle is the Hilbert space $\mathfrak{H}%
={\LARGE L}^{2}\left( \mathbb{R}^{3}\right) $ of square-integrable
four-spinors $\Psi(\mathbf{r})\,$with the scalar product

\begin{equation*}
\left( \Psi _{1},\Psi _{2}\right) =\int d\mathbf{r}\Psi _{1}^{+}(\mathbf{r}%
)\Psi _{2}(\mathbf{r}),\ d\mathbf{r}=dx^{1}dx^{2}dx^{3}=\rho d\rho d\varphi
dz.
\end{equation*}%
\ The Hilbert space $\mathfrak{H}$ can be presented as 
\begin{equation*}
\mathfrak{H}=\sideset{} {^{\,\lower 1mm\hbox{$\oplus$} }}\sum_{\alpha =1}^{4}%
\mathfrak{H}_{\alpha },\ \ \mathfrak{H}_{\alpha }=L^{2}(\mathbb{R}^{3}).
\end{equation*}%
Our first aim is to construct a quantum Dirac Hamiltonian that is a s.a.
operator associated with differential operation $\check{H}$ in this Hilbert
space. A construction is based on the known spatial symmetry in the problem%
\footnote{%
By the spatial symmetry, we mean the invariance under rotations around the
solenoid axis and under the translations along this axis.}, which allows
separating the cylindric coordinates $\rho ,\varphi ,$ and $z$, and on the
s.a. extension theory.

It is convenient to choose the following representation for\ $\gamma$
matrices: \ 

\begin{align*}
\gamma^{0} & =\left( 
\begin{array}{cc}
\sigma^{3} & 0 \\ 
0 & -\sigma^{3}%
\end{array}
\right) ,\ \gamma^{1}=\left( 
\begin{array}{cc}
i\sigma^{2} & 0 \\ 
0 & -i\sigma^{2}%
\end{array}
\right) ,\ \gamma^{2}=\left( 
\begin{array}{cc}
-i\sigma^{1} & 0 \\ 
0 & i\sigma^{1}%
\end{array}
\right) , \\
\ \gamma^{3} & =\left( 
\begin{array}{cc}
0 & I \\ 
-I & 0%
\end{array}
\right) ,\ \gamma^{5}=-i\gamma^{0}\gamma^{1}\gamma^{2}\gamma^{3}=-\left( 
\begin{array}{cc}
0 & I \\ 
I & 0%
\end{array}
\right) ,\ \Sigma^{3}=\left( 
\begin{array}{cc}
\sigma^{3} & 0 \\ 
0 & \sigma^{3}%
\end{array}
\right) .
\end{align*}
Written in the cylindric coordinates, the differential operation $\check{H}$
then becomes%
\begin{align*}
& \check{H}=\left( 
\begin{array}{ll}
Q\left[ \sigma^{3}\partial_{\rho}+\rho^{-1}\left( i\partial_{\varphi
}+\epsilon_{q}\tilde{\phi}\right) \right] +m_{e}\sigma^{3} & -i\sigma
^{3}\partial_{z} \\ 
-i\sigma^{3}\partial_{z} & Q\left[ \sigma^{3}\partial_{\rho}+\rho^{-1}\left(
i\partial_{\varphi}+\epsilon_{q}\tilde{\phi}\right) \right] -m_{e}\sigma^{3}%
\end{array}
\right) , \\
& \epsilon_{q}\tilde{\phi}=\epsilon(\phi_{0}+\mu+\gamma\rho^{2}/2),\;%
\epsilon_{B}\phi=\phi_{0}+\mu,\;\phi_{0}=[\epsilon_{B}\phi],\;0\leq
\mu<1,\;\gamma={e}|B{|}>0, \\
& Q=\sigma^{1}\sin\varphi-\sigma^{2}\cos\varphi=-i\mathrm{antidiag}\left( 
\mathrm{e}^{i\varphi},-\mathrm{e}^{-i\varphi}\right) =-i\left( 
\begin{array}{cc}
0 & -\mathrm{e}^{-i\varphi} \\ 
\mathrm{e}^{i\varphi} & 0%
\end{array}
\right) ,\ Q^{2}=1.
\end{align*}

The formal Dirac Hamiltonian commutes with the s.a. differential operations%
\begin{align}
& \check{p}_{z}=-i\partial _{z}\ ,\ \check{S}_{z}=\gamma ^{5}\left( \gamma
^{3}-m_{e}^{-1}\check{p}_{z}\right) ,  \notag \\
& \check{J}_{z}=\check{L}_{z}+\frac{1}{2}\Sigma ^{3}=-i\partial _{\varphi }+%
\frac{1}{2}\Sigma ^{3}=\mathrm{diag}\left( \check{\jmath}_{z},\check{\jmath}%
_{z}\right) ,\ \check{\jmath}_{z}=-i\partial _{\varphi }+\sigma ^{3}/2. 
\notag
\end{align}

We pass to\ the $p_{z}$ representation for four-spinors, $\Psi(\mathbf{r}%
)\rightarrow\tilde{\Psi}(p_{z},\mathbf{\rho}),$%
\begin{equation*}
\Psi(\mathbf{r})=\frac{1}{\sqrt{2\pi}}\int e^{ip_{z}z}\tilde{\Psi}(p_{z},%
\mathbf{\rho})dp_{z},\;\tilde{\Psi}(p_{z},\mathbf{\rho})=\frac{1}{\sqrt{2\pi}%
}\int e^{-ip_{z}z}\Psi(\mathbf{x})dz.
\end{equation*}
In this representation, the operation $\check{J}_{z}$ is the same, while the
differential operation $\check{H}$ and the operation $\check{S}_{z}$
respectively become%
\begin{align*}
& \check{H}\rightarrow\check{H}\left( p_{z}\right) =\left( 
\begin{array}{ll}
Q\left[ \sigma^{3}\partial_{\rho}+\rho^{-1}\left( i\partial_{\varphi
}+\epsilon_{q}\tilde{\phi}\right) \right] +m_{e}\sigma^{3} & \sigma^{3}p_{z}
\\ 
\sigma^{3}p_{z} & Q\left[ \sigma^{3}\partial_{\rho}+\rho^{-1}\left(
i\partial_{\varphi}+\epsilon_{q}\tilde{\phi}\right) \right] -m_{e}\sigma^{3}%
\end{array}
\right) , \\
& \check{S}_{z}\rightarrow\check{S}_{z}\left( p_{z}\right) =\left( 
\begin{array}{cc}
I & m_{e}^{-1}p_{z} \\ 
m_{e}^{-1}p_{z} & -I%
\end{array}
\right) .
\end{align*}

We decompose the four-spinor $\tilde{\Psi}(p_{z},\mathbf{\rho })$ for a
fixed $p_{z\text{ }}$into two orthogonal components that are the
eigenvectors for the spin matrix $\hat{S}_{z}(p_{z})$:

The space of four-spinors $\tilde{\Psi}(p_{z},\mathbf{\rho})$ with fixed%
\emph{\ }$p_{z}$ is the direct orthogonal sum of two eigenspaces of\emph{\ }$%
\hat {S}_{z}$\emph{\ }$(p_{z})$,%
\begin{equation*}
\tilde{\Psi}(p_{z},\mathbf{\rho})=\tilde{\Psi}_{1}(p_{z},\mathbf{\rho})+%
\tilde{\Psi}_{-1}(p_{z},\mathbf{\rho}),
\end{equation*}
where%
\begin{align*}
& \hat{S}_{z}\left( p_{z}\right) \tilde{\Psi}_{s}(p_{z},\mathbf{\rho })=s%
\frac{M}{m_{e}}\tilde{\Psi}_{s}(p_{z},\mathbf{\rho}),\ M=\sqrt{%
m_{e}^{2}+p_{z}^{2}},\ s=\pm1, \\
& \tilde{\Psi}_{1}(p_{z},\mathbf{\rho})=\left( \frac{M+m_{e}}{2M}\right)
^{1/2}\left( 
\begin{array}{c}
\chi_{1} \\ 
p_{z}\left( m_{e}+M\right) ^{-1}\chi_{1}%
\end{array}
\right) =\chi_{1}(p_{z},\mathbf{\rho})\otimes e_{1}(p_{z}),\  \\
& \tilde{\Psi}_{-1}(p_{z},\mathbf{\rho})=\left( \frac{M+m_{e}}{2M}\right)
^{1/2}\left( 
\begin{array}{c}
-p_{z}\left( m_{e}+M\right) ^{-1}\chi_{-1} \\ 
\chi_{-1}%
\end{array}
\right) =\chi_{-1}(p_{z},\mathbf{\rho})\otimes e_{-1}(p_{z}),
\end{align*}
$e_{s}(p_{z}),s=\pm1,$ are two orthonormalized two-spinors,%
\begin{align}
& e_{1}(p_{z})=\left( 
\begin{array}{c}
(2M)^{-1/2}\left( m_{e}+M\right) ^{1/2} \\ 
p_{z}[2M\left( m_{e}+M\right) ]^{-1/2}%
\end{array}
\right) ,\ e_{-1}(p_{z})=-i\sigma^{2}e_{1}(p_{z}),  \label{9.3.1} \\
& \,e_{r}^{+}(p_{z})\,e_{s}^{+}(p_{z})=\delta_{rs}  \notag
\end{align}
and $\chi_{s}(p_{z},\mathbf{\rho})$ are some doublets. We thus obtain a
one-to-one correspondence between four-spinors $\Psi(\mathbf{r})$ and pairs
of doublets $\chi_{s}(p_{z},\mathbf{\rho),}$%
\begin{equation*}
\Psi(\mathbf{r})\Longleftrightarrow\tilde{\Psi}_{s}(p_{z},\mathbf{\rho }%
)\Longleftrightarrow\chi_{s}(p_{z},\mathbf{\rho}),
\end{equation*}
such that $||\Psi||^{2}=\sum_{s}||\chi_{s}||^{2}=\sum_{s}\int dp_{z}d\mathbf{%
\rho\,}\chi_{s}^{+}(p_{z},\mathbf{\rho})\chi_{s}(p_{z},\mathbf{\rho })$.

The differential operations $\check{H}$ and $\check{J}_{z}$ induce the
differential operations $\mathbf{\check{h}}$ and $\check{\jmath}_{z}$ in the
space of doublets $\chi_{s}(p_{z},\mathbf{\rho})$:%
\begin{align*}
& \check{H}\left( p_{z}\right) \tilde{\Psi}_{s}=\mathbf{\check{h}}\left(
s,p_{z}\right) \chi_{s}\otimes e_{s},\ \ \check{J}_{z}\left( p_{z}\right) 
\tilde{\Psi}_{s}=\check{\jmath}_{z}\chi_{s}\otimes e_{s}\ , \\
& \mathbf{\check{h}}\left( s,p_{z}\right) =Q\left[ \sigma^{3}\partial_{%
\rho}+\rho^{-1}\left( i\partial_{\varphi}+\epsilon_{q}\tilde{\phi }\right) %
\right] +sM\sigma^{3}\ .
\end{align*}

The s.a. operator $\hat{\jmath}_{z}\;$associated with the differential
operation $\check{\jmath}_{z}$ has a discrete spectrum, its eigenvalues are
all half-integers labelled here by integers $l$ as $\epsilon(\phi_{0}-l+1/2)$%
. It is convenient to represent vectors $\xi_{l}(\varphi)\equiv%
\xi_{l}(p_{z},\rho,\varphi)$ of the corresponding eigenspaces,%
\begin{equation*}
\hat{\jmath}_{z}\xi_{l}(\varphi)=[\epsilon(\phi_{0}-l+1/2)]\xi_{l}(\varphi),%
\ l\in\mathbb{Z}\ ,
\end{equation*}
as%
\begin{align}
\xi_{l}(\varphi) & =(2\pi)^{-1/2}\mathrm{e}^{i[\epsilon(\phi_{0}-l+1/2)-%
\sigma^{3}/2]\varphi}\vartheta_{l}=S_{l}(\varphi)\frac{1}{\sqrt {2\pi\rho}}%
F(l,p_{z},\rho),  \notag \\
S_{l}(\varphi) & =\mathrm{e}^{i\epsilon(\phi_{0}-l+1/2)\varphi }\mathrm{%
antidiag}\left( i\mathrm{e}^{i\varphi/2},-\mathrm{e}^{-i\varphi /2}\right)
,\,S_{l}^{+}(\varphi)\,S_{l}(\varphi)=I,  \label{9.3.2}
\end{align}
where $\vartheta_{l}=$ $\vartheta_{l}(p_{z},\rho)$ and $F(l,p_{z},\rho)$ are
arbitrary doublets independent of $\varphi$.

The space of each of doublets $\chi_{s}(p_{z},\mathbf{\rho})$ is a direct
orthogonal sum of the eigenspaces of the operator $\hat{\jmath}_{z}$, which
means that the doublets allows the representations 
\begin{equation*}
\chi_{s}(p_{z},\mathbf{\rho})=\sum_{l\in\mathbb{Z}}\frac{1}{\sqrt{2\pi\rho}}%
S_{l}(\varphi)F(s,l,p_{z},\rho),
\end{equation*}
the factor $1/\sqrt{2\pi\rho}$ is introduced for further convenience\textrm{.%
}

The operation $\mathbf{\check{h}}\left( s,p_{z}\right) $ induces an
operation $\check{h}\left( s,l\right) $ (''radial Hamiltonian'' depending on
parameter $p_{z}$ as well) in the space of doublets $F$, 
\begin{align}
& \mathbf{\check{h}}\left( s,p_{z}\right) \chi_{s}=\sum_{l\in\mathbb{Z}}%
\frac{1}{\sqrt{2\pi\rho}}S_{l}(\varphi)\check{h}\left( s,l\right)
F(s,l,p_{z};\rho),  \notag \\
& \check{h}\left( s,l\right) =i\sigma^{2}\partial_{\rho}+\epsilon
(\gamma\rho/2+\rho^{-1}\varkappa_{l})\sigma^{1}-sM\sigma^{3},  \label{9.3.9}
\end{align}
where $\varkappa_{l}=l+\mu-1/2$.

In the Hilbert space $\mathbb{L}^{2}\left( \mathbb{R}_{+}\right) =L^{2}(%
\mathbb{R}_{+})\oplus L^{2}(\mathbb{R}_{+})$ of doublets $F(\rho )$ (with $%
p_{z}$ fixed), we define the initial symmetric radial Hamiltonian\textrm{\ }$%
\hat{h}\left( s,l\right) $ associated with the s.a. differential operation $%
\check{h}\left( s,l\right) $ by\ 
\begin{equation}
\hat{h}\left( s,l\right) =\left\{ 
\begin{array}{l}
D_{h\left( s,l\right) }=\mathfrak{D}\left( \mathbb{R}_{+}\right) =\mathcal{D}%
(\mathbb{R}_{+})\oplus \mathcal{D}(\mathbb{R}_{+}), \\ 
\hat{h}\left( s,l\right) F(\rho )=\check{h}\left( s,l\right) F(\rho ).%
\end{array}%
\right.  \label{9.3.9a}
\end{equation}

\subsection{Solutions of radial equations}

I. We first consider the homogeneous equation%
\begin{equation}
\left[ \check{h}\left( s,l\right) -W\right] F\left( \rho\right) =0
\label{9.3.10}
\end{equation}
and some of its useful solutions.

We let $f$ and $g$ denote the respective upper and lower components of
doublets $F$, $F=\left( f\diagup g\right) .$ Then eq.(\ref{9.3.10}) is
equivalent to the system of radial equations for the doublet components%
\begin{align}
& f^{\prime }-\epsilon (\gamma \rho /2+\rho ^{-1}\varkappa _{l})f+(W-sM)g=0,
\notag \\
& g^{\prime }+\epsilon (\gamma \rho /2+\rho ^{-1}\varkappa _{l})g-(W+sM)f=0,
\label{9.3.11}
\end{align}%
where the prime denotes the derivatives with respect to $\rho $.

We start with the case of $\epsilon=1$.

The system (\ref{9.3.11}) can be reduced to second-order differential
equations both for $f$ and $g.$ For example, we have the following system
equivalent to (\ref{9.3.11}):%
\begin{align}
& f^{\prime\prime}-\left[ \left( \gamma\rho/2\right) ^{2}+\frac {%
\varkappa_{l}(\varkappa_{l}-1)}{\rho^{2}}-w+\gamma\left( \varkappa_{l}+\frac{%
1}{2}\right) \right] f=0,  \notag \\
& g=\left( W-sM\right) ^{-1}\left[ -f^{\prime}+(\gamma\rho/2+\rho
^{-1}\varkappa_{l})f\right] ,\ w=W^{2}-M^{2}.  \label{9.3.12}
\end{align}
By the substitution%
\begin{equation*}
f(\rho)=z^{a/2}\mathrm{e}^{-z/2}p(z),\ z=\gamma\rho^{2}/2,\;a=1/2\pm
(\varkappa_{l}-1/2),
\end{equation*}
we reduce the first equation (\ref{9.3.12}) to the equation for $p(z)$ that
is the equation for confluent hypergeometric functions,%
\begin{align}
& z\partial_{z}^{2}p+(\beta-z)\partial_{z}p-\alpha p=0,\ \beta =a+1/2,\ 
\label{9.3.14A} \\
& \alpha=a/2+\varkappa_{l}/2+1/2-w/2\gamma.  \notag
\end{align}
Known solutions of eq. (\ref{9.3.14A}) allows obtaining solutions of eqs. (%
\ref{9.3.10}).

In what follows, we use the following solutions $F_{1}(\rho;s,W)$, $%
F_{2}(\rho;s,W)$, and $F_{3}(\rho;s,W)$ of equation (\ref{9.3.10}):%
\begin{align}
& F_{1}=\rho^{1/2-l-\mu}\mathrm{e}^{-z/2}\left( 
\begin{array}{l}
-\left( 2\beta_{1}\right) ^{-1}(W-sM)\rho\Phi(\alpha_{1}+1,\beta_{1}+1;z) \\ 
\Phi(\alpha_{1},\beta_{1};z)%
\end{array}
\right) ,  \notag \\
& F_{2}=\rho^{l+\mu-1/2}\mathrm{e}^{-z/2}\left( 
\begin{array}{l}
\Phi(\alpha_{2},\beta_{2};z) \\ 
\left( 2\beta_{2}\right) ^{-1}(W+sM)\rho\Phi(\alpha_{2},\beta_{2}+1;z)%
\end{array}
\right) ,  \notag \\
& F_{3}=\rho^{1/2-l-\mu}\mathrm{e}^{-z/2}\left( 
\begin{array}{l}
2^{-1}(W-sM)\rho\Psi(\alpha_{1}+1,\beta_{1}+1;z) \\ 
\Psi(\alpha_{1},\beta_{1};z)%
\end{array}
\right) =\omega_{2}F_{1}-\omega_{1}F_{2},{\LARGE \ }  \label{9.3.17}
\end{align}
where%
\begin{align*}
& \beta_{1}=1-l-\mu,\ \alpha_{1}=-w/2\gamma,\ \beta_{2}=l+\mu,\ \alpha
_{2}=l+\mu-w/2\gamma, \\
& \omega_{1}=\omega_{1}(s,W)=\frac{2\left( \gamma/2\right)
^{\beta_{2}}\Gamma(\beta_{1})}{(W+sM)\Gamma(\alpha_{1})},\
\omega_{2}=\omega_{2}(W)=\frac{\Gamma(\beta_{2})}{\Gamma(\alpha_{2})}.
\end{align*}
All the solutions $F_{1}$, $F_{2}$, and $F_{3}$ are real-entire in $W$.

The solutions (\ref{9.3.17}) have the following asymptotic behavior at the
origin and at infinity.

As $\rho\rightarrow0$, we have:%
\begin{align}
& F_{1}=\rho^{1/2-l-\mu}\left( 
\begin{array}{c}
-\left( 2\beta_{1}\right) ^{-1}(W-sM)\rho \\ 
1%
\end{array}
\right) \left( 1+O(\rho^{2})\right) ,  \notag \\
& F_{2}=\rho^{l+\mu-1/2}\left( 
\begin{array}{c}
1 \\ 
\left( 2\beta_{2}\right) ^{-1}(W+sM)\rho%
\end{array}
\right) (1+O(\rho^{2})),  \notag \\
& f_{3}=\frac{(W-sM)\Gamma(\beta_{1})}{2\left( \gamma/2\right)
^{\beta_{1}}\Gamma(\alpha_{1}+1)}\rho^{l+\mu-1/2}\times\left\{ 
\begin{array}{l}
\left( 1+O(\rho^{2})\right) ,\;l\leq-1 \\ 
\left( 1+O(\rho^{2-2\mu})\right) ,\;l=0,\;\mu>0 \\ 
\left( 1+O(\rho^{2}\ln\rho)\right) ,\;l=0,\;\mu=0%
\end{array}
\right. ,\   \notag \\
& g_{3}=\frac{\Gamma(\beta_{2})}{\Gamma(\alpha_{2})}\rho^{1/2-l-\mu}\times%
\left\{ 
\begin{array}{l}
\left( 1+O(\rho^{2})\right) ,\;l\geq1 \\ 
\left( 1+O(\rho^{2\mu})\right) ,\;l=0,\;\mu>0%
\end{array}
\right.  \label{9.3.17a}
\end{align}
where $F_{3}=\left( f_{3}\diagup g_{3}\right) .$

As $\rho\rightarrow\infty$, we have:%
\begin{align*}
& F_{1}=\frac{\left( \gamma/2\right) ^{\alpha_{1}-\beta_{1}}\Gamma
(\beta_{1})}{\Gamma(\alpha_{1})}\rho^{-\varkappa_{l}+2\alpha_{1}-2\beta_{1}}%
\mathrm{e}^{z/2}\left( \gamma\rho(W+sM)^{-1}\diagup1\right) \left(
1+O(\rho^{-2})\right) , \\
& F_{2}=\frac{\left( \gamma/2\right) ^{\alpha_{1}}\Gamma(\beta_{2})}{%
\Gamma(\alpha_{2})}\rho^{\varkappa_{l}+2\alpha_{1}}\mathrm{e}^{z/2}\left(
1\diagup\left( \gamma\rho\right) ^{-1}\left( W+sM\right) \right)
(1+O(\rho^{-2})), \\
& F_{3}=\left( \gamma/2\right) ^{-\alpha_{1}}\rho^{\varkappa_{l}-2\alpha_{1}}%
\mathrm{e}^{-z/2}\left( \left( \gamma\rho\right) ^{-1}\left( W-sM\right)
\diagup1\right) (1+O(\rho^{-2})).
\end{align*}

We define the Wronskian $\mathrm{Wr}(F,\tilde{F})$ of two doublets $F=\left(
f\diagup g\right) $ and $\tilde{F}=\left( \tilde{f}\diagup\tilde{g}\right) $
by%
\begin{equation*}
\mathrm{Wr}(F,\tilde{F})=f\tilde{g}-g\tilde{f}=iF\sigma^{2}\tilde{F}.
\end{equation*}
If $(\check{h}-W)F=(\check{h}-W)\tilde{F}=0,$ then $\mathrm{Wr}(F,\tilde {F}%
)=C=\mathrm{const}$. Solutions $F$ and $\tilde{F}$ are linearly independent
iff $C\neq0$. It is easy to see that $\mathrm{Wr}(F_{1},F_{2})=-1$.

If $\func{Im}W>0$, the solutions $F_{1},F_{2},$ and $F_{4}$ are pairwise
linearly independent,%
\begin{equation*}
\mathrm{Wr}(F_{1},F_{3})=\omega_{1}(W),\ \mathrm{Wr}(F_{2},F_{3})=\omega
_{2}(W).
\end{equation*}

Taking the asymptotics of the linearly independent solutions $F_{1}$ and $%
F_{3}$ into account, we obtain that there are no square integrable solutions
of eq.(\ref{9.3.10}) with $\func{Im}W$ $\neq0$ and $|l|\geq1$ or $l=0$, $%
\mu=0$. This implies that in these cases, the deficiency indices of $\hat{h}%
\left( s,l\right) $ are zero. In the case of $l=0$, $\mu>0$, the solution $%
F_{3}$ is square integrable, which implies that the deficiency indices of $%
\hat{h}\left( s,0\right) $ are equal to $(1,1\dot{)}$.

For any $l$ and $\mu $, the asymptotic behavior of any solution $F$ of eq. (%
\ref{9.3.10}) at the origin, as $\rho \rightarrow 0,$ is not more singular%
\textrm{\ }than $\rho ^{-|\varkappa _{l}|}$, $F(\rho )=O(\rho ^{-|\varkappa
_{l}|})$.

II. We now consider the inhomogeneous equation 
\begin{equation*}
(\check{h}\left( s,l\right) -W)F(\rho )=\Psi (\rho ),\ \forall \Psi \in 
\mathbb{L}^{2}(\mathbb{R}_{+}).
\end{equation*}%
Its general solution allows the representations 
\begin{align}
& F(\rho )=c_{1}F_{d}(\rho ;W)+c_{2}F_{3}(\rho ;W)  \notag \\
& +\omega _{d}^{-1}\left[ F_{d}(\rho ;W)\int_{\rho }^{\infty }F_{3}(r;W)\Psi
(r)dr+F_{3}(\rho ;W)\int_{0}^{\rho }F_{d}(r;W)\Psi (r)dr\right] ,  \notag \\
& \omega _{d}=\mathrm{Wr}(F_{d},F_{3\,}),\ d=1,2,\;l\leq 0\;\mathrm{for}%
\;d=1,\;l\geq 1\;\mathrm{for}\;d=2.  \label{9.3.25}
\end{align}

A simple estimate of the integral terms in the r.h.s. of (\ref{9.3.25})
using the Cauchy-Bunyakovskii inequality shows that they are bounded as $%
\rho\rightarrow\infty$. It follows that $F\in\mathbb{L}^{2}(\mathbb{R}_{+})$
implies $c_{1}=0$.

For $|\varkappa_{l}|\geq1/2$, an evaluation shows\ that as $\rho\rightarrow0$%
, the integral terms are of the order of $O(\rho^{1/2})$ (up to the factor $%
\ln\rho$ for $|\varkappa_{l}|=1/2$) . In this case, $F\in\mathbb{L}^{2}(%
\mathbb{R}_{+})$ implies $c_{2}=0$, and we find 
\begin{equation}
F(\rho)=\omega_{d}^{-1}\left[ F_{d}(\rho;W)\int_{\rho}^{\infty}F_{3}(r;W)%
\Psi(r)dr+F_{3}(\rho;W)\int_{0}^{\rho}F_{d}(r;W)\Psi(r)dr\right] .
\label{9.3.26}
\end{equation}

For $|\varkappa_{l}|\leq1/2$, the doublet $F_{3}(\rho;W)$ is
square-integrable, and a solution $F(\rho)\in\mathbb{L}^{2}(\mathbb{R}_{+})$
allows the representation 
\begin{align}
& F(\rho)=b\omega_{1}^{-1}F_{1}(\rho;W)+c_{2}F_{3}(\rho;W)  \notag \\
& +\omega_{1}^{-1}\left[ F_{3}(\rho;W)\int_{0}^{\rho}F_{1}(r;W)\Psi
(r)dr-F_{1}(\rho;W)\int_{0}^{\rho}F_{3}(r;W)\Psi(r)dr\right] ,  \notag \\
& \,F(\rho)=b\omega_{1}^{-1}F_{1}(\rho;W)+c_{2}F_{3}(\rho;W)+O(\rho
^{1/2}),\;\rho\rightarrow0,  \label{9.3.27}
\end{align}
where%
\begin{equation*}
b=\int_{0}^{\infty}F_{3}(r;W)\Psi(r)dr\ .
\end{equation*}

We use representations. (\ref{9.3.25})-(\ref{9.3.27}) to determine the Green
functions for s.a. radial Hamiltonians.

\subsection{S.a. radial Hamiltonians}

\subsubsection{Generalities}

We proceed to constructing s.a. radial Hamiltonians $\hat{h}_{\mathfrak{e}%
}\left( s,l\right) $ in the Hilbert space $\mathbb{L}^{2}(\mathbb{R}_{+})$
of doublets as s.a. extensions of the initial symmetric radial operators $%
\hat{h}\left( s,l\right) $ (\ref{9.3.9a}) associated with the differential
operations $\check{h}\left( s,l\right) \;$(\ref{9.3.9})\emph{\ }and analyze
the corresponding spectral problems.

The action of all of the following operators associated with the
differential operations $\check{h}\left( s,l\right) $ is given by $\check{h}%
\left( s,l\right) $, therefore we cite only their domains.

We begin with the adjoint $\hat{h}^{+}\left( s,l\right) $ of the initial
symmetric operator $\hat{h}\left( s,l\right) $. Its domain $D_{h^{+}}$ is
the natural domain for $\check{h}\left( s,l\right) ,$%
\begin{equation*}
D_{h^{+}}=D_{\check{h}\left( s,l\right) }^{\ast}\left( \mathbb{R}_{+}\right)
=\left\{ F_{\ast}\left( \rho\right) :F_{\ast}\ \text{\textrm{a.c. in }}%
\mathbb{R}_{+},\ F_{\ast},\check{h}\left( s,l\right) F_{\ast}\in\mathbb{L}%
^{2}(\mathbb{R}_{+})\right\} .
\end{equation*}

The quadratic asymmetry form $\Delta_{h^{+}}\left( F_{\ast}\right) $ of $\ 
\hat{h}^{+}\left( s,l\right) $ is expressed in terms of the local quadratic
form%
\begin{equation*}
\left[ F_{\ast},F_{\ast}\right] \left( \rho\right) =\overline{g(\rho )}%
f(\rho)-\overline{f(\rho)}g(\rho),\ F_{\ast}=\left( f\diagup g\right) ,
\end{equation*}
as follows: 
\begin{equation}
\Delta_{h^{+}}\left( F_{\ast}\right) =\left( F_{\ast},\hat{h}^{+}F_{\ast
}\right) -\left( \hat{h}^{+}F_{\ast},F_{\ast}\right) =-\left. \left[
F_{\ast},F_{\ast}\right] (\rho)\right| _{0}^{\infty}\ .  \notag
\end{equation}

We can prove that $\lim_{\rho\rightarrow\infty}F_{\ast}\left( \rho\right) =0$
for any $F_{\ast}\in D_{\check{h}}^{\ast}\left( \mathbb{R}_{+}\right) $.
Indeed, because $F_{\ast}$ and $\check{h}\left( s,l\right) F_{\ast}$ are
square integrable at infinity, the combination%
\begin{equation*}
F_{\ast}^{\prime}-\left( \gamma\rho/2\right) \sigma^{3}F_{\ast}=-i\sigma
^{2}[\check{h}\left( s,l\right) F_{\ast}-(\varkappa_{l}/\rho)\sigma
^{1}F_{\ast}+sM\sigma^{3}F_{\ast}]
\end{equation*}
is also square-integrable at infinity. This implies that $f$ and $f^{\prime
}-\left( \gamma\rho/2\right) f,$ together with $g$ and $g^{\prime}+\left(
\gamma\rho/2\right) g,$ are square-integrable at infinity. We consider the
identity%
\begin{equation*}
|f(\rho)|^{2}=\int_{a}^{\rho}[\overline{\partial f(r)}f(r)+\overline {f(r)}%
\partial
f(r)]dr+\gamma\int_{a}^{\rho}r|f(r)|^{2}dr+|f(a)|^{2},\;\partial=\partial_{%
\rho}-\gamma\rho/2.
\end{equation*}
The r.h.s. of this identity has a limit (finite or infinite) as $\rho
\rightarrow\infty$. Therefore, $|f(\rho)|$ also has a limit as $\rho
\rightarrow\infty$. This limit has to be zero because $f(\rho)$ is
square-integrable at infinity. In the same way, we can verify that $%
g(\rho)\rightarrow0$ as $\rho\rightarrow\infty$.

To analyze the behavior of $F_{\ast }$ at the origin, we consider the
relation%
\begin{equation}
\Psi =\check{h}\left( s,l\right) F_{\ast },\;\Psi ,F_{\ast }\in \mathbb{L}%
^{2}(\mathbb{R}_{+}),  \label{9.3.44b}
\end{equation}%
or%
\begin{align*}
& f^{\prime }-\left( \gamma \rho /2+\rho ^{-1}\varkappa _{l}\right) f=-\chi
_{2},\;g^{\prime }+\left( \gamma \rho /2+\rho ^{-1}\varkappa _{l}\right)
g=\chi _{1}, \\
& \chi =\left( \chi _{1}\diagup \chi _{2}\right) =\Psi +sM\sigma ^{3}F_{\ast
}\in \mathbb{L}^{2}(\mathbb{R}_{+}),
\end{align*}%
as an equation for $F_{\ast }$ at a given $\chi $. The general solution of
these equations allows the representation%
\begin{align}
f(\rho )& =\rho ^{\varkappa _{l}}\mathrm{e}^{\gamma \rho ^{2}/4}\left[
c_{1}+\int_{\rho }^{\infty }r^{-\varkappa _{l}}e^{-\gamma r^{2}/4}\chi
_{2}(r)dr\right] ,  \notag \\
g(\rho )& =\rho ^{-\varkappa _{l}}\mathrm{e}^{-\gamma \rho ^{2}/4}\left[
c_{2}+\int_{\rho _{0}}^{\rho }r^{\varkappa _{l}}e^{\gamma r^{2}/4}\chi
_{1}(r)dr\right] .  \label{9.3.45}
\end{align}

It turns out that the asymptotic behavior of the functions $f$ and $g$ at
the origin crucially depends on the value of $l.$ Therefore, our exposition
is naturally divided into subsections related to the corresponding regions.
We distinguish three regions of $l$.

\subsubsection{First region: $\varkappa_{l}\leq-1/2\ $}

In this region, we have%
\begin{equation*}
l\leq\left\{ 
\begin{array}{l}
-1,\;\mu>0 \\ 
0,\;\mu=0%
\end{array}
\right. .
\end{equation*}

The representation (\ref{9.3.45}) allows estimating \ an asymptotic behavior
of doublets $F_{\ast}\in D_{\check{h}\left( s,l\right) }^{\ast}\left( 
\mathbb{R}_{+}\right) $ at the origin for the first region.

\begin{align*}
& f(\rho )=\rho ^{-|\varkappa _{l}|}\mathrm{e}^{\gamma \rho ^{2}/4}\left[ 
\tilde{c}_{1}-\int_{0}^{\rho }r^{|\varkappa _{l}|}\mathrm{e}^{-\gamma
r^{2}/4}\chi _{2}(r)dr\right] =\tilde{c}_{1}\rho ^{-|\varkappa _{l}|}+O(\rho
^{1/2}),\;\rho \rightarrow 0, \\
& \tilde{c}_{1}=c_{1}+\int_{0}^{\infty }r^{|\varkappa _{l}|}\mathrm{e}%
^{-\gamma r^{2}/4}\chi _{2}(r)dr.\ 
\end{align*}%
The condition $f\in L^{2}\left( \mathbb{R}_{+}\right) $ implies $\tilde{c}%
_{1}=0$, and therefore, $f(\rho )=O(\rho ^{1/2})$ as$\;\rho \rightarrow 0.$
As to $g(\rho )$, we find%
\begin{equation*}
g(\rho )=\left\{ 
\begin{array}{l}
O(\rho ^{1/2}),\;\varkappa _{l}<-1/2 \\ 
O(\rho ^{1/2}\ln \rho ),\;\varkappa _{l}=-1/2\;(l=0,\mu =0)\;%
\end{array}%
\right. ,\;\rho \rightarrow 0.
\end{equation*}%
We thus obtain that $F_{\ast }\left( \rho \right) \rightarrow 0$ as $\rho
\rightarrow 0$, which implies that $\Delta _{h^{+}}\left( F_{\ast }\right)
=0 $,$\ \forall F_{\ast }\in D_{\check{h}\left( s,l\right) }^{\ast }\left( 
\mathbb{R}_{+}\right) .\ $This means that the deficiency indices of each of
the symmetric operators $\hat{h}\left( s,l\right) $\ in the first region are
zero. Therefore, there exists only one s.a. extension $\hat{h}_{\mathfrak{e}%
}\left( s,l,p_{z}\right) =\hat{h}_{(1)}\left( s,l\right) =\hat{h}^{+}\left(
s,l\right) $ of $\hat{h}\left( s,l\right) $, i.e., a unique s.a. radial
Hamiltonian with given $s$ and $l$, its domain is the natural domain, $D_{%
\hat{h}_{(1)}\left( s,l\right) }=D_{\check{h}\left( s,l\right) }^{\ast
}\left( \mathbb{R}_{+}\right) .$

The representation (\ref{9.3.26}) with $d=1$ implies that the Green function
for the s.a. Hamiltonian $\hat{h}_{(1)}\left( s,l\right) $ is given by 
\begin{equation}
\ G(\rho ,\rho ^{\prime };W)=\frac{1}{\omega _{1}(W)}\left\{ 
\begin{array}{c}
F_{3}(\rho ;W)\otimes F_{1}(\rho ^{\prime };W),\;\rho >\rho ^{\prime } \\ 
F_{1}(\rho ;W)\otimes F_{3}(\rho ^{\prime };W),\;\rho <\rho ^{\prime }%
\end{array}%
\right. .  \label{9.3.36}
\end{equation}%
Unfortunately, we can not use representation (\ref{9.3.17}) for $F_{3}$ as a
sum of two terms directly for all values of $\mu $ because the both are
singular at $\mu =0$ (although the sum is not).

To cover the total range of $\mu,$ we use another representation for $F_{3}$%
..

We let $F_{dl}(\rho ;W)$ denote the functions $F_{d}(\rho ;W)$, $d=1,2,3$,
with a fixed $l$ and represent $F_{3l}$ as 
\begin{align*}
& F_{3l}=\omega _{1}[A_{1l}F_{1l}+F_{4l}],\;A_{1l}=A_{1l}(W)=\Omega
_{1}(W)-\Gamma (\beta _{2})P_{1l}(W), \\
& F_{4l}=F_{4l}(\rho ;W)=\Gamma (\beta _{2})P_{1l}(W)F_{1l}(\rho
;W)-F_{2l}(\rho ;W),\;\Omega _{1}(W)=\frac{\omega _{2}(W)}{\omega _{1}(W)},
\\
& P_{1l}(W)=\frac{(W+sM)(\gamma /2)^{|l|}\Gamma (\alpha _{1})}{2|l|!\Gamma
(\alpha _{1}-|l|)}.
\end{align*}%
Using the relation (see \cite{GraRy94})

\begin{equation}
\lim_{\beta\rightarrow-n}\frac{1}{\Gamma(\beta)}\Phi(\alpha,\beta ;x)=\frac{%
x^{n+1}\Gamma(\alpha+n+1)}{(n+1)!\Gamma(\alpha)}\Phi(\alpha +n+1,n+2;x),
\label{9.3.36b}
\end{equation}
we can verify that 
\begin{equation*}
\left. \Gamma^{-1}(\beta_{2})F_{2l}(\rho;W)\right| _{\mu\rightarrow
0}=\left. P_{1l}(W)F_{1l}(\rho;W)\right| _{\mu=0}\,.
\end{equation*}
Taking the latter relation into account, it is easy to see that in the first
region, $A_{1l}$ and $F_{4l}$ are finite for $\mu\geq0$, as well as $%
\omega_{1}$ and $F_{1l}$, and also that $P_{1l}(E)$ and $F_{4l}(\rho;E)$ are
real.

The Green function is then represented as 
\begin{align}
& \ G(\rho ,\rho ^{\prime };W)=A_{1l}(W)F_{1l}(\rho ;W)\otimes F_{1l}(\rho
^{\prime };W)  \notag \\
& \ +\left\{ 
\begin{array}{c}
F_{4l}(\rho ;W)\otimes F_{1l}(\rho ^{\prime };W),\;\rho >\rho ^{\prime } \\ 
F_{1l}(\rho ;W)\otimes F_{4l}(\rho ^{\prime };W),\;\rho <\rho ^{\prime }%
\end{array}%
\right. .  \label{9.3.36.c}
\end{align}%
for all $\mu \geq 0$.

We choose the guiding functional $\Phi_{1}(F;W)$ for the s.a. operator $\hat{%
h}_{(1)}\left( s,l\right) $\ in the form%
\begin{equation*}
\Phi_{1}(F;W)=\int_{0}^{\infty}F_{1}(\rho;W)F(\rho),\;F(\rho)\in \mathbb{D}%
=D_{r}(\mathbb{R}_{+})\cap D_{\hat{h}_{\left( 1\right) }\left( s,l\right) }.
\end{equation*}
It is easy to prove that the guiding functional is simple. It follows that
the spectrum of $\hat{h}_{(1)}\left( s,l\right) $ is simple.

Using \ representation (\ref{9.3.36.c}) for the Green function , we obtain
that the derivative $\sigma ^{\prime }(E)=[\pi F_{1}^{2}(\rho ;W)]^{-1}\func{%
Im}G(\rho ,\rho ;E+i0)$ of the spectral function is given by%
\begin{equation}
\sigma ^{\prime }(E)=\pi ^{-1}\func{Im}A_{1l}(E+i0).  \label{9.3.36d}
\end{equation}%
It is easy to prove that $\func{Im}A_{1l}(E+i0)$ is continuous in $\mu $ for 
$\mu \geq 0$, such that it is sufficient to find $\sigma ^{\prime }(E)$ only
for\ the case of $\mu >0$ where eq. (\ref{9.3.36d}) is more simple,%
\begin{equation}
\sigma ^{\prime }(E)=\left. \frac{(W+sM)\left( \gamma /2\right) ^{-\beta
_{2}}\Gamma (\beta _{2})}{2\pi \Gamma (\beta _{1})\Gamma (\alpha _{2})}%
\right\vert _{W=E}\left. \func{Im}\Gamma \left( \alpha _{1}\right)
\right\vert _{W=E+i0}\ .  \label{9.3.36a}
\end{equation}

It is easy to see that $\sigma^{\prime}(E)$ may differ from zero only at the
points $E_{k}$ defined by the relation $\alpha_{1}=-k$ ($\Gamma(\alpha
_{1})=\infty$), or $M^{2}-E_{k}^{2}=-2\gamma k,$ which yields%
\begin{equation*}
\ \ E_{k}=\pm M_{k}\ ,\ \ M_{k}=\sqrt{M^{2}+2\gamma k},\ M_{0}=M,\ \ k\in 
\mathbb{Z}_{+}\ .
\end{equation*}
The presence of the factor $(E+sM)$ in the r.h.s. of (\ref{9.3.36a}) implies
that the points $E=-sM=-sM_{0}$ do not belong to the spectrum of $\hat {h}%
_{(1)}\left( s,l\right) .$ In what follows it is convenient to change the
numeration of the spectrum points. Introduce an index $\mathfrak{n}(s)$:%
\begin{equation*}
\mathfrak{n}(s)\in\mathcal{Z}(s)=\left\{ n_{\sigma}(s)\right\} ,\;\sigma
=\pm,\;n_{+}(s)\in\left\{ 
\begin{array}{c}
\mathbb{Z}_{+},\;s=1 \\ 
\mathbb{N},\;s=-1%
\end{array}
\right. ,\;n_{-}(s)\in\left\{ 
\begin{array}{c}
-\mathbb{N},\;s=1 \\ 
\mathbb{Z}_{-},\;s=-1%
\end{array}
\right. .
\end{equation*}
Then we have%
\begin{equation*}
E_{k}=\pm M_{k}\;\Longrightarrow\;E_{\mathfrak{n}(s)}=\sigma M_{|\mathfrak{n}%
(s)|},\;\mathfrak{n}(s)\in\mathcal{Z}(s).
\end{equation*}
Finally, we obtain%
\begin{align*}
& \sigma^{\prime}(E)=\sum_{\mathfrak{n}(s)\in\mathcal{Z}(s)}Q_{\mathfrak{n}%
(s)}^{2}\delta(E-E_{\mathfrak{n}(s)}),\  \\
& Q_{\mathfrak{n}(s)}=\sqrt{\frac{\left( \gamma/2\right)
^{\beta_{1}}\Gamma\left( \beta_{1}+\left| \mathfrak{n}(s)\right| \right)
\left( 1+sME_{k}^{-1}\right) }{|\mathfrak{n}(s)|!\Gamma^{2}(\beta_{1})}}%
,\;\beta _{1}=1+|l|-\mu.
\end{align*}

Thus, the spectrum of the s.a. Hamiltonian $\hat{h}_{(1)}\left( s,l\right) $%
\ is simple and discrete,\ $\mathrm{spec}\hat{h}_{(1)}\left( s,l\right)
=\left\{ E_{\mathfrak{n}(s)},\ \mathfrak{n}(s)\in\mathcal{Z}(s)\right\} .$

The eigenvectors%
\begin{equation}
\overset{I}{U_{\mathfrak{n}(s)}}=\overset{I}{U_{\mathfrak{n}(s)}}\left(
s,l,p_{z};\rho\right) =Q_{\mathfrak{n}(s)}F_{1}(\rho;E_{\mathfrak{n}(s)}),\;%
\mathfrak{n}(s)\in\mathcal{Z}(s),  \label{9.3.37}
\end{equation}
of the Hamiltonian $\hat{h}_{(1)}\left( s,l\right) $ form a complete
orthonormalized set in the space $\mathbb{L}^{2}(\mathbb{R}_{+})$ of
doublets $F\left( \rho\right) $.

\subsubsection{Second region: $\varkappa_{l}\geq1/2$}

In this region, we have $l\geq 1.$

The representation (\ref{9.3.45}) yields the following estimates for an
asymptotic behavior of doublets $F_{\ast }\in D_{\check{h}\left( s,l\right)
}^{\ast }\left( \mathbb{R}_{+}\right) $ at the origin for the second region:%
\begin{equation*}
\left\{ 
\begin{array}{l}
f(\rho )=\left\{ 
\begin{array}{l}
O(\rho ^{1/2}),\;\varkappa _{l}>1/2 \\ 
O(\rho ^{1/2}\ln \rho ),\;\varkappa _{l}=1/2%
\end{array}%
\right. \\ 
g(\rho )=O(\rho ^{1/2})%
\end{array}%
\right. ,\;\rho \rightarrow 0.
\end{equation*}%
It follows that $F_{\ast }\left( \rho \right) \rightarrow 0$ as $\rho
\rightarrow 0$, which implies that $\Delta _{h^{+}}\left( F_{\ast }\right)
=0 $, $\forall F_{\ast }\in D_{\check{h}\left( s,l\right) }^{\ast }\left( 
\mathbb{R}_{+}\right) .$

$\ $This means that the deficiency indices of each of the symmetric
operators $\hat{h}\left( s,l\right) $\ in the second region are also zero.
Therefore, there exists only one s.a. extension $\hat{h}_{\mathfrak{e}%
}\left( s,l,p_{z}\right) =\hat{h}_{(2)}\left( s,l\right) =\hat{h}^{+}\left(
s,l\right) $ of $\hat{h}\left( s,l\right) $, i.e., a unique s.a. radial
Hamiltonian with given $s$ and $l$, its domain is the natural domain, $D_{%
\hat{h}_{(2)}\left( s,l\right) }=D_{\check{h}\left( s,l\right) }^{\ast
}\left( \mathbb{R}_{+}\right) .$

The representation (\ref{9.3.26}) with $d=2$ implies that the Green function
for the s.a. Hamiltonian $\hat{h}_{(2)}\left( s,l\right) $ is given by%
\begin{equation*}
G(\rho ,\rho ^{\prime };W)=\omega _{2}^{-1}(W)\left\{ 
\begin{array}{c}
F_{3l}(\rho ;W)\otimes F_{2l}(\rho ^{\prime };W),\;\rho >\rho ^{\prime } \\ 
F_{2l}(\rho ;W)\otimes F_{3l}(\rho ^{\prime };W),\;\rho <\rho ^{\prime }%
\end{array}%
\right. .
\end{equation*}%
Again, the representation (\ref{9.3.17}) for $F_{3}$ as a sum of two terms
is not applicable directly for $\mu =0$ We therefore use the following
representation for $F_{3}$:%
\begin{align*}
& F_{3l}=\omega _{2}[F_{5l}-A_{2l}F_{2l}],\;A_{2l}=A_{2l}(W)=\Omega
_{2}(W)+\Gamma (\beta _{1})P_{2l}(W), \\
& F_{5l}=F_{5l}(\rho ;W)=F_{1l}(\rho ;W)+\Gamma (\beta
_{1})P_{2l}(W)F_{2l}(\rho ;W),\;\Omega _{2}(W)=\frac{\omega _{1}(W)}{\omega
_{2}(W)}, \\
& P_{2l}(W)=\frac{(W-sM)(\gamma /2)^{l-1}\Gamma (\alpha _{1}+l)}{2(l-1)\dot{!%
}\,\Gamma (\alpha _{1}+1)}.
\end{align*}%
Using relation (\ref{9.3.36b}), we can verify that%
\begin{equation*}
\left. \Gamma ^{-1}(\beta _{1})F_{1l}(\rho ;W)\right\vert _{\mu \rightarrow
0}=-\left. P_{2l}(W)F_{2l}(\rho ;W)\right\vert _{\mu =0}\ .
\end{equation*}%
Taking the latter relation into account, it is easy to see that $A_{2l}$ and 
$F_{5l}$ are finite for $\mu \geq 0$, as well as $\omega _{2}$ and $F_{2l}$%
,\ in the second region, and $P_{2l}(E)$ and $F_{5l}(\rho ;E)$ are real.

The Green function is then represented as 
\begin{align}
& \ G(\rho ,\rho ^{\prime };W)=-A_{2l}(W)F_{2l}(\rho ;W)\otimes F_{2l}(\rho
^{\prime };W)  \notag \\
& \ +\left\{ 
\begin{array}{c}
F_{5l}(\rho ;W)\otimes F_{2l}(\rho ^{\prime };W),\;\rho >\rho ^{\prime } \\ 
F_{2l}(\rho ;W)\otimes F_{5l}(\rho ^{\prime };W),\;\rho <\rho ^{\prime }%
\end{array}%
\right. .  \label{9.3.39}
\end{align}%
for all $\mu \geq 0$.

We choose the guiding functional $\Phi_{2}(F;W)$ for the s.a. operator $\hat{%
h}_{(2)}\left( s,l\right) $\ in the form%
\begin{equation*}
\Phi_{2}(F;W)=\int_{0}^{\infty}F_{2}(\rho;W)F(\rho),\;F(\rho)\in \mathbb{D}%
=D_{r}(\mathbb{R}_{+})\cap D_{\hat{h}_{\left( 2\right) }\left( s,l\right) }.
\end{equation*}
It is easy to prove that the guiding functional is simple. It follows that
the spectrum of $\hat{h}_{(2)}\left( s,l\right) $ is simple.

Using representation (\ref{9.3.39}) for the Green function, we obtain that
the derivative $\sigma^{\prime}(E)$ of the spectral function is given by%
\begin{equation}
\sigma^{\prime}(E)=-\pi^{-1}\func{Im}A_{2l}(E+i0).  \label{9.3.39a}
\end{equation}
It is easy to prove that $\func{Im}A_{2l}(E+i0)$ is continuous in $\mu$\ for 
$\mu\geq0$, such that it is sufficient to find $\sigma^{\prime}(E)$\ only
for the case of $\mu>0$ where eq. (\ref{9.3.39a}) is more simple,%
\begin{equation*}
\sigma^{\prime}(E)=\left. \frac{(W-sM)\left( \gamma/2\right)
^{\beta_{2}}\Gamma(\beta_{1})}{\pi\gamma\Gamma(\beta_{2})\Gamma(1+\alpha_{1})%
}\right| _{W=E}\left. \func{Im}\Gamma\left( \alpha_{2}\right) \right|
_{W=E+i0}\ .
\end{equation*}

It is easy to see that $\sigma^{\prime}(E)$ may differ from zero only at the
points $E_{k}$ defined by the relation $\alpha_{2}=-k$ ($\Gamma\left(
\alpha_{2}\right) =\infty$) \ or%
\begin{equation*}
M^{2}-E_{k}^{2}+2\gamma(l+\mu)=-2\gamma k\ ,\ k\in\mathbb{Z}_{+}\ ,
\end{equation*}
which yields%
\begin{equation*}
E_{k}=\pm\sqrt{M^{2}+2\gamma(k+l+\mu)}=\pm M_{k+l+\mu},\ k\in\mathbb{Z}_{+}\
.
\end{equation*}
All the points $E_{k}$ are the spectrum points.

It is convenient to change indexing $k$ for $\mathfrak{n}(s)$,%
\begin{equation*}
E_{k}\Longrightarrow E_{\mathfrak{n}(s)}=\sigma M_{|\mathfrak{n}(s)|+\mu
},\;\left\{ \mathfrak{n}(s)\in\mathcal{Z}(s),\;|\mathfrak{n}(s)|\geq
l\right\} \;\left( n_{\sigma}(s)=\sigma(k+l),\ k\in\mathbb{Z}_{+}\ \right) .
\end{equation*}

We finally obtain that%
\begin{align*}
& \sigma^{\prime}(E)=\sum_{\mathfrak{n}\in\mathcal{Z},|\mathfrak{n}|\geq
l}Q_{\mathfrak{n}(s)}^{2}\delta(E-E_{\mathfrak{n}}),\  \\
& Q_{\mathfrak{n}(s)}=\sqrt{\frac{\left( \gamma/2\right) ^{l+\mu}\Gamma(|%
\mathfrak{n}(s)|+\mu)(1-sME_{\mathfrak{n}(s)}^{-1})}{(|\mathfrak{n}%
(s)|-l)!\Gamma^{2}(l+\mu)}}\ .
\end{align*}

So, the spectrum of the s.a. Hamiltonian $\hat{h}_{(2)}\left( s,l\right) \ $%
is\ simple and discrete, $\mathrm{spec}\hat{h}_{(2)}\left( s,l\right)
=\left\{ E_{\mathfrak{n}(s)},\ \mathfrak{n}(s)\in\mathcal{Z},\;|\mathfrak{n}%
(s)|\geq l\right\} .$

The eigenvectors 
\begin{equation}
\overset{II}{U_{\mathfrak{n}(s)}}=\overset{II}{U_{\mathfrak{n}(s)}}\left(
s,l,p_{z};\rho\right) =Q_{\mathfrak{n}(s)}F_{2}(\rho;E_{\mathfrak{n}(s)}),%
{\LARGE \ }\mathfrak{n}(s)\in\mathcal{Z}(s),  \label{9.3.40}
\end{equation}
of the Hamiltonian $\hat{h}_{(2)}\left( s,l\right) $ form a complete
orthonormalized set in the space $\mathbb{L}^{2}(\mathbb{R}_{+})$ of
doublets $F\left( \rho\right) $.

\subsubsection{Third region: $\left| \varkappa_{l}\right| <1/2$}

In this region, we have $l=l_{0}=0,$ and $\varkappa _{l}$ reduces to $%
\varkappa _{0}=\mu -1/2$, $\mu >0$.

The representation (\ref{9.3.45}) yields the following asymptotic behavior
of doublets $F_{\ast}\in D_{\check{h}\left( s,l_{0}\right) }^{\ast}\left( 
\mathbb{R}_{+}\right) $ at the origin: 
\begin{equation*}
\left\{ 
\begin{array}{c}
f(\rho)=c_{1}(m_{e}\rho)^{\varkappa_{0}} \\ 
g(\rho)=c_{2}(m_{e}\rho)^{-\varkappa_{0}}%
\end{array}
\right. +O(\rho^{1/2}),\;\rho\rightarrow0.
\end{equation*}
It follows that%
\begin{equation*}
\Delta_{h^{+}}\left( F_{\ast}\right) =\overline{c_{2}}c_{1}-\overline{c_{1}}%
c_{2},\ \ \forall F_{\ast}\in D_{\check{h}(s,l_{0})}^{\ast}\left( \mathbb{R}%
_{+}\right) .
\end{equation*}
Up to the factor $i$, the r.h.s. is a quadratic form in $c_{1}$ and $c_{2}$
with the inertia indices $(1,1)$, which implies that the deficiency indices
of the initial symmetric operator $\hat{h}\left( s,l_{0}\right) $\ are $%
(1,1) $. The additional asymptotic boundary conditions 
\begin{equation}
F(\rho)=c\left( 
\begin{array}{c}
(m_{e}\rho)^{\varkappa_{0}}\cos\lambda \\ 
(m_{e}\rho)^{-\varkappa_{0}}\sin\lambda%
\end{array}
\right) +O(\rho^{1/2}),\;\rho\rightarrow0,  \label{9.3.42}
\end{equation}
with a fixed $\lambda\in\mathbb{S}\left( -\pi/2,\pi/2\right) $ (note that $%
\lambda$ depend on $s$ and $p_{z}$, $\lambda=\lambda(s,p_{z})$) on doublets
define a maximum subspace in $D_{\check{h}(s,l_{0})}^{\ast}\left( \mathbb{R}%
_{+}\right) $ where $\Delta_{h^{+}}=0$. This subspace is the domain of a
s.a. operator that is a s.a. extension of $\hat{h}\left( s,l_{0}\right) $.

We thus obtain that there exists a one-parameter $U(1)$ family of s.a.
radial Hamiltonians $\hat{h}_{\mathfrak{e}}\left( s,l_{0},p_{z}\right) =\hat{%
h}_{\lambda }\left( s,l_{0}\right) $ parametrized by the real parameter $%
\lambda \in \mathbb{S}\left( -\pi /2,\pi /2\right) $. These Hamiltonians are
specified by asymptotic s.a. boundary conditions (\ref{9.3.42}), and their
domains are given by

\begin{equation}
D_{h_{\lambda}\left( s,l_{0}\right) }=\left\{ F(\rho):F(\rho)\in D_{\check{h}%
_{\lambda}\left( s,l_{0}\right) }^{\ast}\left( \mathbb{R}_{+}\right) ,\ F\ 
\mathrm{satisfies\ }\left( \text{\ref{9.3.42}}\right) \right\} .
\label{9.3.43}
\end{equation}

According to representation (\ref{9.3.27}), which certainly holds for the
doublets $F$ belonging to $D_{h_{\lambda}\left( s,l_{0}\right) },$ and (\ref%
{9.3.17a}), the asymptotic behavior of $F$\ at the origin is given by%
\begin{equation*}
F=\left( 
\begin{array}{l}
-c_{2}\omega_{1}\rho^{\varkappa_{0}} \\ 
\left( b\omega_{1}^{-1}+c_{2}\omega_{2}\right) \rho^{-\varkappa_{0}}%
\end{array}
\right) +O(\rho^{1/2}),\;\rho\rightarrow0.
\end{equation*}
On the other hand, $F$ satisfies boundary conditions (\ref{9.3.42}), whence
it follows that there must be 
\begin{equation}
c_{2}=-\frac{b\cos\lambda}{\omega_{1}\omega_{(\lambda)}},\ \ \omega
_{(\lambda)}=\omega_{2}\cos\lambda+m_{e}^{-2\varkappa_{0}}\omega_{1}\sin%
\lambda.  \label{9.3.44}
\end{equation}

Then representation (\ref{9.3.27}) for $F$ with $c_{2}$ given by (\ref%
{9.3.44}) implies that the Green function for the s.a. Hamiltonian $\hat{h}%
_{\lambda}\left( s,l_{0}\right) $ is given by

\begin{align}
& G(\rho,\rho^{\prime};W)=\Omega^{-1}(W)F_{(\lambda)}(\rho;W)\otimes
F_{(\lambda)}(\rho^{\prime};W)  \notag \\
& +\left\{ 
\begin{array}{c}
\tilde{F}_{(\lambda)}(\rho;W)\otimes
F_{(\lambda)}(\rho^{\prime};W),\;\rho>\rho^{\prime} \\ 
F_{(\lambda)}(\rho;W)\otimes\tilde{F}_{(\lambda)}(\rho^{\prime};W),\;\rho
<\rho^{\prime}%
\end{array}
\right. ,  \label{9.3.41}
\end{align}
where 
\begin{align*}
& F_{(\lambda)}(\rho;W)=m_{e}^{-\varkappa_{0}}F_{1}(\rho;W)\sin\lambda
+m_{e}^{\varkappa_{0}}F_{2}(\rho;W)\cos\lambda, \\
& \tilde{F}_{(\lambda)}(\rho;W)=m_{e}^{-\varkappa_{0}}F_{1}(\rho
;W)\cos\lambda-m_{e}^{\varkappa_{0}}F_{2}(\rho;W)\sin\lambda, \\
& \Omega(W)=\frac{\omega_{(\lambda)}(W)}{\tilde{\omega}_{(\lambda)}(W)},\
m_{e}^{\varkappa_{0}}F_{3}=\tilde{\omega}_{(\lambda)}F_{(\lambda
)}+\omega_{(\lambda)}\tilde{F}_{(\lambda)}, \\
& \tilde{\omega}_{(\lambda)}(W)=\omega_{2}\sin\lambda-m_{e}^{-2%
\varkappa_{0}}\omega_{1}\cos\lambda.
\end{align*}

We note that the doublets $F_{(\lambda )}(\rho ;W)$ and $\tilde{F}_{(\lambda
)}(\rho ;W)$ are real-entire in $W,$ and the doublet $F_{(\lambda )}(\rho
;W) $ satisfies asymptotic s.a. boundary conditions (\ref{9.3.42}).

We choose the guiding functional $\Phi_{\lambda}(F;W)$ for the s.a. operator 
$\hat{h}_{\lambda}\left( s,l_{0}\right) $\ in the form%
\begin{equation*}
\Phi_{(\lambda)}(F;W)=\int_{0}^{\infty}F_{(\lambda)}(\rho;W)F(\rho
),\;F(\rho)\in\mathbb{D}=D_{r}(\mathbb{R}_{+})\cap D_{\hat{h}%
_{\lambda}\left( s,l_{0}\right) }.
\end{equation*}
It is easy to prove that the guiding functional is simple. It follows that
the spectrum of $\hat{h}_{\lambda}\left( s,l\right) $ is simple.

Using the representation (\ref{9.3.41}) for the Green function, we obtain
that the derivative $\sigma^{\prime}(E)$ of the spectral function is given
by 
\begin{equation*}
\sigma^{\prime}(E)=\pi^{-1}\func{Im}\Omega^{-1}(E+i0).
\end{equation*}

Because $\Omega(E)$ is real, $\sigma^{\prime}(E)$ differs from zero only at
the zero points $E_{k}$ of the function $\Omega(E)$,\ $\Omega(E_{k})=0$, and
we find. 
\begin{align*}
& \sigma^{\prime}(E)=\sum_{k}Q_{k}^{2}\delta(E-E_{k}),\ \ Q_{k}=\left[
-\Omega^{\prime}(E)\right] ^{-1/2}\ , \\
\ & \mathrm{spec}\hat{h}_{\lambda}\left( s,l_{0}\right) =\left\{
E_{k}\right\} ,\;k\in\mathbb{Z}.
\end{align*}

The eigenvectors%
\begin{equation}
\overset{III}{U_{k}}=\overset{III}{U_{k}}\left( \lambda,s,p_{z};\rho\right)
=Q_{k}F_{(\lambda)}(\rho;E_{k}),\;k\in\mathbb{Z},  \label{9.3.41a}
\end{equation}
of the Hamiltonian $\hat{h}_{\lambda}\left( s,l_{0}\right) $ form a complete
and orthonormalized set in the space $\mathbb{L}^{2}(\mathbb{R}_{+})$ of
doublets $F\left( \rho\right) $.

For $\lambda=0$ and $\lambda=\pm$ $\pi/2$, we can evaluate the spectrum
explicitly.

a) Let $\lambda=0$, and we consider the s.a. Hamiltonian $\hat{h}_{0}\left(
s,l_{0}\right) $. We have%
\begin{equation*}
F_{(0)}(\rho;W)=m_{e}^{\varkappa_{0}}F_{2}(\rho;W),\ \Omega(W)=-\frac {%
m_{e}^{2\varkappa_{0}}\omega_{2}(W)}{\omega_{1}(W)}
\end{equation*}

and find

\begin{equation*}
\sigma^{\prime}(E)=-\left. 
\begin{array}{c}
2m_{e}^{-2\varkappa_{0}}\left( \gamma/2\right) ^{\beta_{2}}\Gamma(\beta _{1})
\\ 
\pi(W+sM)\Gamma(\alpha_{1})\Gamma(\beta_{2})%
\end{array}
\right| _{W=E}\left. \func{Im}\Gamma\left( \alpha_{2}\right) \right|
_{W=E+i0}\ .
\end{equation*}
As in the second region, $\sigma^{\prime}(E)$ differs from zero only at the
points $E_{k}$, defined by the relation $\alpha_{2}=-k$ ($\Gamma\left(
\alpha_{2}\right) =\infty$) \ or%
\begin{equation*}
M^{2}-E_{k}^{2}+2\gamma+\mu=-2\gamma k\ ,\ k\in\mathbb{Z}_{+}\ ,
\end{equation*}
which yields%
\begin{equation*}
E_{k}=\pm M_{k+\mu},\ k\in\mathbb{Z}_{+}\ .
\end{equation*}
All the points $E_{k}$\ are the spectrum points.

It is convenient to introduce an index $\mathfrak{n}$,%
\begin{equation*}
\mathfrak{n}\in\mathcal{Z}=\left\{ n_{\tilde{\sigma}}\in\tilde{\sigma }%
\mathbb{Z}_{+},\;\tilde{\sigma}=\pm\right\} ,
\end{equation*}
where $n_{+}=0$ and $n_{-}=0$ are considered different elements of the set $%
\mathcal{Z}$. Then we can write 
\begin{equation}
E_{k}\Longrightarrow E_{\mathfrak{n}}=\tilde{\sigma}M_{|\mathfrak{n}|+\mu
},\;\mathfrak{n}\in\mathcal{Z}.  \label{9.3.47}
\end{equation}

$.$We finally obtain that%
\begin{equation*}
\sigma^{\prime}(E)=\sum_{\mathfrak{n}\in\mathcal{Z}}m_{e}^{-2%
\varkappa_{0}}Q_{\mathfrak{n}}^{2}\delta(E-E_{\mathfrak{n}}),\ \ Q_{%
\mathfrak{n}}=\sqrt{\frac{\left( \gamma/2\right) ^{\mu}\Gamma(|\mathfrak{n}%
|+\mu)(1-sME_{\mathfrak{n}}^{-1})}{|\mathfrak{n}|!\Gamma^{2}(\mu)}}\ .
\end{equation*}

So, the spectrum of the s.a. Hamiltonians $\hat{h}_{0}\left( s,l_{0}\right) $
is simple and discrete, $\mathrm{spec}\hat{h}_{0}\left( s,l_{0}\right)
=\left\{ E_{\mathfrak{n}},\ \mathfrak{n}\in\mathcal{Z}\right\} .$ The
eigenvectors $\overset{III}{U_{\mathfrak{n}}}=\overset{III}{U_{\mathfrak{n}}}%
\left( 0,s,l,p_{z};\rho\right) =Q_{\mathfrak{n}}F_{2}(\rho;E_{\mathfrak{n}})$%
, $\mathfrak{n}\in\mathcal{Z}$ of the Hamiltonian $\hat{h}_{0}\left(
s,l_{0}\right) $ form a complete orthonormalized set in the space $\mathbb{L}%
^{2}(\mathbb{R}_{+})$ of doublets $F\left( \rho\right) $.

We note that the spectrum, spectral function and eigenfunctions of $\hat {h}%
_{0}\left( s,l_{0}\right) $ are obtained from the respective expressions for
the second region, $\varkappa_{l}\geq1/2$, by the substitution $l=0$. We
also note that for $\mu>1/2$, the function $F_{(0)}(\rho;W)=m_{e}^{%
\varkappa_{0}}F_{2}(\rho;W)$ is minimally singular at the origin among the
functions $F_{(\lambda)}(\rho;W)$, in fact, it is nonsingular.

b) Let $\lambda =\pi /2$, which is equivalent to $\lambda =-\pi /2$, and we
consider the s.a. Hamiltonian $\hat{h}_{\pi /2}\left( s,l_{0}\right) $. We
have%
\begin{equation*}
F_{(\pi /2)}(\rho ;W)=m_{e}^{-\varkappa _{0}}F_{1}(\rho ;W),\ \ \Omega (W)=%
\frac{m_{e}^{-2\varkappa _{0}}\omega _{1}(W)}{\omega _{2}(W)}
\end{equation*}%
and find%
\begin{equation}
\sigma ^{\prime }(E)=\left. \frac{m_{e}^{2\varkappa _{0}}\Gamma (\beta
_{2})(W+sM)}{2\pi \left( \gamma /2\right) ^{\beta _{2}}\Gamma (\beta
_{1})\Gamma (\alpha _{2})}\right\vert _{W=E}\left. \func{Im}\Gamma \left(
\alpha _{1}\right) \right\vert _{W=E+i0}.  \label{9.3.47a}
\end{equation}%
As in the first region, $\sigma ^{\prime }(E)$ differs from zero only at the
points $E_{k}$, defined by the relation $\alpha _{1}=-k$ ($\Gamma (\alpha
_{1})=\infty $), or%
\begin{equation*}
\frac{M^{2}-E_{k}^{2}}{2\gamma }=-k,\;E_{k}=\pm \sqrt{M^{2}+2\gamma k}%
,\;k\in \mathbb{Z}_{+}\ .
\end{equation*}%
The presence of the factor $(E+sM)$ in the r.h.s. of (\ref{9.3.47a}) implies
that the points $E=-sM=-sM_{0}$ do not belong to the spectrum of $\hat{h}%
_{\pi /2}\left( s,l_{0}\right) $. We change the indexing of the spectrum
points : 
\begin{equation}
E_{k}\Longrightarrow E_{\mathfrak{n}}=\sigma M_{|\mathfrak{n}|},\;\mathfrak{n%
}\in \mathcal{Z}(s).  \label{9.3.48}
\end{equation}%
\begin{equation*}
E_{k}=(\mathrm{sign}k)M_{\left\vert k\right\vert },\ |k|\geq 1\ \ E_{0}=sM\
;\ k\in \mathbb{Z\ }.
\end{equation*}

We finally obtain that%
\begin{equation*}
\sigma^{\prime}(E)=\sum_{k\in\mathbb{Z}}m_{e}^{2\varkappa_{0}}Q_{\mathfrak{n}%
}^{2}\delta(E-E_{\mathfrak{n}}),\ Q_{\mathfrak{n}}=\sqrt{\frac{\Gamma (|%
\mathfrak{n}|+1-\mu)\left( 1+sME_{\mathfrak{n}}^{-1}\right) }{\left(
\gamma/2\right) ^{\beta_{2}}|\mathfrak{n}|!\Gamma^{2}(1-\mu)}}\ .
\end{equation*}

So, the spectrum of the s.a. Hamiltonian $\hat{h}_{\pi/2}\left(
s,l_{0}\right) $\ is\ simple and discrete, $\mathrm{spec}\hat{h}_{\pi
/2}\left( s,l_{0}\right) =\left\{ E_{\mathfrak{n}},\ \mathfrak{n}\in\mathcal{%
Z}(s)\right\} .$

The eigenvectors $\overset{III}{U_{\mathfrak{n}}}=\overset{III}{U_{\mathfrak{%
n}}}\left( \pi/2,s,l_{0},p_{z};\rho\right) =Q_{\mathfrak{n}}F_{1}(\rho;E_{%
\mathfrak{n}})$, $\mathfrak{n}\in\mathcal{Z}(s)$, of the Hamiltonian $\hat{h}%
_{\pi/2}\left( s,l_{0}\right) $ form a complete orthonormalized set in the
space $\mathbb{L}^{2}(\mathbb{R}_{+})$ of doublets $F\left( \rho\right) $.

We note that the spectrum, spectral function, and eigenfunctions of $\hat {h}%
_{\pi/2}\left( s,l_{0}\right) $ can be obtained from the respective
expressions for\ the first region, $\varkappa_{l}\leq-1/2$, by the
substitution $l=0$. We also note that for $\mu<1/2$, the function $F_{(\pi
/2)}(\rho;W)=m_{e}^{-\varkappa_{0}}F_{1}(\rho;W)$ is minimally singular at
the origin among the functions $F_{(\lambda)}(\rho;W)$; in fact, it is
nonsingular.

\subsection{Complete spectrum and inversion formulas for Dirac spinors}

In the previous subsubsecs., we constructed all s.a. radial Hamiltonians $%
\hat{h}_{\mathfrak{e}}\left( s,l,p_{z}\right) $ as s.a. extensions of the
symmetric operators $\hat{h}\left( s,l,p_{z}\right) $ for any $s$, $l$, and $%
p_{z}$ and for any values of $\phi _{0}$, $\mu $, and $\gamma $. The total
s.a. operators $\hat{H}_{\mathfrak{e}}\ $associated with the Dirac
differential operation $\check{H}$ in the Hilbert space $\mathfrak{H}=%
{\LARGE L}^{2}\left( \mathbb{R}^{3}\right) $ of Dirac spinors are
constructed from the sets of $\hat{h}_{\mathfrak{e}}\left( s,l,p_{z}\right) $
by means of a procedure of \textquotedblright direct summation over $s$, $l$
and direct integration over $p_{z}$\textquotedblright . Each set of possible
s.a. radial Hamiltonians $\hat{h}_{\mathfrak{e}}\left( s,l,p_{z}\right) $
generates a spatially invariant\footnote{%
I.e., invariant under rotations aroud the $z$ axis and under translations
along the $z$ axis.} s.a. Hamiltonian $\hat{H}_{\mathfrak{e}}$. Namely, let $%
\mathbb{G}$ be the group of the above space transformations $S$: $\mathbf{r}%
\rightarrow S\mathbf{r}$. This group is unitarily represented in $\mathfrak{H%
}$: if $S\in \mathbb{G}$, then the corresponding operator $U_{S}$ is defined
by 
\begin{equation*}
\left( U_{S}\,\psi \right) (\mathbf{r})=e^{-i\theta \Sigma ^{3}/2}\psi
(S^{-1}\mathbf{r}),\forall \psi \in \mathfrak{H,}
\end{equation*}%
where $\theta $ is the rotation angle of vector $\mathbf{\rho }$ around the $%
z$-axis. The operator $\hat{H}$ evidently commutes with $U_{S}$ for any $S$.
We search only for s.a. extensions $\hat{H}_{\mathfrak{e}}$ of $\hat{H}$
that also commutes with $U_{S}$ for any $S$. This condition is the explicit
form of the invariance, or symmetry, of a quantum Hamiltonian under the
space transformations As in classical mechanics, this symmetry allows
separating the cylindrical coordinates $\rho ,$ $\varphi ,$ and $z$ and
reducing the three-dimensional problem to a one-dimensional radial problem.
Let $V$ is the unitary operator defined by the relation 
\begin{equation*}
(Vf)(\rho ,\varphi ,z)=\frac{1}{2\pi \sqrt{\rho }}\int_{\mathbb{R}%
_{z}}dp_{z}\sum_{l\in \mathbb{Z}}\mathrm{e}^{ip_{z}z}\left[ S_{l}(\varphi
)F(s,l,p_{z},\rho )\right] \otimes e_{s}(p_{z}),
\end{equation*}%
where $S_{l}(\varphi )$ and\emph{\ }$e_{s}(p_{z})\;$are given by the
respective (\ref{9.3.1}) and (\ref{9.3.2}). Similarly to the considerations
in subsec. 2 and 3 of sec. 2, it is natural to expect that any s.a.
Hamiltonian $\hat{H}_{\mathfrak{e}}$ can be represented in the form of the
type%
\begin{equation*}
\hat{H}_{\mathfrak{e}}\mathbf{=}V\int_{\mathbb{R}_{z}}dp_{z}\sum_{s=\pm
1}\sum_{l\in \mathbb{Z}}\hat{h}_{\mathfrak{e}}(s,l,p_{z})V^{-1},
\end{equation*}%
where $\hat{h}_{\mathfrak{e}}(s,l,p_{z})$ for fixed $s,l,$ and $p_{z}$ is
s.a. extension of symmetric operator $\hat{h}(s,l,p_{z})$,%
\begin{equation*}
\hat{h}(s,l,p_{z})=\left\{ 
\begin{array}{l}
D_{h(s,l,p_{z})}=\mathcal{D}(\mathbb{R}_{+})\subset \mathbb{L}^{2}(\mathbb{R}%
_{+},d\rho ), \\ 
\hat{h}(s,l,p_{z})=\check{h}(s,l,p_{z})F(s,l,p_{z},\rho ),\ \forall
F(s,l,p_{z},\rho )\in D_{h(s,l,p_{z})},%
\end{array}%
\right.
\end{equation*}%
acting in the Hilbert space $\mathbb{L}^{2}(\mathbb{R}_{+},d\rho )$ of the
functions $f(\rho ,l,p_{z})$ with the scalar product $\left(
F(s,l,p_{z}),G(s,l,p_{z})\right) =\int_{\mathbb{R}_{+}}\overline{%
F(s,l,p_{z},\rho )}G(s,l,p_{z},\rho )d\rho $, $\check{h}(s,l,p_{z})$ is
given by eq. (\ref{9.3.9}). A correct expression for $\hat{H}_{\mathfrak{e}}$
is 
\begin{equation*}
\hat{H}_{\mathfrak{e}}\mathbf{=}V\int_{\mathbb{R}_{z}}^{\oplus }dp_{z}%
\sideset{}{^{\,\lower1mm\hbox{$\oplus$}}}\sum_{s=\pm 1}\sideset{}
{^{\,\lower 1mm\hbox{$\oplus$} }}\sum_{l\in \mathbb{Z}}\hat{h}_{\mathfrak{e}%
}(s,l,p_{z})V^{-1}.
\end{equation*}%
A detailed exposition of the procedure will be published in the short run.

The inversion formulas in the Hilbert space $\mathfrak{H}$ are
correspondingly obtained from the known radial inversion formulas by a
procedure of summation over $s$, $l$ and integration over $p_{z}$ . And we
now must consider the extension parameter $\lambda$ a function of $s$ and $%
p_{z}$, $\lambda =\lambda(s,p_{z})$. In what follows, $\int dp_{z}$ means $%
\int_{-\infty }^{\infty}dp_{z}$.

For $\mu=0$, there is a unique s.a. Dirac Hamiltonian $\hat{H}_{\mathfrak{e}%
}.$

The spectrum of $\hat{H}_{\mathfrak{e}}$ is%
\begin{equation*}
\mathrm{spec}\hat{H}_{\mathfrak{e}}=(-\infty,-m_{e}]\cup\lbrack
m_{e},\infty).
\end{equation*}

A complete set of generalized eigenfunctions of the s.a. Hamiltonian $\hat {H%
}_{\mathfrak{e}}$ is the set $\left\{ \Psi_{s,p_{z},\mathfrak{n}(s),l}(%
\mathbf{r}),\;\mathfrak{n}(s)\in\mathcal{Z}(s),\;l\leq|\mathfrak{n}%
(s)|\right\} $,%
\begin{align*}
& \Psi_{s,p_{z},\mathfrak{n}(s),l}(\mathbf{r})=\frac{1}{2\pi\sqrt{\rho}}%
\mathrm{e}^{ip_{z}z}S_{l}(\varphi)F_{\mathfrak{n}(s)}(s,l,p_{z};\rho)\otimes
e_{s}(p_{z}), \\
& F_{\mathfrak{n}(s)}(s,l,p_{z};\rho)=\left\{ 
\begin{array}{l}
\overset{I}{U_{\mathfrak{n}(s)}}\left( s,l,p_{z};\rho\right) ,\ l\leq0, \\ 
\overset{II}{U_{\mathfrak{n}(s)}}\left( s,l,p_{z};\rho\right) ,\ 1\leq l\leq|%
\mathfrak{n}(s)|,%
\end{array}
\right. ,
\end{align*}
the doublets $\overset{I}{U_{\mathfrak{n}(s)}}\left( s,l,p_{z};\rho\right) $
and $\overset{II}{U_{\mathfrak{n}(s)}}\left( s,l,p_{z};\rho\right) $ are
given by the respective (\ref{9.3.37}) and (\ref{9.3.40}), such that%
\begin{equation*}
\check{H}\Psi_{s,p_{z},\mathfrak{n}(s),l}(\mathbf{r})=E_{s,p_{z},\mathfrak{n}%
(s),l}\Psi_{s,p_{z},\mathfrak{n}(s),l}(\mathbf{r}),\ 
\end{equation*}
where%
\begin{equation*}
E_{s,p_{z},\mathfrak{n}(s),l}=\sigma\sqrt{m_{e}^{2}+p_{z}^{2}+2\gamma\left| 
\mathfrak{n}(s)\right| },\;\mathfrak{n}(s)\in\mathcal{Z}(s),\ l\leq |%
\mathfrak{n}(s)|.
\end{equation*}

The corresponding inversion formulas are%
\begin{align*}
& \Psi(\mathbf{r})=\int dp_{z}\sum_{s=\pm1}\sum_{\mathfrak{n}(s)\in \mathcal{%
Z}(s)}\sum_{l\leq|\mathfrak{n}(s)|}\Phi_{s,p_{z},\mathfrak{n}%
(s),l}\Psi_{s,p_{z},\mathfrak{n}(s),l}(\mathbf{r)}, \\
& \Phi_{s,p_{z},\mathfrak{n}(s),l}=\int\overline{\Psi_{s,p_{z},\mathfrak{n}%
(s),l}(\mathbf{r)}}\Psi(\mathbf{r)}d\mathbf{\mathbf{r},} \\
& \int\left| \Psi(\mathbf{r)}\right| ^{2}d\mathbf{r}=\int dp_{z}\sum
_{s=\pm1}\sum_{\mathfrak{n}(s)\in\mathcal{Z}(s)}\sum_{l\leq|\mathfrak{n}%
(s)|}|\Phi_{s,l,p_{z},n}|^{2},\ \forall\Psi\in{\LARGE L}^{2}\left( \mathbb{R}%
^{3}\right) .
\end{align*}

For $\mu>0$, there is a family of s.a. Dirac Hamiltonians $\hat{H}%
_{\{\lambda(s,p_{z})\}}$ parametrized by two real-valued functions $%
\lambda(s,p_{z})$, $\,\lambda\in\mathbb{S}\left( -\pi/2,\pi/2\right)
,\,s=\pm1$.

The spectrum of $\hat{H}_{\{\lambda(s,p_{z})\}}$ is 
\begin{equation*}
\mathrm{spec}\hat{H}_{\{\lambda(s,p_{z})\}}=(-\infty,-m_{e}]\cup\lbrack
m_{e},\infty).
\end{equation*}

A complete set of generalized eigenfunctions of the s.a. Hamiltonian $\hat {H%
}_{\{\lambda(s,p_{z})\}}$ is the set $\left\{ \Psi_{s,p_{z},\mathfrak{n}%
(s),l}(\mathbf{r}),\;\mathfrak{n}(s)\in\mathcal{Z}(s),\;l\leq|\mathfrak{n}%
(s)|,\;l\neq0\right\} \cup\left\{ \Psi_{s,p_{z},k,l_{0}}^{\lambda(s,p_{z})}(%
\mathbf{r}),\ k\in\mathbb{Z}\right\} $,

\begin{align*}
& \Psi _{s,p_{z},\mathfrak{n}(s),l}(\mathbf{r})=\frac{1}{2\pi \sqrt{\rho }}%
\mathrm{e}^{ip_{z}z}S_{l}(\varphi )F_{\mathfrak{n}(s)}(s,l,p_{z};\rho
)\otimes e_{s}(p_{z}), \\
& F_{n}(s,l,p_{z};\rho )=\left\{ 
\begin{array}{l}
\overset{I}{U_{n}}\left( s,l,p_{z};\rho \right) ,\ l\leq -1,\  \\ 
\overset{II}{U_{n}}\left( s,l,p_{z};\rho \right) ,\ 1\leq l\leq |\mathfrak{n}%
(s)|,%
\end{array}%
\right. ; \\
& \Psi _{s,p_{z},k,l_{0}}^{\lambda (s,p_{z})}(\mathbf{r})=\frac{1}{2\pi 
\sqrt{\rho }}\mathrm{e}^{ip_{z}z}S_{l_{0}}(\varphi )\overset{III}{U_{k}}%
\left( \lambda (s,p_{z}),s,p_{z};\rho \right) \otimes e_{s}(p_{z}),
\end{align*}%
where $\overset{III}{U_{n}}\left( \lambda (s,p_{z}),s,p_{z};\rho \right) $
are given by (\ref{9.3.41a}) with the substitution $\lambda \rightarrow
\lambda (s,p_{z})$, such that%
\begin{equation*}
\check{H}\Psi _{s,p_{z},\mathfrak{n}(s),l}(\mathbf{r})=E_{s,p_{z},\mathfrak{n%
}(s),l}\Psi _{s,p_{z},\mathfrak{n}(s),l}(\mathbf{r}),
\end{equation*}%
where,%
\begin{align*}
& E_{s,p_{z},\mathfrak{n}(s),l}=\sigma \sqrt{m_{e}^{2}+p_{z}^{2}+2\gamma
\lbrack \left\vert \mathfrak{n}(s)\right\vert +\theta (l)]},\;l\leq |%
\mathfrak{n}(s)|,\;l\neq 0, \\
& \theta (l)=\left\{ 
\begin{array}{l}
0,\;l\leq 0\  \\ 
1,\;l\geq 1%
\end{array}%
\right. ,
\end{align*}%
and%
\begin{equation*}
\check{H}\Psi _{s,p_{z},k,l_{0}}^{\lambda \left( s,p_{z}\right) }(\mathbf{r}%
)=E_{s,p_{z},k,l_{0}}^{\lambda \left( s,p_{z}\right) }\Psi
_{s,p_{z},k,l_{0}}^{\lambda \left( s,p_{z}\right) }(\mathbf{r}),
\end{equation*}%
where%
\begin{align*}
& E_{s,p_{z},k,l_{0}}^{\lambda \left( s,p_{z}\right) }:\ \Omega \left(
\lambda ,E_{s,p_{z},k,l_{0}}^{\lambda }\right) =0,\;\Omega \left( \lambda
,W\right) =\frac{\cos \lambda +a\left( W\right) \sin \lambda }{\sin \lambda
-a\left( W\right) \cos \lambda }, \\
& a\left( W\right) =\frac{2m_{e}^{2\mu -1}\left( \gamma /2\right) ^{1-\mu
}\Gamma \left( \mu \right) \Gamma \left( 1-\mu -w/2\gamma \right) }{\left(
W+sM\right) \Gamma \left( 1-\mu \right) \Gamma \left( -w/2\gamma \right) }.
\end{align*}

We recall that for $\lambda\left( s,p_{z}\right) =0$ and $\lambda\left(
s,p_{z}\right) =\pm\pi/2,$ the eigenvalues $E_{s,l_{1},p_{z},n}^{\lambda%
\left( s,p_{z}\right) }$ can be found explicitly, see the respective eqs. (%
\ref{9.3.47}) and (\ref{9.3.48}).

The corresponding inversion formulas are%
\begin{align*}
& \Psi(\mathbf{r})=\int dp_{z}\sum_{s=\pm1}\left[ \sum_{\mathfrak{n}(s)\in
Z(s)}\sum_{l\leq|\mathfrak{n}(s)|,l\neq0}\Phi_{s,p_{z},\mathfrak{n}%
(s),l}\Psi_{s,p_{z},\mathfrak{n}(s),l}(\mathbf{r)}+\sum_{k}%
\Phi_{s,p_{z},k,l_{0}}\Psi_{s,p_{z},k,l_{0}}^{\lambda\left( s,p_{z}\right) }(%
\mathbf{r)}\right] , \\
& \Phi_{s,p_{z},\mathfrak{n}(s),l}=\int\overline{\Psi_{s,p_{z},\mathfrak{n}%
(s),l}(\mathbf{r)}}\Psi(\mathbf{r)}d\mathbf{\mathbf{r}},\ l\neq0,\ \Phi
_{s,p_{z},k,l_{0}}=\int\overline{\Psi_{s,p_{z},k,l_{0}}^{\lambda\left(
s,p_{z}\right) }(\mathbf{r)}}\Psi(\mathbf{r)}d\mathbf{\mathbf{r}}, \\
& \int\left| \Psi(\mathbf{r)}\right| ^{2}d\mathbf{r}=\int dp_{z}\sum
_{s=\pm1}\left[ \sum_{\mathfrak{n}(s)\in Z(s)}\sum_{l\leq|\mathfrak{n}%
(s)|,l\neq0}\left| \Phi_{s,p_{z},\mathfrak{n}(s),l}\right| ^{2}+\sum
_{k}\left| \Phi_{s,p_{z},k,l_{0}}\right| ^{2}\right] ,\ \forall\Psi \in%
{\LARGE L}^{2}\left( \mathbb{R}^{3}\right) ,
\end{align*}

\subsection{The case $\protect\epsilon =-1$.}

We let $\check{h}_{+}=\check{h}_{+}(s,l)$ and $\check{h}_{-}=\check{h}%
_{-}(s,l)$ denote the differential operation $\check{h}$ with the respective 
$\epsilon=1$ and $\epsilon=-1$. We then have%
\begin{align*}
& \check{h}_{-}(s,l)=i\sigma^{2}\partial_{\rho}-\left( \gamma\rho
/2+\rho^{-1}\varkappa_{l}\right) \sigma^{1}-sM\sigma^{3}= \\
& \,=i\sigma^{2}\left[ i\sigma^{2}\partial_{\rho}+(\gamma\rho/2+\rho
^{-1}\varkappa)\sigma^{1}+sM\sigma^{3}\right] \left( i\sigma^{2}\right)
^{+}=i\sigma^{2}\check{h}_{+}(-s,l)\left( i\sigma^{2}\right) ^{+}.
\end{align*}
It follows that solutions $F_{-}=F_{-}(s,l,E_{-}(s);\rho)$ of the equation $%
\left( \check{h}_{-}-E_{-}(s)\right) F_{-}=0$ are bijectively related to
solutions $F_{+}=F_{+}(s,l,,E_{+}(s);\rho)$ of the equation $\left( \check {h%
}_{+}-E_{+}(s)\right) F_{+}=0$ by 
\begin{equation*}
F_{-}(s,l,E_{-}(s);\rho)=i\sigma^{2}F_{+}(-s,l,E_{+}(-s);\rho),%
\;E_{-}(s)=E_{+}(-s).
\end{equation*}

\subparagraph{Acknowledgement}

Gitman is grateful to the Brazilian foundations FAPESP and CNPq for
permanent support; Tyutin thanks FAPESP and RFBR, grant 08-01-00737;
Smirnov, Tyutin, and Voronov thank LSS-1615.2008.2 for partial support.

\end{document}